\documentclass[a4paper,onecolumn,superscriptaddress,11pt]{quantumarticle}
\usepackage[utf8]{inputenc}
\usepackage[blocks,affil-it]{authblk}
\usepackage{geometry}
\usepackage{color,tikz,float}
\usetikzlibrary{positioning,shapes.geometric,decorations.markings,arrows,knots,hobby}
\usepackage{amsmath,amssymb,amsthm,braket,bbm,mathtools}
\usepackage{capt-of,hyperref}
\usepackage{caption}
\usepackage{subcaption}
\usepackage{color}
\usepackage{enumitem}
\usepackage{todonotes}
\usepackage{cleveref}

\crefname{defi}{Definition}{Definitions}
\Crefname{defi}{Definition}{Definitions}
\crefname{thm}{Theorem}{Theorems}
\Crefname{thm}{Theorem}{Theorems}
\crefname{lem}{Lemma}{Lemmas}
\Crefname{lem}{Lemma}{Lemmas}
\crefname{rem}{Remark}{Remarks}
\Crefname{rem}{Remark}{Remarks}
\crefname{prop}{Proposition}{Propositions}
\Crefname{prop}{Proposition}{Propositions}
\crefname{cor}{Corollary}{Corollaries}
\Crefname{cor}{Corollary}{Corollaries}
\crefname{section}{Section}{Sections}
\Crefname{section}{Section}{Sections}
\crefname{equation}{}{}
\Crefname{equation}{}{}
\crefname{figure}{Figure}{Figures}
\Crefname{figure}{Figure}{Figures}
\crefname{ex}{Example}{Examples}
\Crefname{ex}{Example}{Examples}
\crefname{appendix}{Appendix}{Appendices}
\Crefname{appendix}{Appendix}{Appendices}

\labelcrefformat{subequation}{#2(#1)#3}
\labelcrefrangeformat{subequation}{#3(#1)#4 to #5(#2)#6}

\theoremstyle{definition}
\newtheorem{defi}{Definition}[section]
\newtheorem{ex}[defi]{Example}
\theoremstyle{plain}

\newtheorem{rem}[defi]{Remark}

\DeclareMathOperator{\tr}{tr}

\newcommand{\myemph}[1]{\textit{#1}}
\newcommand{\obj}{\mathrm{ob}}
\newcommand{\myhom}{\mathrm{hom}}

\begin{document}

\title{From categories to anyons: a travelogue}

\author{Kerstin~Beer}
\email{kerstin.beer@itp.uni-hannover.de}
\author{Dmytro~Bondarenko}
\author{Alexander~Hahn}
\author{Maria~Kalabakov}
\author{Nicole~Knust}
\author{Laura~Niermann}
\author{Tobias~J.~Osborne}
\author{Christin~Schridde}
\author{Stefan~Seckmeyer}
\author{Deniz~E.~Stiegemann}
\author{Ramona~Wolf}
\email{ramona.wolf@itp.uni-hannover.de}
\affiliation{\normalsize Institut f{\"u}r Theoretische Physik, Appelstra{\ss}e 2, 30167 Hannover, Germany}

\date{\today}
\maketitle

\begin{abstract}
	In this paper we provide an overview of category theory, focussing on applications in physics. The route we follow is motivated by the final goal of understanding anyons and topological QFTs using category theory. This entails introducing modular tensor categories and fusion rings. Rather than providing an in-depth mathematical development we concentrate instead on presenting the ``highlights for a physicist''.
\end{abstract}

\def\fcobo {\filldraw[draw=black,fill=gray,fill opacity=0.8] (0,0) ellipse (1.25 and 0.5);
\filldraw[draw=black,fill=gray,fill opacity=0.5] (-1.25,0) -- (-1.25,-3.5) arc (180:360:1.25 and 0.5) -- (1.25,0) arc (0:180:1.25 and -0.5);
\draw [dashed] (-1.25,-3.5) arc (180:360:1.25 and -0.5);
\begin{scope}[shift={(4,0)}]
\filldraw [draw=black,fill=gray,fill opacity=0.5] (-1.25,-2) -- (-1.25,-3.5) arc (180:360:1.25 and 0.5) -- (1.25,-2) arc (0:180:1.25 and 1.5);
\draw [dashed] (-1.25,-3.5) arc (180:360:1.25 and -0.5);
\end{scope}}
\def\gcobo {\filldraw[draw=black,fill=gray,fill opacity=0.8] (-2,0) ellipse (1.25 and 0.5);
\filldraw[draw=black,fill=gray,fill opacity=0.8] (2,0) ellipse (1.25 and 0.5);
\filldraw[draw=black,fill=gray,fill opacity=0.5] (-3.25,0) to[in=90,out=270] (-1.25,-3.5) arc (180:360:1.25 and 0.5) to[in=270,out=90] (3.25,0) arc (0:180:1.25 and -0.5)-- (0.75,0) arc (0:180:0.75 and -1)-- (-0.75,0) arc (0:180:1.25 and -0.5);
\draw [dashed] (-1.25,-3.5) arc (180:360:1.25 and -0.5);}

\tableofcontents

\newpage

\section{Introduction}

This paper is an attempt to give an introduction to category theory for physicists. Therefore, our focus is on the techniques and concepts of category theory that could be useful in the everyday work of a physicist rather than a comprehensive development of the mathematical foundations. The origins of this work are notes from a seminar called ``Quantum Information, Complexity and Categories'', held by Tobias J.\ Osborne at the Leibniz Universit\"at Hannover during the summer term of 2018. The purpose of the seminar was to introduce students to category theory in general and to discuss the complete definition of modular tensor categories, illustrated throughout by physically relevant examples.

The most important physical example studied here is the theory of \myemph{anyons}, which are mathematically described by \emph{modular tensor categories}. Anyons are indistinguishable particles which usually only occur in two dimensions and whose properties are less restricted than those of fermions and bosons: The exchange of fermions or bosons results in a phase factor $e^{i\varphi}$ with $\varphi_{\text{Bos}}=0$ (bosons) or $\varphi_{\text{Ferm}}=\pi$ (fermions). However, in two dimensions indistinguishable particles can have more complicated statistics. It is possible to assign an \myemph{arbitrary} phase factor to the exchange of two anyons or even a unitary matrix. The fact that these particles can have any such statistics is the origin of the name \myemph{anyons} \cite{AnyonName}. In this paper we study one particular anyon model in detail, namely,  \myemph{Fibonacci anyons}. Such particles are understood to be a good effective model for the description of the physics of the fractional quantum Hall effect at a certain filling factor and, additionally, can be exploited to carry out topological quantum computation (\cite{GolCh,FQHFib,LectureNotes,FreedmanStuff}).

Since the definition of modular tensor categories contains multiple adjectives, we commence with an introduction to basic category theory in \cref{Ch2}. We state the main definitions and illustrate them in terms of several physically relevant examples, for instance, the category of Hilbert spaces. We also introduce a graphical visualisation of objects, morphisms and other relevant concepts in terms of string diagrams, which is also of considerable help for illustrating the examples mentioned in this paper. In \cref{Ch3}, we focus on monoidal and braided categories and introduce further terminology required for the definition of modular tensor categories. In \cref{Ch4} we finally state the complete definition of a modular tensor category and point out its connection to fusion rings. Furthermore, as mentioned above, we study the example of Fibonacci anyons in considerable detail. In the last chapter, \cref{Ch5}, we briefly discuss literature lying at the crossroads between category theory and physics.

One of the initial motivations for this paper was that in articles and textbooks on category theory in general, and modular tensor categories in particular, several different notations are employed. This makes it difficult for the beginner to compare definitions from different sources and to gain an overview of this theory. In this paper, we made an effort to stick to one notation and to highlight different notations which are also used throughout the literature. We followed several articles rather closely: We first studied the approach to categories presented in the excellent article \cite{Baez2011}, but also complemented this presentation with references to mathematical textbooks on the topic, for example \cite{Leinster} and \cite{Awodey}. Furthermore, for several definitions, and especially the chapter about modular tensor categories, we referred to \cite{Rowell04} and \cite{QGroup}.

\newpage

\section{Basic Category Theory}
\label{Ch2}

	A category is a mathematical construct built from \myemph{objects} and \myemph{morphisms}. Roughly speaking, these represent \myemph{things} and \myemph{ways to go between things}. This chapter follows the presentation of \cite{Baez2011} complemented by several examples described in~\cite{Leinster}.

\subsection{Definitions and Examples}

	We begin by giving the mathematical definition of a category, illustrating it with several examples from a diversity of fields.

		\begin{defi}[Category]
			A \textbf{category} $\mathcal{C}$ consists of a collection of \textbf{objects} $\obj(\mathcal{C})$ and, for each pair of objects $X,Y\in \obj(\mathcal{C})$, a set of \textbf{morphisms} $\myhom(X,Y)$ (also called a homset), and such that for every object $X\in\obj(\mathcal{C})$ there is an identity morphism
			\begin{equation}
			\mathrm{id}_X\in \myhom(X,X).\label{Def_Id-Morphism}
			\end{equation}	
			Moreover, there is a function called composition:
			\begin{align}
			\begin{split}
			\circ: \myhom(Y,Z)\times \myhom(X,Y) &\to \myhom(X,Z)\\
			(g,f) &\mapsto g\circ f,
			\end{split}\label{Def_Comp}
			\end{align}
			where \cref{Def_Id-Morphism} and \cref{Def_Comp} satisfy the following properties:
			\begin{itemize}
				\item associativity: for each $f\in \myhom(X,Y)$, $g\in \myhom(Y,Z)$ and $h\in \myhom(Z,W)$, it holds that $(h \circ g)\circ f=h\circ (g\circ f)$.
				\item identity laws: for $f\in \myhom(X,Y)$ the identity morphisms $\mathrm{id}_X$ and $\mathrm{id}_Y$ satisfy: $f\circ \mathrm{id}_X=\mathrm{id}_Y\circ f=f$.
			\end{itemize}
		\end{defi}

	Within the literature, different notations are used. Beside the one we choose in these notes, it is also common to write $X\in\mathcal{C}$ instead of $X\in \obj(\mathcal{C})$ and $f:X\to Y$ instead of $f\in \myhom(X,Y)$. Sometimes $\mathcal{C}(X,Y)$ instead of $\myhom(X,Y)$ is used to denote the set of all morphisms from $X$ to $Y$ in $\mathcal{C}$. In these notes we will stick to the notation of $\myhom(X,Y)$ denoting the set of morphisms from $X$ to $Y$ with $X,Y\in\obj(\mathcal{C})$ and $f:X\to Y$ denoting a morphism in this set.

	A practical visualisation of these things is provided by \emph{string diagrams}, which were originally described in~\cite{JoyalStreet} and widely used afterwards (see e.g.~\cite{QGroup,Kitaev2005}). The idea is that objects in the category label (directed) strings, e.g.\ an object $X$ is depicted here as a label
		
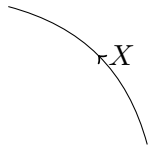
\begin{center}
			\begin{tikzpicture}[scale=2,decoration={markings,mark=at position 0.5 with {\arrow[scale=1,thick]{>}}}] 
				\node (end) at (0,1) {};
				\node (start) at (1,0) {};
				\draw[postaction={decorate}] (start) to [bend right] node [right,midway] {$X$} (end);
			\end{tikzpicture}
		\end{center}
	and morphisms $f: X\to Y$ are represented by black boxes with an input string of type $X$ and an output string of type $Y$:
	
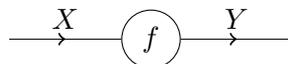
		\begin{figure}[H]
		\centering
		\begin{tikzpicture}[scale=2,decoration={markings,mark=at position 0.5 with {\arrow[scale=1,thick]{>}}}] 
			\node (start) at (0,0) {};
			\node (f) at (1,0) [circle,draw] {$f$};
			\node (end) at (2,0) {};
			\draw[postaction={decorate}] (start) -- (f) node[midway,above] {$X$};
			\draw[postaction={decorate}] (f) -- (end) node[midway,above] {$Y$};
		\end{tikzpicture}
		\end{figure}
	\noindent
	The composition of two morphisms $g:Y\to Z$ and $f:X\to Y$ is $g\circ f:X\to Z$, i.e.\ the composition of two functions is visualised by:
	
		
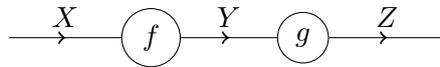
\begin{figure}[H]
		\centering
		\begin{tikzpicture}[scale=2,decoration={markings,mark=at position 0.5 with {\arrow[scale=1,thick]{>}}}] 
			\node (start) at (0,0) {};
			\node (f) at (1,0) [circle,draw] {$f$};
			\node (g) at (2,0) [circle,draw] {$g$};
			\node (end) at (3,0) {};
			\draw[postaction={decorate}] (start) -- (f) node[midway,above] {$X$};
			\draw[postaction={decorate}] (f) -- (g) node[midway,above] {$Y$};
			\draw[postaction={decorate}] (g) -- (end)node[midway,above] {$Z$};
		\end{tikzpicture}
		\end{figure}
	\noindent
	In this visualisation, associativity of morphisms is implicit since our notation for both $h\circ (g\circ f)$ and $(h\circ g)\circ f$ is
	
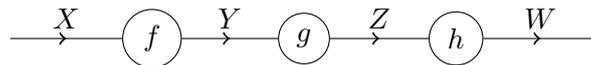
\begin{center}
		\begin{tikzpicture}[scale=2,decoration={markings,mark=at position 0.5 with {\arrow[scale=1,thick]{>}}}] 
		\node (start) at (0,0) {};
		\node (f) at (1,0) [circle,draw] {$f$};
		\node (g) at (2,0) [circle,draw] {$g$};
		\node (h) at (3,0) [circle,draw] {$h$};
		\node (end) at (4,0) {};
		\draw[postaction={decorate}] (start) -- (f) node[midway,above] {$X$};
		\draw[postaction={decorate}] (f) -- (g) node[midway,above] {$Y$};
		\draw[postaction={decorate}] (g) -- (h) node[midway,above] {$Z$};
		\draw[postaction={decorate}] (h) -- (end) node[midway,above] {$W$};
		\end{tikzpicture}
	\end{center}		
	Also, the left and right identity is implicit since we visualise $\mathbbm{1}_X:X\to X$ as
	\begin{center}
		\begin{tikzpicture}[scale=2,decoration={markings,mark=at position 0.5 with {\arrow[scale=1,thick]{>}}}] 
		\node (start) at (0,0) {};
		\node (end) at (1,0) {};
		\draw[postaction={decorate}] (start) -- (end) node[midway,above] {$X$};
		\end{tikzpicture}
	\end{center} 
	Note that here we have drawn arrows from the left to the right. In physical examples, we mostly use the notation where arrows go from bottom to top. However, whenever the directions appear in the notation used above it is purely for legibility and does not imply any direct physical interpretation.

	When studying category theory we often want to identify more structure in the collections of objects and morphisms to capture more specific features of an example or class of examples. The first such notion is that of \emph{isomorphisms}:
	
	\begin{defi}[Isomorphism]
		We say a morphism $f:X\to Y$ is an \textbf{isomorphism} if it has an inverse, i.e.~there exists a morphism $g:Y\to X$ such that $g\circ f = \mathbbm{1}_X$ and $f\circ g = \mathbbm{1}_Y$.
	\end{defi}

	For a better understanding of the given definitions, we consider some familiar examples.
	
	\begin{ex}[Sets]
		As a first example we consider the category of sets, denoted \textbf{Set}. In this category the objects are all possible sets, which implies that the collection of objects itself does \myemph{not} form a set. Instead of a set, these objects form a class (see also \cref{rem24}).
		
		Morphisms in this category are the homsets $\myhom(X,Y)$ given by all maps from $X$ to $Y$. Composition in this category then corresponds to ordinary composition of maps.
		
		For the category of sets, the identity morphism $\mathbbm{1}_X:X\rightarrow X$ exists for every $X\in \obj(\textbf{Set})$: It is $\mathbbm{1}(a)=a$ for all $a\in X$. Isomorphisms in this category are the bijections.
		\label{ex22}
	\end{ex}

	\begin{ex}[Groups]
		The category of groups \textbf{Grp} consists of  the class of all groups as objects and the group homomorphisms as morphisms. Isomorphisms in this category are the isomorphisms of groups.
		\label{ex:Grp}
	\end{ex}

	\begin{ex}[A group as a category]
		\label{ex:bullet}
		A group $G$ can be described as a category that has only one object and with all its morphisms being isomorphisms. To make this more clear, consider a category $\mathcal{G}$ with only one object, which is denoted $\obj(\mathcal{G})=\{\bullet\}$. Hence, $\mathcal{G}$ consists of a set (or class) $\myhom(\bullet,\bullet)$, an associative composition function
			\begin{equation*}
				\circ:\myhom(\bullet,\bullet)\times\myhom(\bullet,\bullet)\to\myhom(\bullet,\bullet)
			\end{equation*}
		and a two-sided unit $1_\bullet\in\myhom(\bullet,\bullet)$.	This corresponds to the group structure in the following way: The maps in $\mathcal{G}$, i.e.\ $\myhom(\bullet,\bullet)$, are the elements of the corresponding group $G$. The composition operation corresponds to the group operation and the identity morphism $1_\bullet$ to the unit object of the group. Saying that every element of $\myhom(\bullet,\bullet)$ is an isomorphism is the same as saying that every element of $\myhom(\bullet,\bullet)$ has an inverse with respect to $\circ$, which ensures that that every element of the group $G$ has an inverse. This example illustrates a key guiding principle in category theory, namely, one should try to formulate categories in such a way that the interesting structure is captured by the homsets $\myhom(X,Y)$ rather than in terms of complicated object classes. Here the objects are as simple as possible.
	\end{ex}

	\begin{ex}[Natural numbers]
		Next, we consider a category whose objects are the natural numbers, i.e.~$\obj(\mathcal{C})=\mathbb{N}$. We define a morphism between two objects by the set of the two numbers: $\myhom(j,k)=\{j,k\}$. The composition of two morphisms $f:j\rightarrow k$ and $g:k\rightarrow l$ is a morphism $h:j\rightarrow l$ so that $h=\{j,l\}$. The identity is defined for every $j\in \obj(\mathcal{C})$ by $\mathbbm{1}_j=\{j,j\}$.
	\end{ex}

	Another important example which departs from the picture of morphisms being ``just functions'' is the category of a partially ordered set. It emphasizes that homsets of morphisms may be far more abstract than functions:
	
	\begin{ex}[Partially ordered set]
	A partially ordered set $\mathcal{P}$ is a set $\{a,b,c,...\}$ with one relation $\le$. When $a\le b$, we say that $a$ is related to $b$. The relation has to fulfil the following properties:
		\begin{enumerate}
			\item reflexivity: $a\le a$.
			\item antisymmetry: if $a\le b$ and $b \le a$, then $a=b$.
			\item transitivity: if $a\le b$ and $b \le c$, then $a\le c$.
		\end{enumerate}
	
	A set with a partial order is a category in the following sense: The objects are the elements of the set. The morphisms are constructed in the following way: If $a \le b$ for $a,b\in\mathcal{P}$, $\myhom(a,b)$ is non-empty. If there is no relation between $a$ and $b$, $\myhom(a,b)=\emptyset$, e.g.
		\begin{align*}
			\obj(\mathcal{P})&=\{a,b,c|a\le b,a\le c\}\\
			\myhom(b,c)&=\emptyset.
		\end{align*}
		
	Morphisms are composable: If $\myhom(a,b)\neq\emptyset$ and $\myhom(b,c)\neq\emptyset$, $\myhom(a,c)\neq\emptyset$ due to transitivity of the relation.
	
	The existence of an identity morphism for every element is ensured by the condition of reflexivity: $a\le a\Rightarrow\myhom(a,a)\neq \emptyset$. There is only one isomorphism in this category, which is the identity. Since there is no relation $\geq$, all the other morphisms have no inverse and therefore are not isomorphisms.
	\label{ex:poset}
	\end{ex}

	The next example, namely Hilbert spaces, plays a major role in quantum mechanics. Hilbert spaces are used to describe quantum mechanical systems using states and operators.
	\begin{ex}[Hilbert spaces]
	The category \textbf{Hilb} consists of
	\begin{itemize}
		\item a collection of objects $X\in \obj(\textbf{Hilb})$, where $X$ is a finite dimensional Hilbert space and
		\item morphisms $f\in\myhom(X,Y)$, which are linear operators.
	\end{itemize}
	In order to avoid technical issues infinite dimensional Hilbert spaces may cause, we restrict this example to finite dimensional spaces.
	\label{ex:Hilb}
	\end{ex}

	In physics one often encounters categories in which the objects represent choices of \emph{space} (i.e., timelike slice), and the morphisms represent choices of \emph{spacetime}. The simplest example is the category \textbf{$n$Cob}:
	\begin{ex}[Cobordisms]
		In the category \textbf{$n$Cob}
		\begin{itemize}
			\item the objects $X\in \obj(\textbf{$n$Cob})$ are $(n-1) $-dimensional manifolds and
			\item the morphisms $f:X\to Y$ are $n$-dimensional manifolds, whose boundary is the disjoint union of the $(n-1)$-dimensional manifolds $X$ and $Y$. These manifolds are called \myemph{cobordisms}.
		\end{itemize}
		\Cref{subfig:fcmpg} depicts an example for the composition of two cobordisms.
		\begin{figure}[H]
			\centering
			\begin{subfigure}[b]{0.3\textwidth}
				\centering
				\begin{tikzpicture}[scale=0.5]
				\gcobo
				\end{tikzpicture}
				\begin{tikzpicture}[scale=1.1,decoration={markings,mark=at position 0.5 with {\arrow[scale=1,thick]{<}}}] 
				\node (start) at (0,0) {};
				\node (f) at (0,1) [circle,draw] {$f$};
				\node (g) at (0,2) [] {};
				\draw[postaction={decorate}] (f) -- (start) node[midway,left] {$X$};
				\draw[postaction={decorate}] (g) -- (f) node[midway,left] {$Y$};
				\end{tikzpicture}
				\caption{morphism $f\in\hom(X,Y)$}
			\end{subfigure}
			\hfill
			\begin{subfigure}[b]{0.3\textwidth}
				\centering
				\begin{tikzpicture}[scale=0.5]
				\fcobo
				\end{tikzpicture}
				\begin{tikzpicture}[scale=1.1,decoration={markings,mark=at position 0.5 with {\arrow[scale=1,thick]{<}}}] 
				\node (start) at (0,0) {};
				\node (f) at (0,1) [circle,draw] {$g$};
				\node (g) at (0,2) [] {};
				\draw[postaction={decorate}] (f) -- (start) node[midway,left] {$Y$};
				\draw[postaction={decorate}] (g) -- (f) node[midway,left] {$Z$};
				\end{tikzpicture}
				\caption{morphism $g\in\hom(Y,Z)$}
			\end{subfigure}
			\hfill
			\begin{subfigure}[b]{0.3\textwidth}
				\centering
				\begin{tikzpicture}[scale=0.5]
				\gcobo
				\begin{scope}[shift={(-2,3.5)}]
				\fcobo
				\end{scope}
				\end{tikzpicture}
				\begin{tikzpicture}[scale=1.1,decoration={markings,mark=at position 0.5 with {\arrow[scale=1,thick]{<}}}] 
				\node (start) at (0,0) {};
				\node (f) at (0,2) [circle,draw] {$gf$};
				\node (g) at (0,4) [] {};
				\draw[postaction={decorate}] (f) -- (start) node[midway,left] {$X$};
				\draw[postaction={decorate}] (g) -- (f) node[midway,left] {$Z$};
				\end{tikzpicture}
				\caption{composition $g\circ f$}
				\label{subfig:fcmpg}
			\end{subfigure}		
			\caption{Examples for cobordisms in the case $n=2$.}
			\label{fig:3Cobs}
		\end{figure}
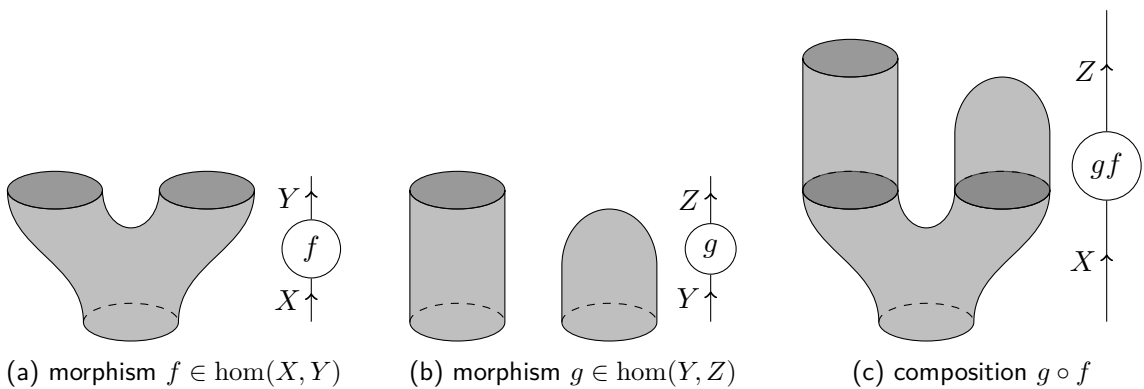
		\label{ex:nCob}
	\end{ex}
	
	Other  typical examples motivated by physics are categories where the objects represent \myemph{collections of particles} and the morphisms represent their \myemph{worldlines and interactions}. An example of this kind of category is the category of tangles, \textbf{Tang$_{k}$}:
	\begin{ex}[Tangles]
	In the category \textbf{Tang$_{k}$}
		\begin{itemize}
			\item an object $X\in \obj(\textbf{Tang}_{k})$ is a collection of particles in a $k$-dimensional cube and
			\item the morphisms $f:X\to Y$ are tangles between two objects.
		\end{itemize}
	More precisely, a tangle is a collection of arcs and circles smoothly embedded in a $(k + 1)$-dimensional cube. The arcs' endpoints touch the boundaries of the cube only at the top and bottom, while the circles lie in the interior only. An example for such a tangle is depicted in \cref{fig:tangk}.
	
	Tangles can be framed and \textit{oriented}. An orientation can be indicated by attaching arrows to each arc and circle. Physically speaking, the arrows indicate whether the particle is moving forward or backward in time (following Feynman's idea of antiparticles being particles that move backwards in time). In a framed tangle, the curves are replaced by ribbons to keep track of twists of particles. A \myemph{composition} of tangles corresponds to attaching two cubes top to bottom, see \cref{subfig:fcmpg}. The corresponding parallelepiped can be smoothly deformed into a cube.
	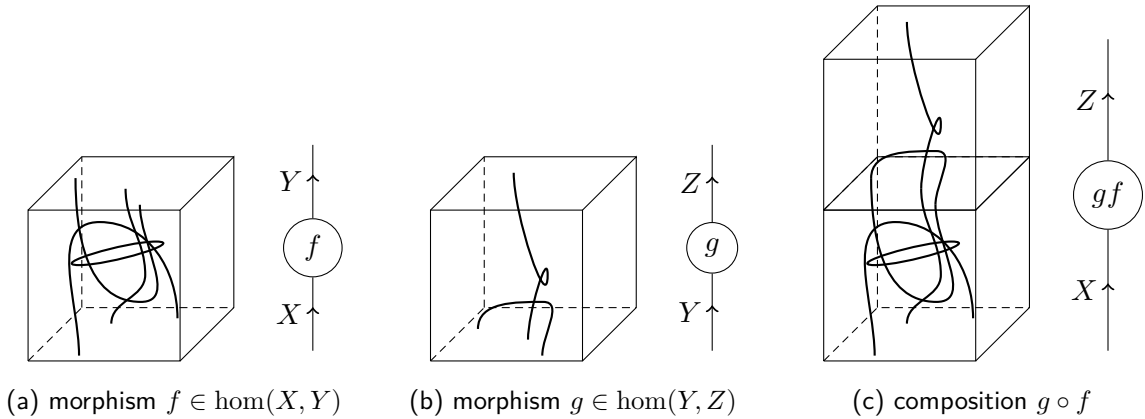
\begin{figure}[H]
\centering
\begin{subfigure}[b]{0.3\textwidth}
\centering
\begin{tikzpicture}[scale=2,%
x={(1cm,0cm)},
y={(0cm,1cm)},
z={({0.5*cos(45)},{0.5*sin(45)})},
]
\coordinate (A) at (0,0,0); 
\coordinate (B) at (1,0,0) ;
\coordinate (C) at (1,1,0); 
\coordinate (D) at (0,1,0); 
\coordinate (E) at (0,0,1); 
\coordinate (F) at (1,0,1); 
\coordinate (G) at (1,1,1); 
\coordinate (H) at (0,1,1);
\node[left= 1pt of A]{};
\node[right= 1pt of B]{};
\node[right= 1pt of C]{};
\node[left= 1pt of D]{};
\node[left= 1pt of E]{};
\node[right= 1pt of F]{};
\node[right= 1pt of G]{};
\node[left= 1pt of H]{};

\draw[] (A)-- (B) -- (C) -- (D) -- (A);
\draw[] (B) -- (F) -- (G) -- (C);
\draw[] (G) -- (H) -- (D);
\draw[densely dashed] (A) -- (E) -- (F);
\draw[densely dashed] (E) -- (H);
\draw[thick] (0.3,0,.1)to[out=90,in=180](0.4,0.85,.2)to[out=0,in=90](0.7,0,.8);
\draw[thick] (0.7,1,.1)to[out=270,in=0](0.6,0.25,.4)to[out=180,in=270](0.1,1,.6);
\draw[thick] (0.5,1,.4)to[out=270,in=90](0.7,0.5,.2)to[out=270,in=90](0.3,0,.7);
\draw[thick] (0.3,0.6,.1)to[out=180,in=180](0.7,0.65,.4)to[out=0,in=0](0.3,0.6,.1);
\end{tikzpicture}
\begin{tikzpicture}[scale=1.5,decoration={markings,mark=at position 0.5 with {\arrow[scale=1,thick]{<}}}] 
\node (start) at (0,0) {};
\node (f) at (0,1) [circle,draw] {$f$};
\node (g) at (0,2) [] {};
\draw[postaction={decorate}] (f) -- (start) node[midway,left] {$X$};
\draw[postaction={decorate}] (g) -- (f) node[midway,left] {$Y$};
\end{tikzpicture}
\caption{morphism $f\in\hom(X,Y)$}
\end{subfigure}
\hfill
\begin{subfigure}[b]{0.3\textwidth}
\centering
\begin{tikzpicture}[scale=2,%
x={(1cm,0cm)},
y={(0cm,1cm)},
z={({0.5*cos(45)},{0.5*sin(45)})},
]
	\coordinate (A) at (0,0,0); 
\coordinate (B) at (1,0,0) ;
\coordinate (C) at (1,1,0); 
\coordinate (D) at (0,1,0); 
\coordinate (E) at (0,0,1); 
\coordinate (F) at (1,0,1); 
\coordinate (G) at (1,1,1); 
\coordinate (H) at (0,1,1);
\node[left= 1pt of A]{};
\node[right= 1pt of B]{};
\node[right= 1pt of C]{};
\node[left= 1pt of D]{};
\node[left= 1pt of E]{};
\node[right= 1pt of F]{};
\node[right= 1pt of G]{};
\node[left= 1pt of H]{};
\draw[] (A)-- (B) -- (C) -- (D) -- (A);
\draw[] (B) -- (F) -- (G) -- (C);
\draw[] (G) -- (H) -- (D);
\draw[densely dashed] (A) -- (E) -- (F);
\draw[densely dashed] (E) -- (H);
\draw[thick] (0.7,0,.1)to[out=90,in=0](0.6,0.25,.4)to[out=180,in=90](0.1,0,.6);
\draw[thick] (0.5,0,.4)to[out=90,in=90](0.7,0.5,.2)to[out=270,in=270](0.3,1,.7);
\end{tikzpicture}
\begin{tikzpicture}[scale=1.5,decoration={markings,mark=at position 0.5 with {\arrow[scale=1,thick]{<}}}] 
\node (start) at (0,0) {};
\node (f) at (0,1) [circle,draw] {$g$};
\node (g) at (0,2) [] {};
\draw[postaction={decorate}] (f) -- (start) node[midway,left] {$Y$};
\draw[postaction={decorate}] (g) -- (f) node[midway,left] {$Z$};
\end{tikzpicture}
\caption{morphism $g\in\hom(Y,Z)$}
\end{subfigure}
\hfill
\begin{subfigure}[b]{0.3\textwidth}
\centering
\begin{tikzpicture}[scale=2,%
x={(1cm,0cm)},
y={(0cm,1cm)},
z={({0.5*cos(45)},{0.5*sin(45)})},
]
\coordinate (A) at (0,0,0); 
\coordinate (B) at (1,0,0) ;
\coordinate (C) at (1,1,0); 
\coordinate (D) at (0,1,0); 
\coordinate (E) at (0,0,1); 
\coordinate (F) at (1,0,1); 
\coordinate (G) at (1,1,1); 
\coordinate (H) at (0,1,1);
\node[left= 1pt of A]{};
\node[right= 1pt of B]{};
\node[right= 1pt of C]{};
\node[left= 1pt of D]{};
\node[left= 1pt of E]{};
\node[right= 1pt of F]{};
\node[right= 1pt of G]{};
\node[left= 1pt of H]{};

\draw[] (A)-- (B) -- (C) -- (D) -- (A);
\draw[] (B) -- (F) -- (G) -- (C);
\draw[] (G) -- (H) -- (D);
\draw[densely dashed] (A) -- (E) -- (F);
\draw[densely dashed] (E) -- (H);
\draw[thick] (0.3,0,.1)to[out=90,in=180](0.4,0.85,.2)to[out=0,in=90](0.7,0,.8);
\draw[thick] (0.7,1,.1)to[out=270,in=0](0.6,0.25,.4)to[out=180,in=270](0.1,1,.6);
\draw[thick] (0.5,1,.4)to[out=270,in=90](0.7,0.5,.2)to[out=270,in=90](0.3,0,.7);
\draw[thick] (0.3,0.6,.1)to[out=180,in=180](0.7,0.65,.4)to[out=0,in=0](0.3,0.6,.1);
\begin{scope}[shift={(0,1)}]
	\coordinate (A) at (0,0,0); 
\coordinate (B) at (1,0,0) ;
\coordinate (C) at (1,1,0); 
\coordinate (D) at (0,1,0); 
\coordinate (E) at (0,0,1); 
\coordinate (F) at (1,0,1); 
\coordinate (G) at (1,1,1); 
\coordinate (H) at (0,1,1);
\node[left= 1pt of A]{};
\node[right= 1pt of B]{};
\node[right= 1pt of C]{};
\node[left= 1pt of D]{};
\node[left= 1pt of E]{};
\node[right= 1pt of F]{};
\node[right= 1pt of G]{};
\node[left= 1pt of H]{};
\draw[] (A)-- (B) -- (C) -- (D) -- (A);
\draw[] (B) -- (F) -- (G) -- (C);
\draw[] (G) -- (H) -- (D);
\draw[densely dashed] (A) -- (E) -- (F);
\draw[densely dashed] (E) -- (H);
\draw[thick] (0.7,0,.1)to[out=90,in=0](0.6,0.25,.4)to[out=180,in=90](0.1,0,.6);
\draw[thick] (0.5,0,.4)to[out=90,in=90](0.7,0.5,.2)to[out=270,in=270](0.3,1,.7);
\end{scope}
\end{tikzpicture}
\begin{tikzpicture}[scale=1.1,decoration={markings,mark=at position 0.5 with {\arrow[scale=1,thick]{<}}}] 
\node (start) at (0,0) {};
\node (f) at (0,2) [circle,draw] {$gf$};
\node (g) at (0,4) [] {};
\draw[postaction={decorate}] (f) -- (start) node[midway,left] {$X$};
\draw[postaction={decorate}] (g) -- (f) node[midway,left] {$Z$};
\end{tikzpicture}
\caption{composition $g\circ f$}
\end{subfigure}			\caption{Example for \textbf{Tang}$_{2}$, which yields tangles in a $3$-dimensional cube.}
			\label{fig:tangk}
		\end{figure}	\label{ex:Tangk}	
	\end{ex}

	\begin{rem}[Big and small categories]
	We distinguish between big and small categories, because not all constructions that work for small categories work for big ones. A category is considered \textbf{small} if the class of objects and the class of morphisms are both sets, otherwise it is considered \textbf{big}. We have already seen an example for a big category; the category of sets (\cref{ex22}) is big, because the class of all sets is not a set itself.
	\label{rem24}
	\end{rem}

\subsection{Counterexamples}

	Because of the generality of the structure, it is difficult to think of a counterexample to a category. One approach is to take any category and simply remove one of its properties.	
	
	\begin{ex}
		In the example of a partially ordered set (see \cref{ex:poset}), one can choose the natural numbers $\mathbb{N}$ as objects and define $3\nleq5$. Hence, transitivity is not fulfilled any more (because $3\le4$ and $4\le5$, but $3\nleq5$) and therefore, morphisms are not composable in general.
	\end{ex}
	
	\begin{ex}
		Another idea is to take a partially ordered set and simply remove the property of reflexivity. In that case we have no identities and therefore not all the conditions of a category are fulfilled.
	\end{ex}

\subsection{Functors}

	Now that we have defined categories and have gained some experience with them by studying examples, we can add a level of abstraction and consider maps between categories, which are called \textit{functors}:
	
	\begin{defi}[Functor]
		A \textbf{functor} $F:\mathcal{C}\to\mathcal{D}$ from a category $\mathcal{C}$ to a category $\mathcal{D}$ is a map that sends
		\begin{itemize}
			\item any object $X\in\obj(\mathcal{C})$ to an object $F(Y)\in\obj(\mathcal{D})$,
			\item any morphism $f:X\to Y$ in $\mathcal{C}$ to a morphism $F(f):F(X)\to F(Y)$ in $\mathcal{D}$,
		\end{itemize}
		such that the following properties are fulfilled:
		\begin{itemize}
			\item $F$ preserves identities: for any object $X\in\obj(\mathcal{C})$, $F(\text{id}_X)=\text{id}_{F(X)}$,
			\item $F$ preserves composition: for any two morphisms $f:X\to Y$, $g:Y\to Z$ in $\mathcal{C}$, $F(g\circ f)=F(g)\circ F(f)$.
		\end{itemize}
		Note that these axioms imply that $F$ is invertible whenever $f$ is invertible: Let $f^{-1}:Y\to X$ be the inverse of $f:X\to Y$, then
			\begin{align*}
				\text{id}_{F(X)}&=F(\text{id}_X)\\
				&=F(f\circ f^{-1})\\
				&=F(f)\circ F(f^{-1}),
			\end{align*}
		hence $F(f)^{-1}=F(f^{-1})$. Graphically, functors are depicted in the following way:
		
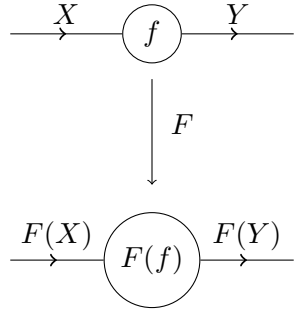
\begin{figure}[H]
			\centering
			\begin{tikzpicture}
			[scale=2,decoration={markings,mark=at position 0.5 with {\arrow[scale=1,thick]{>}}}] 
			\node (start) at (0,0) {};
			\node (f) at (1,0) [circle,draw] {$f$};
			\node (end) at (2,0) {};
			\node (start2) at (0,-1.5) {};
			\node (Ff) at (1,-1.5) [circle,draw] {$F(f)$};
			\node (end2) at (2,-1.5) {};
			\node (fi) at (1.2,-0.6) {$F$};
			\draw[postaction={decorate}] (start) -- (f) node[midway,above] {$X$};
			\draw[postaction={decorate}] (f) -- (end) node[midway,above] {$Y$};
			\draw [->] (1,-0.3) -- (1,-1);
			\draw[postaction={decorate}] (start2) -- (Ff) node[midway,above] {$F(X)$};
			\draw[postaction={decorate}] (Ff) -- (end2) node[midway,above] {$F(Y)$};
			\end{tikzpicture}
		\end{figure}
	\end{defi}

	\begin{ex} Consider an arbitrary category $\mathcal{C}$ and a category $\mathcal{D}$ that has only one object $X\in\obj(\mathcal{D})$ and one morphism $\mathbbm{1}_X$. Then, a functor $F:\mathcal{C}\to\mathcal{D}$ maps any object of $\mathcal{C}$ to the one object $X\in\obj(\mathcal{D})$. Analogously, every morphism in $\mathcal{C}$ is mapped to $\mathbbm{1}_X$ in $\mathcal{D}$:
		\begin{itemize}
			\item $A\in\obj(\mathcal{C})$ is mapped to $F(A)=X\in\obj(\mathcal{D})$.
			\item $f:A\to B$ is sent to $F(f)=\mathbbm{1}_X$.
		\end{itemize}
	\end{ex}

	One of the easiest examples are those of \myemph{forgetful functors}. Roughly speaking, these are functors that forget some of the structure of the input category:
	
	\begin{ex}
		Consider the category of sets, \textbf{Set} (\cref{ex22}), and the category of groups, \textbf{Grp} (\cref{ex:Grp}). There is a functor $F:\textbf{Grp}\to\textbf{Set}$ that is defined in the following way: For a group $G\in\obj(\mathbf{Grp})$, $F(G)$ is the underlying set of elements and for a group homomorphism $f:G\to H$, $F(f)$ is the function $f$ itself. Hence, $F$ forgets the group structure and forgets that group homomorphisms are homomorphisms.
	\end{ex}
	
	Functors can also allow us to formulate a way for a category to act on a system:
	\begin{ex}[Representations]
		\label{ex:reps}
		A functor can also be furnished by a \myemph{representation} of a group of symmetries. Recall that, in categorical terms, a group is a category with only one object, denoted $\bullet$, and all morphisms are isomorphisms (see \cref{ex:bullet}). A representation is a way of implementing symmetries on a system: For instance, in quantum mechanics it is a functor mapping from a category $\mathcal{G}$ (which is the group of symmetries we consider) to the category \textbf{Hilb},  $F:\mathcal{G}\rightarrow \mathbf{Hilb}$. 
		\begin{itemize}
			\item The object $\bullet\in\mathcal{G}$ is mapped to a Hilbert space via $F$:
				\begin{equation}
					F(\bullet)= \mathbb{C}^{d}\in \obj(\mathbf{Hilb}).
				\end{equation}
			\item The morphisms of $\mathcal{G}$ are mapped to linear operators between Hilbert spaces: $F(f)=$ a linear 	operator on $\mathbb{C}^{d}$.
		\end{itemize}
		The functor $F$ fulfills the following axioms:
		\begin{align*}
			F(\mathbbm{1}_{\bullet})&=\mathbbm{1}_X, \\
			F(f\circ g)&= F(f)\circ F(g),\\
			F(f)\circ F(f^{-1})&=\mathbbm{1}_X.
		\end{align*}
	\end{ex}

	\begin{ex}[Partially ordered sets]
		Consider two categories $\mathcal{P}$ and $\mathcal{Q}$ which are two partially ordered sets together with a functor $F:\mathcal{P}\rightarrow \mathcal{Q}$.
		Let $X, Y\in\mathcal{P}$ with $X\le Y$, i.e.~$\myhom(X,Y)\neq\emptyset$. These objects are mapped to objects $R, S\in\mathcal{Q}$ that fulfil $R\le S$ and therefore $\myhom(R,S)\neq\emptyset$.
	\end{ex}

\subsection{Natural Transformations}

	We now know about categories and about maps between them, i.e.\ functors. There is also a certain notion for maps between functors, which are called \myemph{natural transformations}. 
	
	In the following, we sometimes use the terminology of a \myemph{commuting} diagram. In general this means the following: whenever there are two paths from an object $X$ to an object $Y$, the map from $X$ to $Y$ obtained by composing the maps along one path equals the map we get by composing along the other path. For example, consider the diagram
		
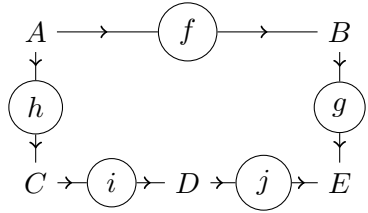
\begin{center}
		\begin{tikzpicture}
			[scale=2,decoration={markings,mark=at position 0.5 with {\arrow[scale=1, thick]{>}}}] 
			\node (A) at (0,0) {$A$};
			\node (B) at (2,0) {$B$};
			\node (C) at (0,-1) {$C$};
			\node (D) at (1,-1) {$D$};
			\node (E) at (2,-1) {$E$};
			\node (f) at (1,0) [circle, draw] {$f$};
			\node (h) at (0,-0.5) [circle, draw] {$h$};
			\node (g) at (2,-0.5) [circle, draw] {$g$};
			\node (i) at (0.5,-1) [circle, draw] {$i$};
			\node (j) at (1.5,-1) [circle, draw] {$j$};
			\draw[postaction={decorate}] (A) -- (f);
			\draw[postaction={decorate}] (f) -- (B);
			\draw[postaction={decorate}] (A) -- (h);
			\draw[postaction={decorate}] (h) -- (C);
			\draw[postaction={decorate}] (C) -- (i);
			\draw[postaction={decorate}] (i) -- (D);
			\draw[postaction={decorate}] (D) -- (j);
			\draw[postaction={decorate}] (j) -- (E);
			\draw[postaction={decorate}] (B) -- (g);
			\draw[postaction={decorate}] (g) -- (E);
		\end{tikzpicture}
		\end{center}
	This diagram is said to commute if $g\circ f=j\circ i\circ h$.
	
	\begin{defi}[Natural transformation]
		\label{def:nattrans}
		Let $\mathcal{C}$ and $\mathcal{D}$ be categories and $F,G:\mathcal{C}\to\mathcal{D}$ be functors between these categories. A natural transformation $\alpha:F\to G$ is a family of maps
			\begin{equation*}
				\left(F(X)\xrightarrow{\alpha_X} G(X)\right)_{X\in\obj(\mathcal{C})}
			\end{equation*}
		in $\mathcal{D}$ such that for every morphism $X\xrightarrow{f} X'$ in $\mathcal{C}$, the following diagram commutes:
	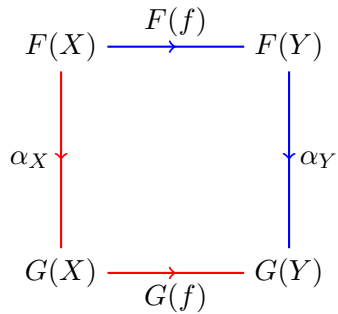
\begin{figure}[H]
	\centering
		\begin{tikzpicture}[scale=3,decoration={markings,mark=at position 0.5 with {\arrow[scale=1,thick]{>}}}] 
		\node (e) at (0,0) [] {$G(X)$};
		\node (f) at (0,1) [] {$F(X)$};
		\node (g) at (1,1) [] {$F(Y)$};
		\node (h) at (1,0) [] {$G(Y)$};
		\draw[postaction={decorate},red,thick] (f) -- (e) node[black,midway,left] {$\alpha_X$};
		\draw[postaction={decorate},blue,thick] (f) -- (g) node[black,midway,above] {$F(f)$};
		\draw[postaction={decorate},blue,thick] (g) -- (h) node[black,midway,right] {$\alpha_Y$};
		\draw[postaction={decorate},red,thick] (e) -- (h) node[black,midway,below] {$G(f)$};
		\end{tikzpicture}
	\caption[naturaltrafo]{Natural transformation. Going the red path yields the same result as going the blue one.}
	\label{fig:naturaltrafo}
	\end{figure}
	\noindent		
		The maps $\alpha_X$ are the \myemph{components} of $\alpha$.
	\end{defi}

	Analogous to the way we defined isomorphisms for morphisms, we can define natural isomorphisms:
	\begin{defi}[Natural isomorphism]
		Let $\mathcal{C}$ and $\mathcal{D}$ be categories and $F,G:\mathcal{C}\to\mathcal{D}$ be functors between these categories. A natural isomorphism is a natural transformation $\alpha:F\to G$ such that $\alpha_X$ is an isomorphism for every $X\in\obj(\mathcal{C})$.
	\end{defi}
	
	\begin{ex}
		Let us consider an extension of \cref{ex:reps}. We take a group as a category $\mathcal{C}$ with only one object $\bullet$, and the category of Hilbert spaces \textbf{Hilb} and two functors $F, G: \mathcal{C}\rightarrow \mathbf{Hilb}$. Hence, as we have seen before, the functors $F$ and $G$ are two different representations of $\mathcal{C}$ on the Hilbert spaces $F(\bullet)$ and $G(\bullet)$.
		The corresponding natural transformation between these two functors then has to fulfil
			\begin{align*}
				\underbrace{\alpha_{Y}}_{\in M_{d',d(\mathbb{C})}}\underbrace{F(f)}_{\in M_{d,d(\mathbb{C})}} 		=\underbrace{G(f)}_{\in M_{d',d'(\mathbb{C})}}\underbrace{\alpha_{X}}_{\in M_{d',d(\mathbb{C})}},
			\end{align*}
		where $M_{d,d'(\mathbb{C})}$ is the set of $d\times d'$ matrices over $\mathbb{C}$.
		Therefore, is a collection of rectangular matrices $\alpha_{X}$ that translate between the two ways of implementing a symmetry. They are called \myemph{intertwiners} or \myemph{intertwining operators}. Diagrammatically, this is depicted as
			
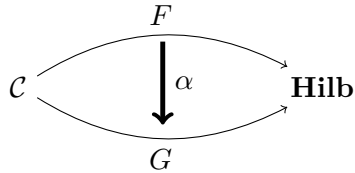
\begin{center}
		\begin{tikzpicture}
			[scale=2,decoration={markings,mark=at position 0.999 with {\arrow[scale=1]{>}}}]
			\node (C) at (0,0) {$\mathcal{C}$};
			\node (Hilb) at (2,0) {$\mathbf{Hilb}$};
			\draw[postaction={decorate}] (C) to [bend left] node [right,above] {$F$} (Hilb);
			\draw[postaction={decorate}] (C) to [bend right] node [right,below] {$G$} (Hilb);
			\draw[->,line width=0.65mm] (0.95,0.3)--(0.95,-0.25) node[midway,right] {$\alpha$};
		\end{tikzpicture}
		\end{center}
	\end{ex}


\tikzset{->-/.style={decoration={
			markings,
			mark=at position #1 with {\arrow[scale=1,thick]{>}}},postaction={decorate}}}

\section{Monoidal Categories and Braiding}
\label{Ch3}

	So far, we have focused on understanding categories by illustrating the definitions in terms of several different examples. We have also seen several physically motivated examples. We now want to concentrate on the description of physical systems and processes: In a category, objects can be seen as \myemph{physical systems} and morphisms as \myemph{processes}. According to this analogy, composition corresponds to performing one process after another.

	For example, consider the process of preparing a system $X$ from the vacuum or a thermal state $\mathbf{1}$ may be visualised as
		
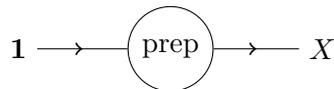
\begin{figure}[H]
		\centering
		\begin{tikzpicture}[scale=2,decoration={markings,mark=at position 0.5 with {\arrow[scale=1,thick]{>}}}] 
			\node (1) at (0,0) {$\mathbf{1}$};
			\node (prep) at (1,0) [circle, draw] {prep};
			\node (X) at (2,0) {$X$};
			\draw[postaction={decorate}] (1) -- (prep) node[midway,above] {};
			\draw[postaction={decorate}] (prep) -- (X) node[midway,above] {};
		\end{tikzpicture}
		\end{figure}
	\noindent
	Another typical example that occurs in the description of physical systems is the evolution $f$ of one system to another. Here we visualise an evolution via:
		
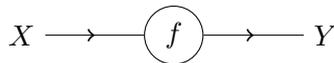
\begin{figure}[H]
		\centering
		\begin{tikzpicture}[scale=2,decoration={markings,mark=at position 0.5 with {\arrow[scale=1,thick]{>}}}] 
			\node (start) at (0,0) {$X$};
			\node (f) at (1,0) [circle,draw] {$f$};
			\node (end) at (2,0) {$Y$};
			\draw[postaction={decorate}] (start) -- (f);
			\draw[postaction={decorate}] (f) -- (end);
		\end{tikzpicture}
		\end{figure}
	\noindent
	And, finally, we can also visualise a measurement process $M$ with classical output $m$ as:
		
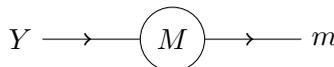
\begin{figure}[H]
		\centering
		\begin{tikzpicture}[scale=2,decoration={markings,mark=at position 0.5 with {\arrow[scale=1,thick]{>}}}] 
			\node (x) at (0,0) {$Y$};
			\node (M) at (1,0)  [circle,draw]  {$M$};
			\node (cb) at (2,0) {$m$};
			\draw[postaction={decorate}] (x) -- (M) node[midway,above] {};
			\draw[postaction={decorate}] (M) -- (cb) node[midway,above] {};
		\end{tikzpicture}
		\end{figure}
	\noindent Using these three processes, and composition, we can build an entire experimental setup, where first a system $X$ is prepared which then undergoes an evolution $f$ and is measured afterwards:
	
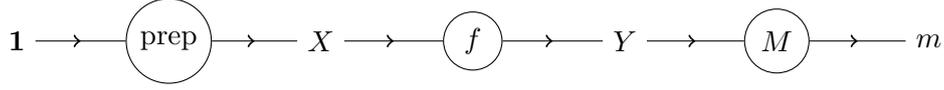
\begin{figure}[H]
		\centering
		\begin{tikzpicture}[scale=2,decoration={markings,mark=at position 0.5 with {\arrow[scale=1,thick]{>}}}] 
			\node (1) at (0,0) {$\mathbf{1}$};
			\node (prep) at (1,0)  [circle,draw]  {prep};
			\node (x) at (2,0) {$X$};
			\node (f) at (3,0)  [circle,draw]  {$f$};
			\node (y) at (4,0) {$Y$};
			\node (M) at (5,0)  [circle,draw]  {$M$};
			\node (m) at (6,0) {$m$};
			\draw[postaction={decorate}] (1) -- (prep) node[midway,above] {};
			\draw[postaction={decorate}] (prep) -- (x) node[midway,above] {};
			\draw[postaction={decorate}] (x) -- (f) node[midway,above] {};
			\draw[postaction={decorate}] (f) -- (y) node[midway,above] {};
			\draw[postaction={decorate}] (y) -- (M) node[midway,above] {};
			\draw[postaction={decorate}] (M) -- (m) node[midway,above] {};
		\end{tikzpicture}
		\end{figure}

	But what if we want to describe experiments that consist of multiple subsystems or where several processes happen in \emph{parallel}? In physics, we use \myemph{tensor products} to describe these setups. For example, the state space of a quantum harmonic oscillator is $L^2(\mathbb{R})$ (i.e.~the set of square integrable functions $\psi:\mathbb{R}\to\mathbb{C}$ on the real line), so the state space of two oscillators is given by $L^2(\mathbb{R})\otimes L^2(\mathbb{R})$. Following the example of the experimental setup above, we can also ask how to describe an experiment where a measurement on a system consisting of several subsystems is described in categorical terms. This leads us to the topic of

\subsection{Monoidal Categories}	
	
	To be able to talk about a collection of systems, we need to introduce the Cartesian product of two categories:

	\begin{defi}[Cartesian product] The Cartesian product $\mathcal{C}\times \mathcal{C}'$ of categories $\mathcal{C}$ and $\mathcal{C}'$ is a category where
	\begin{itemize}
		\item an object is a pair $(X,X')$ with $X\in \obj(\mathcal{C})$ and $X'\in \obj(\mathcal{C}')$,
		\item a morphism from $(X,X')$ to $(Y,Y')$ is a pair $(f,f')$ consisting of $f:X\rightarrow Y$ and $f':X'\rightarrow Y'$,
		\item composition is done componentwise: $(g,g')\circ(f,f')=(g\circ f,g'\circ f')$ and
		\item identity morphisms are also defined componentwise: $1_{(X,X')}=(1_X,1_{X'})$
	\end{itemize}
	\end{defi}

	Now we demonstrate how the tensor product enters the categorical formalism and which additional properties are required in order to get a reasonable definition of a category with a tensor product. We need to be aware that the tensor product is only associative up to an isomorphism and that tensoring units to the left or right side of an object does not have to be equal in the first place.

	\begin{defi}[Monoidal category]
		\label{def:monoidalcat}
		A monoidal category consists of a category $\mathcal{C}$ equipped with a functor 
			\begin{equation}
				\otimes:\mathcal{C}\times\mathcal{C}\to\mathcal{C}
			\end{equation}
		(the tensor product) and an unit object $\mathbb{I}\in\mathcal{C}$. In addition, there are natural isomorphisms
			\begin{align*}
				\alpha_{X,Y,Z}:\left(X\otimes Y\right)\otimes Z&\to X\otimes\left(Y\otimes Z\right)\\
				l_X:X&\to\mathbb{I}\otimes X\\
				r_X:X&\to X\otimes\mathbb{I}
			\end{align*}
		such that the following diagrams commute:
			\begin{figure}[H]
				\centering
				\begin{tikzpicture}[scale=2.5,decoration={markings,mark=at position 1 with {\arrow[scale=1,thick]{>}}}] 
				\node(up) at (0,0) {$X\otimes W$};
				\node(left) at (-1,-1) {$(X\otimes\mathbb{I})\otimes W$};
				\node(right) at (1,-1) {$X\otimes(\mathbb{I}\otimes W)$};
				\draw[postaction={decorate}] (up)--(left) node[midway,left] {$r_X\otimes\mathrm{id}_W$};
				\draw[postaction={decorate}] (up)--(right) node[midway,right] {$\mathrm{id}_X\otimes l_W$};
				\draw[postaction={decorate}] (left)--(right) node[midway,above] {$\alpha_{X,\mathbb{I},W}$};
				\end{tikzpicture}
				\caption{Triangle equation.}
			\end{figure}
			\begin{figure}[H]
			\centering
					\begin{tikzpicture}[scale=2,decoration={markings,mark=at position 1 with {\arrow[scale=1,thick]{>}}}] 
				\node(up) at (0,1) {$(X\otimes Y)\otimes(Z\otimes W)$};
				\node(left) at (-3,0) {$((X\otimes Y)\otimes Z)\otimes W$};
				\node(right) at (3,0) {$X\otimes (Y\otimes (Z\otimes W))$};
				\node(downleft) at (-1.5,-1) {$(X\otimes (Y\otimes Z))\otimes W$};
				\node(downright) at (1.5,-1) {$X\otimes ((Y\otimes Z)\otimes W)$};
				\draw[postaction={decorate}] (left) -- (up) node[midway,left=0.3cm] {$\alpha_{X\otimes Y,Z,W}$};
				\draw[postaction={decorate}] (up) -- (right) node[midway,right=0.3cm] {$\alpha_{X,Y,Z\otimes W}$};
				\draw[postaction={decorate}] (left) -- (downleft) node[midway,left=0.1cm] {$\alpha_{X,Y,Z}\otimes\mathrm{id}_W$};
				\draw[postaction={decorate}] (downleft) -- (downright) node[midway,above] {$\alpha_{X, Y\otimes Z,W}$};
				\draw[postaction={decorate}] (downright) -- (right) node[midway,right=0.3cm] {$\mathrm{id}_X\otimes \alpha_{Y,Z,W}$};
			\end{tikzpicture}
			\caption{Pentagon equation.}
			\label{fig:pentagon}
			\end{figure}
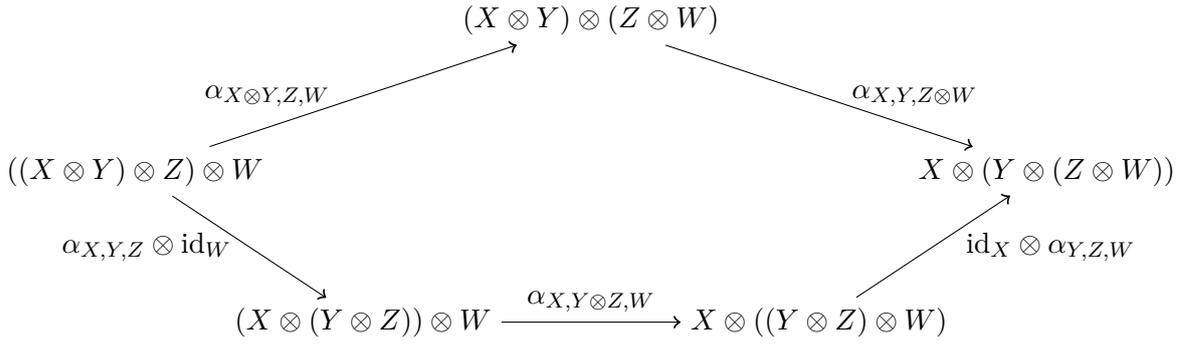
	\label{def:monoidal}
	\end{defi}

	As depicted in \cref{fig:pentagon}, there are five different ways to parenthesize the tensor product of four objects and it seems that the natural isomorphism $\alpha$ lets us construct two different isomorphisms going from $((X\otimes Y)\otimes Z)\otimes W$ to $X\otimes (Y\otimes (Z\otimes W))$ (following either the upward or downward direction of the arrows). The pentagon equation ensures that both of these isomorphisms are equal. Furthermore, when we have tensor products of more than four objects, there are even more ways to parenthesize them and therefore more isomorphisms that go from one side to another. However, in \cite{MacLane} it was shown that the pentagon equation implies that all these isomorphisms are the same, independent of the number of objects we consider. Additionally, if we consider the triangle equation, all isomorphisms with the same source and target constructed from the natural isomorphism $\alpha$ and the left and right units $r$ and $l$ are equal.
	
	\begin{ex} Several of the examples mentioned in the previous chapter can be made into a monoidal category:
		\begin{itemize}
			\item In the category \textbf{Set} (\cref{ex22}), the tensor product of two sets, $X\otimes Y$, can be chosen to be the Cartesian product $X\times Y$. In this case, the unit object is any one-element set. As mentioned previously, this tensor product is not associative, but there is a natural isomorphism $(X\times Y)\times Z\to X\times (Y\times Z)$ (similar for the left and right units). The tensor product of two morphisms $f:X\to Y$ and $f':X'\to Y'$ is given by
				\begin{align*}
					f\otimes f':X\times X'&\to Y\times Y'\\
					(x,x')&\mapsto (f(x),f'(x')).
				\end{align*}
			\item In \textbf{$n$Cob} (\cref{ex:nCob}), the tensor product is just the disjoint union, for example for two morphisms it is 
				
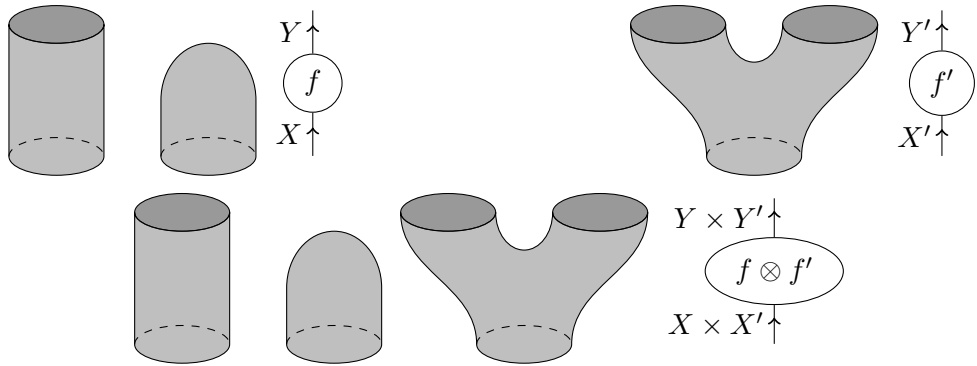
\begin{figure}[H]
					\centering
				\begin{subfigure}[b]{0.45\textwidth}
\centering
\begin{tikzpicture}[scale=0.5]
\fcobo
\end{tikzpicture}
\begin{tikzpicture}[scale=1.1,decoration={markings,mark=at position 0.5 with {\arrow[scale=1,thick]{<}}}] 
\node (start) at (0,0) {};
\node (f) at (0,1) [circle,draw] {$f$};
\node (g) at (0,2) [] {};
\draw[postaction={decorate}] (f) -- (start) node[midway,left] {$X$};
\draw[postaction={decorate}] (g) -- (f) node[midway,left] {$Y$};
\end{tikzpicture}
\end{subfigure}
\hfill
\begin{subfigure}[b]{0.45\textwidth}
\centering
\begin{tikzpicture}[scale=0.5]
\gcobo
\end{tikzpicture}
\begin{tikzpicture}[scale=1.1,decoration={markings,mark=at position 0.5 with {\arrow[scale=1,thick]{<}}}] 
\node (start) at (0,0) {};
\node (f) at (0,1) [circle,draw] {$f'$};
\node (g) at (0,2) [] {};
\draw[postaction={decorate}] (f) -- (start) node[midway,left] {$X'$};
\draw[postaction={decorate}] (g) -- (f) node[midway,left] {$Y'$};
\end{tikzpicture}
\end{subfigure}
\hfill
\begin{subfigure}[b]{0.7\textwidth}
\centering
\begin{tikzpicture}[scale=0.5]
\gcobo
\begin{scope}[shift={(-9,0)}]
\fcobo
\end{scope}
\end{tikzpicture}
\begin{tikzpicture}[scale=1.1,decoration={markings,mark=at position 0.5 with {\arrow[scale=1,thick]{<}}}] 
\node (start) at (0,0) {};
\node (f) at (0,1) [ellipse,draw] {$f\otimes f'$};
\node (g) at (0,2) [] {};
\draw[postaction={decorate}] (f) -- (start) node[midway,left] {$X\times X'$};
\draw[postaction={decorate}] (g) -- (f) node[midway,left] {$Y\times Y'$};
\end{tikzpicture}
\end{subfigure}					\end{figure}
			\item It is similar for the category of tangles \textbf{Tang}$_k$ (\cref{ex:Tangk}) for $k\ge 1$: here, we also choose the disjoint union as tensor product, e.g.~
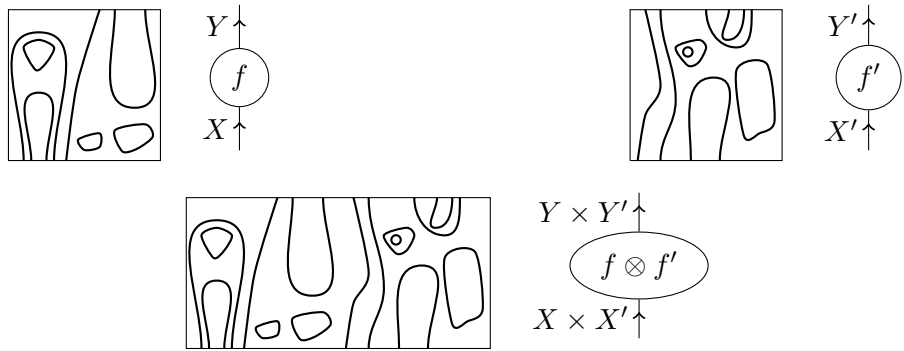
	\begin{figure}[H]
\centering
\begin{subfigure}[b]{0.45\textwidth}
\centering
\begin{tikzpicture}[scale=2,%
x={(1cm,0cm)},
y={(0cm,1cm)},
z={({0.5*cos(45)},{0.5*sin(45)})},
]
\coordinate (A) at (0,0); 
\coordinate (B) at (1,0) ;
\coordinate (C) at (1,1); 
\coordinate (D) at (0,1); 
\node[left= 1pt of A]{};
\node[right= 1pt of B]{};
\node[right= 1pt of C]{};
\node[left= 1pt of D]{};

\draw[] (A)-- (B) -- (C) -- (D) -- (A);
\draw[thick] (0.1,0)to[out=90,in=180](0.2,0.85)to[out=0,in=90](0.3,0);
\draw[thick] (0.15,0)to[out=90,in=180](0.2,0.45)to[out=0,in=90](0.25,0);
\draw[thick] (0.9,1)to[out=270,in=0](0.83,0.35)to[out=180,in=270](0.7,1);
\draw[thick] plot [smooth] coordinates{(.6,1) (.45,.5) (.38,0)};
\draw [thick] plot [smooth cycle] coordinates {(0.85,0.08) (.95,.16) (.9,.24) (.7,0.20) (.75,.06)};
\draw [thick] plot [smooth cycle] coordinates {(0.60,0.09) (.60,.18) (.46,0.14) (.5,.07)};
\draw[thick] plot [smooth cycle] coordinates {(0.1,.75) (0.2,.8) (0.3,.75) (0.25,.65) (0.18,.6)};
\end{tikzpicture}
\begin{tikzpicture}[scale=1.1,decoration={markings,mark=at position 0.5 with {\arrow[scale=1,thick]{<}}}] 
\node (start) at (0,0) {};
\node (f) at (0,1) [circle,draw] {$f$};
\node (g) at (0,2) [] {};
\draw[postaction={decorate}] (f) -- (start) node[midway,left] {$X$};
\draw[postaction={decorate}] (g) -- (f) node[midway,left] {$Y$};
\end{tikzpicture}
\end{subfigure}
\hfill
\begin{subfigure}[b]{0.45\textwidth}
\centering
\begin{tikzpicture}[scale=2,%
x={(1cm,0cm)},
y={(0cm,1cm)},
z={({0.5*cos(45)},{0.5*sin(45)})},
]
\coordinate (A) at (0,0); 
\coordinate (B) at (1,0) ;
\coordinate (C) at (1,1); 
\coordinate (D) at (0,1); 
\node[left= 1pt of A]{};
\node[right= 1pt of B]{};
\node[right= 1pt of C]{};
\node[left= 1pt of D]{};
\draw[] (A)-- (B) -- (C) -- (D) -- (A);
\draw[thick] (0.4,0)to[out=90,in=180](0.55,0.55)to[out=0,in=90](0.59,0);
\draw[thick] (0.75,1)to[out=270,in=0](0.64,0.80)to[out=180,in=270](0.65,1);
\draw[thick] (0.8,1)to[out=270,in=0](0.7,0.75)to[out=180,in=270](0.5,1);
\draw[thick] plot [smooth] coordinates {(.1,1) (.2,.5) (.1,.3) (.05,0)}; 
\draw[thick] (0.2,1)to[out=270,in=90](0.3,0.4)to[out=270,in=90](0.2,0);
\draw [thick] plot [smooth cycle] coordinates {(0.85,0.18) (.95,.26) (.9,.64) (.7,0.60) (.75,.16)};
\draw[thick] plot [smooth cycle] coordinates {(0.3,.75) (0.45,.8) (0.5,.75) (0.45,.65) (0.38,.63)};
\draw[thick] (.38,.72) circle (.031);
\end{tikzpicture}
\begin{tikzpicture}[scale=1.1,decoration={markings,mark=at position 0.5 with {\arrow[scale=1,thick]{<}}}] 
\node (start) at (0,0) {};
\node (f) at (0,1) [circle,draw] {$f'$};
\node (g) at (0,2) [] {};
\draw[postaction={decorate}] (f) -- (start) node[midway,left] {$X'$};
\draw[postaction={decorate}] (g) -- (f) node[midway,left] {$Y'$};
\end{tikzpicture}
\end{subfigure}
\hfill
\begin{subfigure}[b]{0.7\textwidth}
\centering
\begin{tikzpicture}[scale=2,%
x={(1cm,0cm)},
y={(0cm,1cm)},
z={({0.5*cos(45)},{0.5*sin(45)})},
]
\coordinate (A) at (0,0); 
\coordinate (B) at (2,0) ;
\coordinate (C) at (2,1); 
\coordinate (D) at (0,1); 
\node[left= 1pt of A]{};
\node[right= 1pt of B]{};
\node[right= 1pt of C]{};
\node[left= 1pt of D]{};

\draw[] (A)-- (B) -- (C) -- (D) -- (A);
\draw[thick] (0.1,0)to[out=90,in=180](0.2,0.85)to[out=0,in=90](0.3,0);
\draw[thick] (0.15,0)to[out=90,in=180](0.2,0.45)to[out=0,in=90](0.25,0);
\draw[thick] (0.9,1)to[out=270,in=0](0.83,0.35)to[out=180,in=270](0.7,1);
\draw[thick] plot [smooth] coordinates{(.6,1) (.45,.5) (.38,0)};
\draw [thick] plot [smooth cycle] coordinates {(0.85,0.08) (.95,.16) (.9,.24) (.7,0.20) (.75,.06)};
\draw [thick] plot [smooth cycle] coordinates {(0.60,0.09) (.60,.18) (.46,0.14) (.5,.07)};
\draw[thick] plot [smooth cycle] coordinates {(0.1,.75) (0.2,.8) (0.3,.75) (0.25,.65) (0.18,.6)};
\begin{scope}[shift={(1,0)}]
\draw[thick] (0.4,0)to[out=90,in=180](0.55,0.55)to[out=0,in=90](0.59,0);
\draw[thick] (0.75,1)to[out=270,in=0](0.64,0.80)to[out=180,in=270](0.65,1);
\draw[thick] (0.8,1)to[out=270,in=0](0.7,0.75)to[out=180,in=270](0.5,1);
\draw[thick] plot [smooth] coordinates {(.1,1) (.2,.5) (.1,.3) (.05,0)}; 
\draw[thick] (0.2,1)to[out=270,in=90](0.3,0.4)to[out=270,in=90](0.2,0);
\draw [thick] plot [smooth cycle] coordinates {(0.85,0.18) (.95,.26) (.9,.64) (.7,0.60) (.75,.16)};
\draw[thick] plot [smooth cycle] coordinates {(0.3,.75) (0.45,.8) (0.5,.75) (0.45,.65) (0.38,.63)};
\draw[thick] (.38,.72) circle (.031);
\end{scope}
\end{tikzpicture}
\begin{tikzpicture}[scale=1.1,decoration={markings,mark=at position 0.5 with {\arrow[scale=1,thick]{<}}}] 
\node (start) at (0,0) {};
\node (f) at (0,1) [ellipse,draw] {$f\otimes f'$};
\node (g) at (0,2) [] {};
\draw[postaction={decorate}] (f) -- (start) node[midway,left] {$X\times X'$};
\draw[postaction={decorate}] (g) -- (f) node[midway,left] {$Y\times Y'$};
\end{tikzpicture}
\end{subfigure}
\label{fig:3Cobs}
\end{figure}

		\end{itemize}
	\end{ex}

	\begin{ex}[Hilbert spaces]
		There is one example from the previous chapter that we are going to investigate a little further because of its importance to quantum physics, namely, the category \textbf{Hilb} (\cref{ex:Hilb}). This category can be made into a monoidal category by using the usual tensor product of Hilbert spaces. E.g., suppose $\mathcal{H}$ is a Hilbert space, then the total Hilbert space for the composite system of two copies is then given by $\mathcal{H}\otimes \mathcal{H}$. 

		Let's now use composition and the tensor product operation to build some complicated morphisms. Suppose you have the following processes:
			
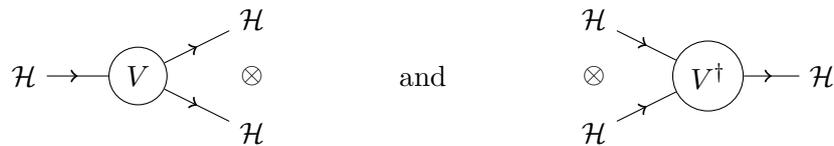
\begin{figure}[H]
				\centering
				\begin{tikzpicture}[scale=1.5,decoration={markings,mark=at position 0.5 with {\arrow[scale=1,thick]{>}}}] 
					\node (H) at (0,0) {$\mathcal{H}$};
					\node (V) at (1,0) [circle, draw] {$V$};
					\node (H1) at (2,0.5) {$\mathcal{H}$};
					\node (H2) at (2,-0.5) {$\mathcal{H}$};
					\node (otimes1) at (2,0) {$\otimes$};
					\node (and) at (3.5,0) {and};
					\node (H3) at (5,0.5) {$\mathcal{H}$};
					\node (H4) at (5,-0.5) {$\mathcal{H}$};
					\node (otimes2) at (5,0) {$\otimes$};
					\node (Vt) at (6,0) [circle, draw] {$V^\dagger$};
					\node (H5) at (7,0) {$\mathcal{H}$};
					\draw[{postaction=decorate}] (H)--(V);
					\draw[{postaction=decorate}] (V)--(H1);
					\draw[{postaction=decorate}] (V)--(H2);
					\draw[{postaction=decorate}] (H3)--(Vt);
					\draw[{postaction=decorate}] (H4)--(Vt);
					\draw[{postaction=decorate}] (Vt)--(H5);
				\end{tikzpicture}
			\end{figure}
		\noindent
		For $\mathcal{H}=\mathbb{C}^2$, the action of $V$ could be, for example,
			\begin{align*}
				V\ket{0}&=\ket{01},\\
				V\ket{1}&=\frac{\ket{00}+\ket{11}}{\sqrt{2}}.
			\end{align*}
		We can use this tensor network formalism to describe various ``atomic'' processes by composing string diagrams in series (composition) or parallel (tensoring), e.g.\
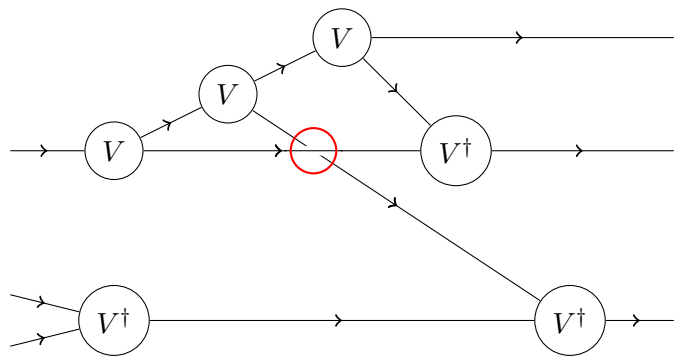
\begin{figure}[H]
\centering
\begin{tikzpicture}[scale=1.5,decoration={markings,mark=at position 0.5 with {\arrow[scale=1,thick]{>}}}]
\node (start1) at (0,0) {};
\node (start2) at (0,-1.25) {};
\node (start3) at (0,-1.75) {};
\node (V1) at (1,0) [circle, draw] {$V$};
\node (Vt1) at (1,-1.5) [circle, draw] {$V^\dagger$};
\node (V2) at (2,0.5) [circle, draw] {$V$};
\node (V3) at (3,1) [circle, draw] {$V$};
\node (Vt2) at (4,0) [circle, draw] {$V^\dagger$};
\node (Vt3) at (5,-1.5) [circle, draw] {$V^\dagger$};
\node (end1) at (6,1) {};
\node (end2) at (6,0) {};
\node (end3) at (6,-1.5) {};
\draw[{postaction=decorate}] (start1)--(V1);
\draw[{postaction=decorate}] (start2)--(Vt1);
\draw[{postaction=decorate}] (start3)--(Vt1);
\draw[{postaction=decorate}] (V1)--(V2);
\draw[{postaction=decorate}] (V2)--(V3);
\draw[{postaction=decorate}] (Vt1)--(Vt3);
\draw[{postaction=decorate}] (V3)--(Vt2);
\draw[{postaction=decorate}] (V3)--(end1);
\draw[{postaction=decorate}] (Vt2)--(end2);
\draw[{postaction=decorate}] (Vt3)--(end3);
\begin{knot}[
flip crossing/.list={}  ]
\strand[{postaction=decorate}] (V1)--(Vt2);
\strand[{postaction=decorate}] (V2)--(Vt3);
\end{knot}
\draw[red, thick] (2.75,0) circle (0.2cm);
\end{tikzpicture}
\end{figure}
\noindent
The point marked by the red circle is where two lines cross each other. However, we have not yet defined what it means when two lines in such a diagram cross; this the topic of the next subsection.
	\end{ex}

\subsection{Braided Monoidal Categories}

	In physics, we are familiar with the process of \myemph{swapping systems}. In category theory, a monoidal category in which we can swap things is called \myemph{braided}:
	
	\begin{defi}[Braided monoidal category] A braided monoidal category consists of a monoidal category $\mathcal{C}$ and a family of isomorphisms called braiding
		\begin{align*}
			b_{X,Y}:X\otimes Y \rightarrow Y\otimes X,
		\end{align*}
	which is natural with respect to $X$ and $Y$, i.e.~the following diagram commutes for every choice of morphisms $f:X \to X'$, $g:Y\to Y'$:
		\begin{figure}[H]
			\centering
			\begin{tikzpicture} [scale=2,decoration={markings,mark=at position 1 with {\arrow[scale=1,thick]{>}}}]
				\node (AB) at (0,0) {$X\otimes Y$};
				\node (ABp) at (2,0) {$X'\otimes Y'$};
				\node (BA) at (0,-1) {$Y\otimes X$};
				\node (BAp) at (2,-1) {$Y'\otimes X'$};
				\draw[{postaction=decorate}] (AB) -- (BA) node[midway,left] {$b_{X,Y}$};
				\draw[{postaction=decorate}] (AB) -- (ABp) node[midway,above] {$f\otimes g$};
				\draw[{postaction=decorate}] (BA) -- (BAp) node[midway,above] {$g\otimes f$};
				\draw[{postaction=decorate}] (ABp) -- (BAp) node[midway,right] {$b_{X',Y'}$};
			\end{tikzpicture}
		\end{figure}
	\noindent
	Additionally, it is required that the so-called hexagon equation is fulfilled, i.e.~the following diagrams commute:
		\begin{figure}[H]
			\centering
			\begin{tikzpicture} [scale=1,decoration={markings,mark=at position 1 with {\arrow[scale=1,thick]{>}}}] 
			\node (XY-Z) at (0,0) {$(X\otimes Y)\otimes Z$};
			\node (X-YZ) at (4.5,0) {$X\otimes (Y\otimes Z)$};
			\draw[postaction={decorate}] (XY-Z) -- (X-YZ) node[midway,above] {$\alpha_{X,Y,Z}$};
			\node (X-ZY) at (9,0) {$(Y\otimes Z)\otimes X$};
			\draw[postaction={decorate}] (X-YZ) -- (X-ZY) node[midway,above] {$b_{X,Y\otimes Z}$};
			\node (XZ-Y) at (9,-3) {$Y\otimes (Z\otimes X)$};
			\draw[->,postaction={decorate}] (X-ZY) -- (XZ-Y) node[midway,right] {$\alpha_{Y,Z,X}$};
			\node (ZX-Y) at (4.5,-3) {$Y\otimes (X\otimes Z)$};
			\draw[postaction={decorate}] (ZX-Y) -- (XZ-Y) node[midway,below] {$\mathrm{id}_Y\otimes b_{X,Z}$};
			\node (Z-XY) at (0,-3) {$(Y\otimes X)\otimes Z$};
			\draw[postaction={decorate}] (Z-XY) -- (ZX-Y) node[midway,below] {$\alpha_{Y,X,Z}$};
			\draw[postaction={decorate}] (XY-Z) -- (Z-XY) node[midway,left] {$b_{X, Y} \otimes \mathrm{id}_Z$};
			\end{tikzpicture}
			\\[0.3cm]
			\begin{tikzpicture} [scale=1,decoration={markings,mark=at position 1 with {\arrow[scale=1,thick]{>}}}] 
			\node (X-YZ) at (0,0) {$X\otimes (Y\otimes Z)$};
			\node (XY-Z) at (4.5,0) {$(X\otimes Y)\otimes Z$};
			\draw[->,postaction={decorate}] (X-YZ) -- (XY-Z) node[midway,above] {$\alpha^{-1}_{X,Y,Z}$};
			\node (YX-Z) at (9,0) {$Z\otimes (X\otimes Y)$};
			\draw[->,postaction={decorate}] (XY-Z) -- (YX-Z) node[midway,above] {$b_{X\otimes Y,Z}$};
			\node (Y-XZ) at (9,-3) {$(Z\otimes X)\otimes Y$};
			\draw[->,postaction={decorate}] (YX-Z) -- (Y-XZ) node[midway,right] {$\alpha^{-1}_{Z,X,Y}$};
			\node (Y-ZX) at (4.5,-3) {$(X\otimes Z) \otimes Y$};
			\draw[->,postaction={decorate}] (Y-ZX) -- (Y-XZ) node[midway,below] {$b_{X,Z}\otimes \mathrm{id}_Y$};
			\node (YZ-X) at (0,-3) {$X\otimes (Z\otimes Y)$};
			\draw[->,postaction={decorate}] (YZ-X) -- (Y-ZX) node[midway,below] {$\alpha^{-1}_{X,Z,Y}$};
			\draw[->,postaction={decorate}] (X-YZ) -- (YZ-X) node[midway,left] {$\mathrm{id}_X\otimes b_{Y,Z}$};
			\end{tikzpicture}
			\caption{Hexagon equation.}
			\label{fig:hex}
		\end{figure}
	\noindent where $\alpha_{X,Y,Z}$ is the isomorphism between different ways to parenthesize the tensor product of multiple objects as introduced in \cref{def:monoidal}. The hexagon equation ensures that braiding is consistent with going between different ways of parenthesizing the tensor product.
	\label{def:braided}
	\end{defi}

	In string diagrams, the braiding isomorphism $b_{X,Y}$ is drawn as
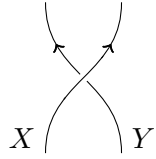
\begin{figure}[H]
\centering
\begin{tikzpicture}[decoration={markings,mark=at position .75 with {\arrow[scale=1,thick]{>}}}] 
\node (Y) at (1,.2)[right] {$Y$};
\node (X) at (0,.2)[left] {$X$};
\begin{knot}[
flip crossing/.list={}  ]
\strand[{postaction=decorate}] (0,0) to[out=90,in=225](.5,1)to[out=45,in=270](1,2);
\strand[{postaction=decorate}] (1,0)to[out=90,in=315](.5,1)to[out=135,in=270](0,2);
\end{knot}
\end{tikzpicture}
\end{figure}
\noindent
and its inverse $b^{-1}_{X,Y}$ is 
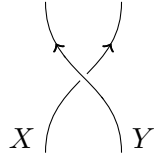
\begin{figure}[H]
\centering
\begin{tikzpicture}[decoration={markings,mark=at position .75 with {\arrow[scale=1,thick]{>}}}] 
\node (Y) at (1,.2)[right] {$Y$};
\node (X) at (0,.2)[left] {$X$};
\begin{knot}[
flip crossing/.list={1}  ]
\strand[{postaction=decorate}] (0,0) to[out=90,in=225](.5,1)to[out=45,in=270](1,2);
\strand[{postaction=decorate}] (1,0)to[out=90,in=315](.5,1)to[out=135,in=270](0,2);
\end{knot}
\end{tikzpicture}
\end{figure}
	
	\noindent
	Composing the braiding isomorphism and its inverse yields the identity morphism (as expected):

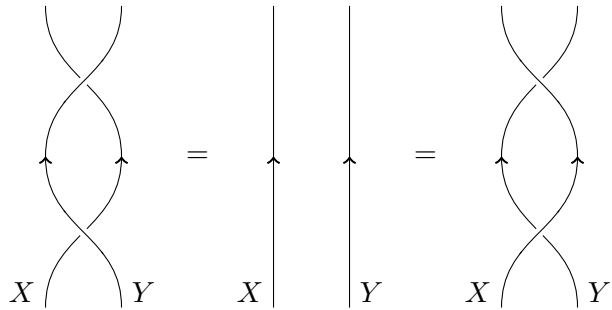
\begin{figure}[H]
\centering
\begin{tikzpicture}[decoration={markings,mark=at position .5 with {\arrow[scale=1,thick]{>}}}] 
\node (Y) at (1,.2)[right] {$Y$};
\node (X) at (0,.2)[left] {$X$};
\begin{knot}[flip crossing/.list={1,2}]
\strand[{postaction=decorate}] (0,0)to[out=90,in=225](.5,1)to[out=45,in=270](1,2)to[out=90,in=315](.5,3)to[out=135,in=270](0,4);
\strand[{postaction=decorate}] (1,0)to[out=90,in=315](.5,1)to[out=135,in=270](0,2)to[out=90,in=225](.5,3)to[out=45,in=270](1,4);
\end{knot}
\begin{scope}[shift={(3,0)}]
\node (a) at (-1,2) {$=$};
\node (b) at (2,2) {$=$};
\node (Y) at (1,.2)[right] {$Y$};
\node (X) at (0,.2)[left] {$X$};
\begin{knot}[flip crossing/.list={}]
\strand[{postaction=decorate}] (0,0)--(0,4);
\strand[{postaction=decorate}] (1,0)--(1,4);\
\end{knot}
\end{scope}
\begin{scope}[shift={(6,0)}]
\node (Y) at (1,.2)[right] {$Y$};
\node (X) at (0,.2)[left] {$X$};
\begin{knot}[flip crossing/.list={}]
\strand[{postaction=decorate}] (0,0)to[out=90,in=225](.5,1)to[out=45,in=270](1,2)to[out=90,in=315](.5,3)to[out=135,in=270](0,4);
\strand[{postaction=decorate}] (1,0)to[out=90,in=315](.5,1)to[out=135,in=270](0,2)to[out=90,in=225](.5,3)to[out=45,in=270](1,4);
\end{knot}
\end{scope}
\end{tikzpicture}
\end{figure}
Let $X,Y,Z$ be objects in a braided monoidal category. Using the hexagon equation in combination with the naturality of braiding, we find the so-called \myemph{Yang-Baxter equation} also known as the type III Reidemeister move:
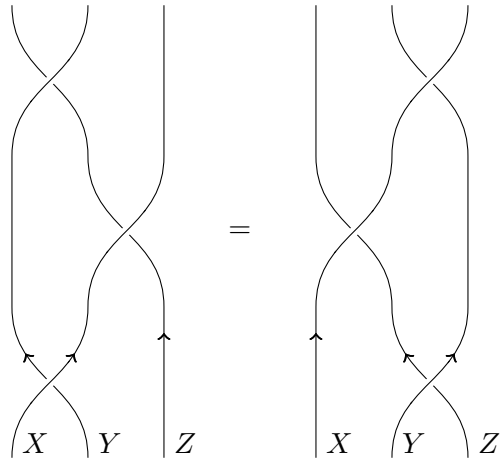
\begin{figure}[H]
\centering
\begin{tikzpicture}[decoration={markings,mark=at position .25 with {\arrow[scale=1,thick]{>}}}] 
\node (X) at (0,.2)[right] {$X$};
\node (Y) at (1,.2)[right] {$Y$};
\node (Z) at (2,.2)[right] {$Z$};
\begin{knot}[flip crossing/.list={}]
\strand[{postaction=decorate}] (0,0)to[out=90,in=225](.5,1)to[out=45,in=270](1,2)to[out=90,in=225](1.5,3)to[out=45,in=270](2,4)--(2,6);
\strand[{postaction=decorate}] (1,0)to[out=90,in=315](.5,1)to[out=135,in=270](0,2)--(0,4)to[out=90,in=225](.5,5)to[out=45,in=270](1,6);
\strand[{postaction=decorate}] (2,0)--(2,2)to[out=90,in=315](1.5,3)to[out=135,in=270](1,4)to[out=90,in=315](.5,5)to[out=135,in=270](0,6);
\end{knot}
\begin{scope}[shift={(4,0)}]
\node (X) at (0,.2)[right] {$X$};
\node (Y) at (1,.2)[right] {$Y$};
\node (Z) at (2,.2)[right] {$Z$};
\node (a) at (-1,3) {$=$};
\begin{knot}[flip crossing/.list={}]
	\strand[{postaction=decorate}] (0,0)--(0,2)to[out=90,in=225](.5,3)to[out=45,in=270](1,4)to[out=90,in=225](1.5,5)to[out=45,in=270](2,6);
\strand[{postaction=decorate}] (1,0)to[out=90,in=225](1.5,1)to[out=45,in=270](2,2)--(2,4)to[out=90,in=315](1.5,5)to[out=135,in=270](1,6);
\strand[{postaction=decorate}] (2,0)to[out=90,in=315](1.5,1)to[out=135,in=270](1,2)to[out=90,in=315](.5,3)to[out=135,in=270](0,4)--(0,6);
\end{knot}
\end{scope}
\end{tikzpicture}
\caption{Yang-Baxter equation.}
\end{figure}
The Yang-Baxter equation plays a crucial role in knot theory as well as in the study of exactly solvable problems in condensed matter physics (\cite{QGroup,Knots,Rowell04,Kitaev2005}). Equivalently, we can express this equation in terms of a commuting diagram, which includes the associator (which cannot be shown in terms of string diagrams):
	\begin{figure}[H]
		\centering
		\begin{tikzpicture}[scale=1.5,decoration={markings,mark=at position 1 with {\arrow[scale=1,thick]{>}}}] 
		\node (start) at (0,0) {$\left(X\otimes Y\right)\otimes Z$};
		\node (left1) at (-2,1) {$\left(Y\otimes X\right)\otimes Z$};
		\node (left2) at (-2,2) {$Y\otimes\left(X\otimes Z\right)$};
		\node (left3) at (-2,3) {$Y\otimes\left(Z\otimes X\right)$};
		\node (left4) at (-2,4) {$\left(Y\otimes Z\right)\otimes X$};
		\node (left5) at (-2,5) {$\left(Z\otimes Y\right)\otimes X$};
		\node (right1) at (2,1) {$X\otimes \left(Y\otimes Z\right)$};
		\node (right2) at (2,2) {$X\otimes\left(Z\otimes Y\right)$};
		\node (right3) at (2,3) {$\left(X\otimes Z\right)\otimes Y$};
		\node (right4) at (2,4) {$\left(Z\otimes X\right)\otimes Y$};
		\node (right5) at (2,5) {$Z\otimes\left( X\otimes Y\right)$};
		\node (end) at (0,6) {$Z\otimes\left(Y\otimes X\right)$};
		\draw[{postaction=decorate}] (start) -- (left1) node[midway,left] {\small $b_{X,Y}\otimes\mathrm{id}_Z$};
		\draw[{postaction=decorate}] (left1) -- (left2) node[midway,left] {\small $\alpha_{Y,X,Z}$};
		\draw[{postaction=decorate}] (left2) -- (left3) node[midway,left] {\small $\mathrm{id}_Y\otimes b_{X,Z}$};
		\draw[{postaction=decorate}] (left3) -- (left4) node[midway,left] {\small $\alpha_{Y,Z,X}^{-1}$};
		\draw[{postaction=decorate}] (left4) -- (left5) node[midway,left] {\small $b_{Y,Z}\otimes\mathrm{id}_X$};
		\draw[{postaction=decorate}] (start) -- (right1) node[midway,right] {\small $\ \alpha_{X,Y,Z}$};
		\draw[{postaction=decorate}] (right1) -- (right2) node[midway,right] {\small $\mathrm{id}_X\otimes b_{Y,Z}$};
		\draw[{postaction=decorate}] (right2) -- (right3) node[midway,right] {\small $\alpha_{X,Z,Y}^{-1}$};
		\draw[{postaction=decorate}] (right3) -- (right4) node[midway,right] {\small $b_{X,Z}\otimes Y$};
		\draw[{postaction=decorate}] (right4) -- (right5) node[midway,right] {\small $\alpha_{Z,X,Y}$};
		\draw[{postaction=decorate}] (left5) -- (end) node[midway,left] {\small $\alpha_{Z,Y,X}$};
		\draw[{postaction=decorate}] (right5) -- (end) node[midway,right] {\small $\mathrm{id}_Z\otimes b_{X,Y}$};
		\end{tikzpicture}
	\end{figure}
		
	In general, there are multiple ways to assign a braiding to a given monoidal category. However, most of the examples we have seen before come with an obvious standard braiding:
		\begin{ex}We can easily assign standard braidings to some categories:
			\begin{itemize}
				\item In the category \textbf{Hilb} (the category of Hilbert spaces) with its standard tensor product, the standard braiding is given by 
					\begin{align*}
						b_{X,Y}:X\otimes Y&\to Y\otimes X\\
						x\otimes y&\mapsto y\otimes x.
					\end{align*}
				\item In \textbf{Tang}$_2$, the standard braiding for $X$ and $Y$, each being a single point, looks like 
					\begin{figure}[H]
						\centering
						\begin{tikzpicture}[scale=2,%
						x={(1cm,0cm)},
						y={(0cm,1cm)},
						z={({0.5*cos(45)},{0.5*sin(45)})},
						]
						\coordinate (A) at (0,0,0); 
						\coordinate (B) at (1,0,0) ;
						\coordinate (C) at (1,1,0); 
						\coordinate (D) at (0,1,0); 
						\coordinate (E) at (0,0,1); 
						\coordinate (F) at (1,0,1); 
						\coordinate (G) at (1,1,1); 
						\coordinate (H) at (0,1,1);
						\node[left= 1pt of A]{};
						\node[right= 1pt of B]{};
						\node[right= 1pt of C]{};
						\node[left= 1pt of D]{};
						\node[left= 1pt of E]{};
						\node[right= 1pt of F]{};
						\node[right= 1pt of G]{};
						\node[left= 1pt of H]{};
						
						\draw[] (A)-- (B) -- (C) -- (D) -- (A);
						\draw[] (B) -- (F) -- (G) -- (C);
						\draw[] (G) -- (H) -- (D);
						\draw[densely dashed] (A) -- (E) -- (F);
						\draw[densely dashed] (E) -- (H);
						\begin{knot}[flip crossing/.list={1}]						\strand[thick] plot[smooth] coordinates {(.3,1,.5) (.3,0.7,.5)(.7,0.3,.5)(.7,0,.5)};			
					\strand[thick] plot[smooth] coordinates {(.7,1,.5) (.7,0.7,.6)(.3,0.3,.6)(.3,0,.5)};\end{knot}						\end{tikzpicture}
						\begin{tikzpicture}[scale=1.5,decoration={markings,mark=at position 0.5 with {\arrow[scale=1,thick]{<}}}] 
						\node (start) at (0,0) {};
						\node (f) at (0,1) [circle,draw] {$b_{X,Y}$};
						\node (g) at (0,2) [] {};
						\draw[postaction={decorate}] (f) -- (start) node[midway,left] {$X\otimes Y$};
						\draw[postaction={decorate}] (g) -- (f) node[midway,left] {$Y\otimes X$};
						\end{tikzpicture}
						\caption{Morphism $b_{X,Y}\in\hom(X\otimes Y,Y\otimes X)$.}
					\end{figure}
			\end{itemize}
		\end{ex}
	In practice, braided monoidal categories have numerous applications in both mathematics and physics. In the following, we describe two examples: The first one concerns anyons, which are identical particles existing in two dimensions, whose exotic properties are exploited in topological quantum computation (see e.g.~\cite{Wang10,PachosTop}). The second example is from the theory of knots, which is also closely related to quantum information theory (see e.g.~\cite{KauffmanComputing}).
	
	\begin{ex}[Anyons]
		\label{ex:Anyons}
		It is well known that in three spatial dimensions there are two types of indistinguishable particles, bosons and fermions. They correspond to the possible types of spins, either half-integer spin ($\frac{1}{2}$, $\frac{3}{2}$, $\frac{5}{2}$, \dots) or integer spin ($0$, $1$, $2$, \dots). In two dimensions, indistinguishable particles can have arbitrary spin, which leads to exotic statistics: The exchange of two particles gives rise to an arbitrary phase factor (in contrast to the boson/fermion case, where the factor is always $\pm 1$). This implies that, in the anyonic case, the world lines of particles depicted in \cref{fig:worldlines} are not equal.
			\begin{figure}[H]
				\centering
				\begin{tikzpicture}[scale=2,%
				x={(1cm,0cm)},
				y={(0cm,1cm)},
				z={({0.5*cos(45)},{0.5*sin(45)})},
				]
				\draw[->,densely dashed] (-1,0,0) -- (1,0,0);
				\draw[->,densely dashed] (0,-0.7,0) -- (0,1,0);
				\draw[->,densely dashed] (0,0,-1) -- (0,0,1);
				\node[] (u) at (1,.15,0) {$x$};
				\node[] (u) at (.15,1,0) {$t$};
				\node[] (u) at (.15,0,1) {$y$};
				\begin{knot}[consider self intersections=true,
ignore endpoint intersections=false,
flip crossing/.list={2}]				\strand[thick,red] plot [smooth] coordinates{(-.55,-.4,0)(-.5,-.3,0) (-.6,0,0) (-.56,.2,0)(.25,.4,0) (.38,.5,0) (-.30,.6,0)(-.55,.9,0)};
							\node[thick,red] (u) at (-.56,.2,0) {\textbullet};
				\strand[thick,blue] plot [smooth] coordinates{(.3,-.4,0)(.7,-.23,0) (.6,0,0) (.56,.2,0) (-.25,.3,0) (-.48,.45,0) (-.48,.60,0) (-.63,.60,0) (-.79,.52,0) (-.63,.38,0) (-.48,.45,0) (.38,.6,0)(.3,.9,0)};
				\node[thick,blue] (u) at (-.79,.52,0) {\textbullet};
			\end{knot}
				\node[] (u) at (1.5,.25,0) {$\neq$};
				\begin{scope}[shift={(3,0)}]
				\draw[->,densely dashed] (-1,0,0) -- (1,0,0);
				\draw[->,densely dashed] (0,-0.7,0) -- (0,1,0);
				\draw[->,densely dashed] (0,0,-1) -- (0,0,1);
				\node[] (u) at (1,.15,0) {$x$};
				\node[] (u) at (.15,1,0) {$t$};
				\node[] (u) at (.15,0,1) {$y$};
				\draw[thick,red] plot [smooth] coordinates{(-.55,-.4,0)(-.55,.9,0)};
				\draw[thick,blue] plot [smooth] coordinates{(.3,-.4,0)(.3,.9,0)};
				\node[thick,red] (u) at (-.55,.2,0) {\textbullet};
				\node[thick,blue] (u) at (.3,.52,0) {\textbullet};
				\end{scope}
				\end{tikzpicture}
				\caption{World lines of a system of two anyons.}
				\label{fig:worldlines}
			\end{figure}
		\noindent
		In the laboratory, these particles exist as quasiparticles arising as effective excitations in many-body systems which can be observed in, e.g., the fractional quantum hall effect (see e.g.\ \cite{FQH} for a review on this topic).
		How can these particles be detected? To measure their exotic statistics, one can measure scattering amplitudes:
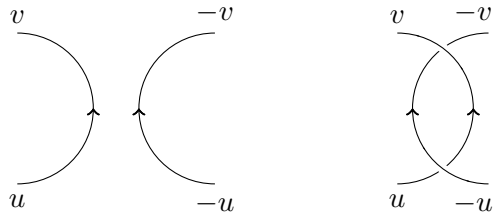
\begin{figure}[H]
\centering
\begin{tikzpicture}[scale=2,decoration={markings,mark=at position .5 with {\arrow[scale=1,thick]{>}}}] 
\node (u) at (0,-1) [below]{$u$};
\node (v) at (0,-0) [above]{$v$};
\node (mu) at (1.3,-1)[below]{$-u$};
\node (mv) at (1.3,-0)[above]{$-v$};
\begin{knot}
\strand[{postaction=decorate}] (0,-1) arc(270:450:.5);
\strand[{postaction=decorate}] (1.3,-1) arc(270:90:.5);
\end{knot}
\begin{scope}[shift={(2.5,0)}]
\begin{knot}[
flip crossing/.list={1} ]
\strand[{postaction=decorate}] (0,-1) arc(270:450:.5);
\strand[{postaction=decorate}] (0.6,-1) arc(270:90:.5);
\end{knot}
\node (u) at (0,-1) [below]{$u$};
\node (v) at (0,-0) [above]{$v$};
\node (mu) at (.5,-1)[below]{$-u$};
\node (mv) at (.5,-0)[above]{$-v$};\end{scope}
\end{tikzpicture}
\end{figure}

		The scattering amplitude $m(u\to v)$ has a dynamical part (which is intrinsic to each particle) and a topological part (that contains the information about the exchange statistics). By engineering the experiment in the right way, one can get rid of the dynamical part, such that the scattering amplitude only depends on the topological properties of worldlines of the anyons. Theoretically, these exchange statistics (or \myemph{braiding}) may be described by a braided monoidal category. More details about anyons can be found in~\cite{Kitaev2005}, which also includes a short discussion about their connection to categories, and in \cite{Rao17}. This example will be further analysed later in these notes, see \cref{ex:FibCat}.
	\end{ex}

	\begin{ex}[Topology of knots]
		Around the end of the 19th century, Sir William Thomson (1st Baron Kelvin) came up with the theory of vortex atoms \cite{Vortex1867}, hypothesising that an atom was a vortex in the aether. This theory was so popular by that time, a whole new branch of topology was developed, called knot theory. The theory of vortex atoms was discarded some decades later, but the theory of knots remained an interesting field of research until today. It has now applications in topological quantum field theory and quantum information theory (see e.g.~\cite{Knots} for an introduction to knot theory and applications in physics).
	
		Some interesting (yet simple) knots are listed below, for example the \myemph{unknot}, which is the neutral element in knot theory.
\begin{figure}[H]
\begin{minipage}[t]{.25\textwidth}
	\centering
\begin{tikzpicture}
    \draw [ultra thick] (0,0) circle (1.0cm);
    \draw[white](0,-1.5)--(0,-2);
\end{tikzpicture}				\subcaption{Unknot}
\end{minipage}
\hfill
\begin{minipage}[t]{.25\textwidth}
\centering
\begin{tikzpicture}[scale=0.75]
\begin{knot}[
 consider self intersections=true,
 flip crossing/.list={1,3}  ]
  \strand[ultra thick] (0,2) .. controls +(2.2,0) and +(120:-2.2) .. (210:2) .. controls +(120:2.2) and +(60:2.2) .. (-30:2) .. controls +(60:-2.2) and +(-2.2,0) .. (0,2);
\end{knot}
\end{tikzpicture}				\subcaption{Trefoil knot}
\end{minipage}
\hfill
\begin{minipage}[t]{.25\textwidth}
\centering
\begin{tikzpicture}[use Hobby shortcut]
\begin{knot}[
  consider self intersections=true,
  ignore endpoint intersections=false,
  flip crossing/.list={1,4,5}]
  \strand[ultra thick] ([closed]0,0) .. (1.5,-1) .. (.5,-2) .. (-.5,-1) .. (.5,0) .. (0,.5) .. (-.5,0) .. (.5,-1) .. (-.5,-2) .. (-1.5,-1) ;
\end{knot}
\draw[white](0,-2.5)--(0,-3);
\end{tikzpicture}
\subcaption{Figure-eight knot}				
\end{minipage}
		\caption{Typical knots.}
		\end{figure}
	
	How do braided monoidal categories come into play in knot theory? A typical question one can ask is whether two knots are equal. Topologically, the deformation of knots does not change the knot itself, which seems like a lot of freedom and so it should be easy to tell if two knots are the same. However, it is not trivial to tell if two knots are equivalent. In fact it is known to be \textbf{NP}-hard \cite{knotNP}, i.e.~the answer is unlikely, in general, to be found in a reasonable time as the size of the knot increases. Nevertheless, we can still find algorithms that can sometimes tell us whether two knots are equal (or unequal, respectively). Such algorithms can exploit the theory of \myemph{knot polynomials}, i.e.~they takes advantage of the fact that we can assign a polynomial to every knot (e.g.~the Jones polynomial~\cite{Jones85}). This works as follows:
		\begin{enumerate}
			\item Project the 3-dimensional knot onto the plane:
			
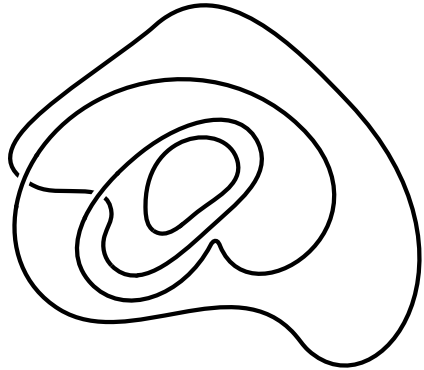
\begin{figure}[H]
\centering
\begin{tikzpicture}[use Hobby shortcut,scale=.9,
	rotate=-50
	]
\begin{knot}[
consider self intersections=true,
ignore endpoint intersections=false,
flip crossing/.list={}]
\strand[ultra thick] ([closed]0,0.4) .. (.2,1)..(0,2) .. (-.3,0.5);
\strand[ultra thick]
([closed]0,-.4) .. (.5,1)..(0,2.4) .. (-1,1)..
(0,-.8) .. (.7,0.9).. (.7,1).. (.8,1)..(0,3) .. (-2,1)..
(0,-1.4) .. (2.5,1).. (2.8,1)..(0.4,3.8)..(-2.3,2.5)..
(-2.3,2.2)..(-1.6,-.5)..(-.6,0.1)..(-.4,-0.2)
;
\end{knot}
\draw[white](0,-2.5)--(0,-3);
\end{tikzpicture}
\end{figure}
			\item Represent the knot as a \myemph{braid}, i.e.~an element of $\myhom(\mathbb{I},\mathbb{I})$:
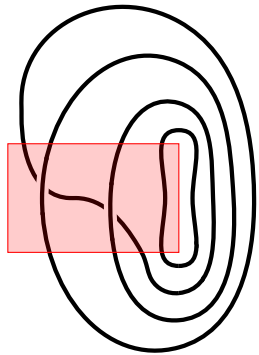
\begin{figure}[H]
\centering
\begin{tikzpicture}[use Hobby shortcut,scale=.9]
\begin{knot}[
consider self intersections=true,
ignore endpoint intersections=false,
flip crossing/.list={}]
\strand[ultra thick] ([closed]0,0) .. (.2,1)..(0,2) .. (-.2,1);
\strand[ultra thick]
([closed]0,-.4) .. (.5,1)..(0,2.4) .. (-1,1)..
(0,-.8) .. (.8,1)..(0,3) .. (-2,1)..
(0,-1.2) .. (1.1,1)..(0,3.6)..(-2.3,2.5)..
(-2.3,2.3)..(-2.3,2.2)..(-1.6,1)..(-1.5,1)..(-.6,0.3)..(-.5,0.1)..(-.4,-0.2)
;
\end{knot}
\draw [fill=red,red,fill opacity=0.2] (0,0.2) rectangle (-2.5,1.8); 
\end{tikzpicture}
\end{figure}

			\item Calculate the value of the diagram in your favourite braided monoidal category.
			\item Compare the values for the two knots in question. If you get different numbers, the knots are definitely unequal.
		\end{enumerate}
	\end{ex}

	\begin{defi}[Symmetric monoidal categories] A symmetric monoidal category is a braided monoidal category where the braiding satisfies $b_{X,Y} = b_{Y,X}^{-1}$.
	\end{defi}

	The definition of symmetric monoidal categories implies the following:
\begin{figure}[H]
\centering
\begin{tikzpicture}[ use Hobby shortcut,
decoration={markings,mark=at position .5 with {\arrow[scale=1,thick]{>}}}] 
\node (Y) at (1,.2)[right] {$Y$};
\node (X) at (0,.2)[left] {$X$};
\begin{knot}[flip crossing/.list={2}]
\strand[{postaction=decorate}] (0,0)to[out=90,in=225](.5,1)to[out=45,in=270](1,2)to[out=90,in=315](.5,3)to[out=135,in=270](0,4);
\strand[{postaction=decorate}] (1,0)to[out=90,in=315](.5,1)to[out=135,in=270](0,2)to[out=90,in=225](.5,3)to[out=45,in=270](1,4);
\end{knot}
\begin{scope}[shift={(3,0)}]
\node (a) at (-1,1.5) {$=$};
\node (Y) at (1,.2)[right] {$Y$};
\node (X) at (0,.2)[left] {$X$};
\begin{knot}[flip crossing/.list={}]
\strand[{postaction=decorate}] (0,0)--(0,4);
\strand[{postaction=decorate}] (1,0)--(1,4);\
\end{knot}
\end{scope}
\end{tikzpicture}
\end{figure}
\begin{figure}[H]
\centering
\begin{tikzpicture}[decoration={markings,mark=at position .75 with {\arrow[scale=1,thick]{>}}}] 
\node (Y) at (1,.2)[right] {$Y$};
\node (X) at (0,.2)[left] {$X$};
\begin{knot}[
flip crossing/.list={}  ]
\strand[{postaction=decorate}] (0,0) to[out=90,in=225](.5,1)to[out=45,in=270](1,2);
\strand[{postaction=decorate}] (1,0)to[out=90,in=315](.5,1)to[out=135,in=270](0,2);
\end{knot}
\begin{scope}[shift={(3,0)}]
\node (a) at (-1,1) {$=$};
\node (Y) at (1,.2)[right] {$Y$};
\node (X) at (0,.2)[left] {$X$};
\begin{knot}[
flip crossing/.list={1}  ]
\strand[{postaction=decorate}] (0,0) to[out=90,in=225](.5,1)to[out=45,in=270](1,2);
\strand[{postaction=decorate}] (1,0)to[out=90,in=315](.5,1)to[out=135,in=270](0,2);
\end{knot}
\end{scope}
\end{tikzpicture}
\end{figure}

	Regarding \cref{ex:Anyons}, these categories correspond to Bosons or Fermions (i.e.~particles with trivial braiding), so it is a notion for things that are \myemph{not} anyons or, in other words, trivial anyons.

\subsection{Compact Categories}

	In quantum mechanics, we are familiar with the concept of duals, e.g.~the corresponding dual space of a vector space. We can also talk about duals in categories: Consider for example \textbf{Hilb}, the category of finite-dimensional Hilbert spaces. Here, each object $X\in\mathbf{Hilb}$ has a dual object $X^*$ consisting of all linear operators $f:X\to\mathbb{I}$ (with the inner product given by $(A,B)\equiv \mathrm{tr}(A^* B)$), where the unit object $\mathbb{I}$ is $\mathbb{C}$. Hence, there is a linear operator 
		\begin{align*}
			e_X: X^*\otimes X&\to\mathbb{I}\\
			x\otimes f&\mapsto f(x)
		\end{align*}
	which we call the \myemph{counit} of $X$. This implies that the space of all linear operators $g:X\to Y\in \mathbf{Hilb}$ can be identified with $X^*\otimes Y$. Additionally, we can define an operator called the \myemph{unit} of $X$:
		\begin{align*}
			i_X:\mathbb{I}&\to X\otimes X^*\\
			c&\mapsto c \  \mathrm{id}_X
		\end{align*}
	that maps any complex number $c$ to the corresponding multiple of the identity operator over $X$.
	
	In quantum mechanics, say we have a Hilbert space $X$ that describes the states of a particle, then the dual space $X^*$ is the Hilbert space for the corresponding antiparticle. Feynman proposed the idea that antiparticles are particles going backwards in time. In terms of string diagrams, we can illustrate this by declaring that a string labelled by $X$ is equal to a string labelled by $X^*$ that goes \myemph{backwards in time}, i.e.~points down instead of up:
		\begin{figure}[H]
			\centering
			\begin{tikzpicture}[scale=1,decoration={markings,mark=at position .5 with {\arrow[scale=1,thick]{>}}}] 
				\draw[{postaction=decorate}] (0,0) to (0,-3) {};
				\draw[{postaction=decorate}] (2,-3) to (2,0) {};
				\node (=) at (1,-1.5) {$=$};
				\node (X*) at (-0.5,-1.5) {$X^*$};
				\node (X) at (2.5,-1.5) {$X$};
			\end{tikzpicture}
		\end{figure}
	\noindent
	Using this depiction, the unit and counit can be drawn as \myemph{cups} and \myemph{caps}:
		\begin{figure}[H]
\centering
\begin{tikzpicture}[scale=2, use Hobby shortcut,
  decoration={markings,mark=at position .5 with {\arrow[scale=1,thick]{>}}}] 
\node (u) at (0,.5) [left]{$X$};
\node (mv) at (1,.5)[right]{$X$};
\begin{knot}
\strand[{postaction=decorate}](1,0.5)arc(360:180:.5);
\end{knot}
\begin{scope}[shift={(2.5,0)}]
\begin{knot}
\strand[{postaction=decorate}](1,0)arc(0:180:.5);
\end{knot}
\node (u) at (0,0) [left]{$X$};
\node (mv) at (1,-0)[right]{$X$};
\end{scope}
\end{tikzpicture}
\end{figure}
		
	\noindent
	which can be thought of annihilation and creation of particle-antiparticle pairs. The unit and counit satisfy two equations that are called the \myemph{zig-zag equations}, which are fulfilled since the unit and counit are dual to each other:
\begin{figure}[H]
\centering
\begin{tikzpicture} 
\node (X) at (0,.2)[right] {$X$};
\node (x) at (4,.2)[right] {$X$};
\node (a) at (3,1) {$=$};
\begin{knot}[flip crossing/.list={1}, use Hobby shortcut]
\strand[{postaction=decorate},
  decoration={markings,mark=at position .25 with {\arrow[scale=1,thick]{<}},mark=at position .5 with {\arrow[scale=1,thick]{<}},mark=at position .75 with {\arrow[scale=1,thick]{<}}}] (0,0)--(0,1)arc(180:0:.5)arc(180:360:.5)--(2,2);
\strand[{postaction=decorate},
  decoration={markings,mark=at position .5 with {\arrow[scale=1,thick]{<}}}] (4,0)--(4,2);
\end{knot}
 \begin{scope}[shift={(6,0)}]
\node (Z) at (2,.2)[right] {$X$};
\node (x) at (4,.2)[right] {$X$};
\node (a) at (3,1) {$=$};
\begin{knot}[flip crossing/.list={}, use Hobby shortcut]
\strand[{postaction=decorate},
  decoration={markings,mark=at position .25 with {\arrow[scale=1,thick]{<}},mark=at position .5 with {\arrow[scale=1,thick]{<}},mark=at position .75 with {\arrow[scale=1,thick]{<}}}] (0,2)--(0,1)arc(180:360:.5)arc(180:0:.5)--(2,0);
\strand[{postaction=decorate},
  decoration={markings,mark=at position .5 with {\arrow[scale=1,thick]{<}}}] (4,2)--(4,0);
\end{knot}
\end{scope}
\end{tikzpicture}
\caption{Zig-zag equation.}
\label{fig:zigzag}
\end{figure}
	\noindent
	Writing these equations in terms of commutative diagrams with maps $l_X$ and $r_X$ and the associator $\alpha_{X,Y,Z}$ from \cref{def:monoidalcat} yields
		\begin{figure}[H]
			\centering
			\begin{tikzpicture} 
			\node (X-I) at (0,0) {$X^*\otimes \mathbb{I}$};
			\node (X-XX) at (4.5,0) {$X^*\otimes (X\otimes X^*)$};
			\draw[->,postaction={decorate}] (X-I) -- (X-XX) node[midway,above] {$\mathrm{id}_{X^*} \otimes i_X$};
			\node (XX-X) at (9,0) {$(X^*\otimes X)\otimes X^*$};
			\draw[->,postaction={decorate}] (X-XX) -- (XX-X) node[midway,above] {$\alpha^{-1}_{X^*,X,X^*}$};
			\node (I-X) at (9,-3) {$\mathbb{I}\otimes X^*$};
			\draw[->,postaction={decorate}] (XX-X) -- (I-X) node[midway,right] {$e_X \otimes \mathrm{id}_{X^*}$};
			\node (X) at (0,-3) {$X^*$};
			\draw[->,postaction={decorate}] (X-I) -- (X) node[midway,left] {$r_{X^*}^{-1}$};
			\draw[->,postaction={decorate}] (X) -- (I-X) node[midway,above] {$l_{X^*}$};
			\end{tikzpicture}
		\end{figure}
		\begin{figure}[H]
			\centering
			\begin{tikzpicture} 
			\node (X-I) at (0,0) {$\mathbb{I}\otimes X$};
			\node (X-XX) at (4.5,0) {$(X\otimes X^*) \otimes X$};
			\draw[->,postaction={decorate}] (X-I) -- (X-XX) node[midway,above] {$i_X \otimes \mathrm{id}_X$};
			\node (XX-X) at (9,0) {$X\otimes (X^* \otimes X)$};
			\draw[->,postaction={decorate}] (X-XX) -- (XX-X) node[midway,above] {$\alpha_{X,X^*,X}$};
			\node (I-X) at (9,-3) {$X \otimes \mathbb{I}$};
			\draw[->,postaction={decorate}] (XX-X) -- (I-X) node[midway,right] {$\mathrm{id}_{X} \otimes e_X$};
			\node (X) at (0,-3) {$X$};
			\draw[->,postaction={decorate}] (X-I) -- (X) node[midway,left] {$l_{X}^{-1}$};
			\draw[->,postaction={decorate}] (X) -- (I-X) node[midway,above] {$r_{X}$};
			\end{tikzpicture}
		\end{figure}
	\noindent
	Note that you cannot pinpoint each operation in the commutative diagrams to a definite location in the graphical representation. The graphical representation for example represents $I \otimes X$, $X \otimes I$ and $X$ in the same way.
	
	The unit and counit exist not only in the example of \textbf{Hilb}. We can also define them, for example, in the category \textbf{$2$Cob}: For $X$ being the circle, the unit and counit are
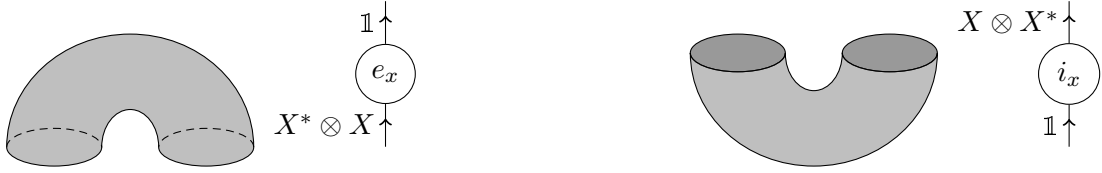
\begin{figure}[H]
\centering
\begin{subfigure}[b]{0.4\textwidth}
\centering
\begin{tikzpicture}[use Hobby shortcut,rotate=180,scale=0.5]
\filldraw[draw=black,fill=gray,fill opacity=0.5] (-3.25,0) arc (0:180:-3.25 and -3) --  (3.25,0) arc (0:180:1.25 and 0.5)-- (0.75,0) arc (0:180:0.75 and -1)-- (-0.75,0) arc (0:180:1.25 and 0.5);
\draw[densely dashed] (3.25,0) arc (0:180:1.25 and -0.5);
\draw[densely dashed] (-3.25,0) arc (0:180:-1.25 and -0.5);
\end{tikzpicture}
\begin{tikzpicture}[scale=1.1,decoration={markings,mark=at position 0.5 with {\arrow[scale=1,thick]{<}}}] 
\node (start) at (0,0) {};
\node (f) at (0,1) [circle,draw] {$e_x$};
\node (g) at (0,2) [] {};
\draw[postaction={decorate}] (f) -- (start) node[midway,left] {$X^*\otimes X$};
\draw[postaction={decorate}] (g) -- (f) node[midway,left] {$\mathbbm{1}$};
\end{tikzpicture}
\end{subfigure}
\hfill
\begin{subfigure}[b]{0.4\textwidth}
\centering
\begin{tikzpicture}[use Hobby shortcut,scale=0.5]
\filldraw[draw=black,fill=gray,fill opacity=0.8] (-2,0) ellipse (1.25 and 0.5);
\filldraw[draw=black,fill=gray,fill opacity=0.8] (2,0) ellipse (1.25 and 0.5);
\filldraw[draw=black,fill=gray,fill opacity=0.5] (-3.25,0) arc (0:180:-3.25 and -3) --  (3.25,0) arc (0:180:1.25 and -0.5)-- (0.75,0) arc (0:180:0.75 and -1)-- (-0.75,0) arc (0:180:1.25 and -0.5);
\end{tikzpicture}
\begin{tikzpicture}[scale=1.1,decoration={markings,mark=at position 0.5 with {\arrow[scale=1,thick]{<}}}] 
\node (start) at (0,0) {};
\node (f) at (0,1) [circle,draw] {$i_x$};
\node (g) at (0,2) [] {};
\draw[postaction={decorate}] (f) -- (start) node[midway,left] {$\mathbbm{1}$};
\draw[postaction={decorate}] (g) -- (f) node[midway,left] {$X\otimes X^*$};
\end{tikzpicture}
\end{subfigure}
\end{figure}
	
	\begin{defi}[Left duality]
		Let $\mathcal{C}$ be a monoidal category with unit object $\mathbb{I}$. It is a monoidal category with left duality if for each object $X\in\mathcal{C}$ there exists an object $X^*\in\mathcal{C}$ and morphisms	
			\begin{align*}
				i_X&:\mathbb{I}\to X\otimes X^*,\\
				e_X&:X^*\otimes X\to \mathbb{I},
			\end{align*}
		called the unit and counit, respectively, that satisfy the zig-zag equation (see \cref{fig:zigzag}):
			\begin{align*}
				\left(\mathrm{id}_{X}\otimes e_{X}\right)\circ	\left(i_{X}\otimes \mathrm{id}_{X}\right)&=\mathrm{id}_{X},\\
				\left(e_X\otimes\mathrm{id}_{X^*}\right)\circ\left(\mathrm{id}_{X^*}\otimes i_X\right)&=\mathrm{id}_{X^*}.
			\end{align*}
	\end{defi}	
	
	\begin{defi}[Right duality]
		There is also a notion of right duality: Let $\mathcal{C}$ again be a monoidal category with unit object $\mathbb{I}$, then we say that it exhibits right duality if for each object $X\in\mathcal{C}$ there exists an object $X^*\in\mathcal{C}$ and morphisms
			\begin{align*}
				i_X'&:\mathbb{I}\to X^*\otimes X,\\
				e_X'&:X\otimes X^*\to \mathbb{I},
			\end{align*}
		that satisfy
			\begin{align}
			\label{eq:conds}
				\begin{split}
				\left(e_X'\otimes\mathrm{id}_X\right)\circ \left(\mathrm{id}_X\otimes i_X'\right)&=\mathrm{id}_X,\\
				\left(\mathrm{id}_{X^*}\otimes e_X'\right)\circ \left(i_{X}'\otimes \mathrm{id}_{X^*}\right)&=\mathrm{id}_{X^*}.
				\end{split}
			\end{align}
	\end{defi}	
	
	\noindent
	There is a simple way to connect the two notions of duality: we can express the morphisms that appear in the definition of right duality as
		\begin{align}
		\label{eq:leftrightduality}
		\begin{split}
			i_X'&=\left(\mathrm{id}_{X^*}\otimes \theta_X\right)\circ b_{X,X^*}\circ i_X,\\
			e_X'&=e_X\circ b_{X,X^*}\circ\left(\theta_X\otimes \mathrm{id}_{X^*}\right).
		\end{split}
		\end{align}
	Using this definition, it can be shown that the conditions for right duality in \cref{eq:conds} are fulfilled (see \cite{QGroup}, Lemma XIV.3.4 and Proposition XIV.3.5).
	
	Finally, we can give the general definition of a compact category:

	\begin{defi}[Compact categories]
		\label{def:compact}
		A monoidal category is compact if every object $X\in\mathcal{C}$ has both a left and a right dual. These dualities let us also define duals of morphisms: for any morphism $f\in\myhom(X,Y)$ we define its dual $f^*\in\myhom(Y^*,X^*)$ by
			\begin{equation*}
				f^*=\left(e_Y\otimes\mathrm{id}_{X^*}\right)\circ\left(\mathrm{id}_{Y^*}\otimes f\otimes \mathrm{id}_{X^*}\right)\circ\left(\mathrm{id}_{Y^*}\otimes i_X\right).
			\end{equation*}
		Graphically, this is depicted as
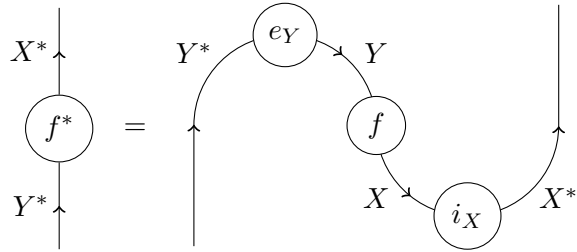
\begin{figure}[H]
\centering
\begin{tikzpicture}[use Hobby shortcut, scale=0.8,decoration={markings,mark=at position 0.25 with {\arrow[scale=1,thick]{>}},mark=at position 0.75 with {\arrow[scale=1,thick]{>}}}] 
\node (f*) at (0,0) [circle, draw] {$f^*$};
\draw[{postaction=decorate}]  (0,-2) -- (f*) node[midway,left] {$Y^*$}--(0,2) node[midway,left] {$X^*$};
\node (=) at (1.25,0) {$=$};
\end{tikzpicture}
\begin{tikzpicture}[use Hobby shortcut, scale=0.8,decoration={markings,mark=at position 0.15 with {\arrow[scale=1,thick]{>}},mark=at position 0.4 with {\arrow[scale=1,thick]{>}},mark=at position 0.6 with {\arrow[scale=1,thick]{>}},mark=at position 0.85 with {\arrow[scale=1,thick]{>}}}] 
\node (X) at (6,-1.2) {$X$};
\node (X*) at (9,-1.2) {$X^*$};
\node (Y) at (6,1.2) {$Y$};
\node (Y*) at (3,1.2) {$Y^*$};
\draw[{postaction=decorate}] (3,-2)--(3,0)arc(180:0:1.5)arc(180:360:1.5)--(9,2);
\node (f) at (6,0) [fill=white,circle,draw] {$f$};
\node (i) at (4.5,1.5) [fill=white,circle,draw] {$e_Y$};
\node (e) at (7.5,-1.5) [fill=white,circle,draw] {$i_X$};
\end{tikzpicture}
\end{figure}

	\end{defi}

	\subsection{Dagger Categories}

	So far, every statement we have made about \textbf{Hilb} did not require the existence of an inner product; indeed everything we said is also true for \textbf{Vect}, the category of all finite-dimensional \myemph{vector spaces} (and linear operators). Both of these categories are compact symmetric monoidal categories. However, the difference between \textbf{Hilb} and \textbf{Vect} will become clear after we have defined what is called a \myemph{dagger category}:

	\begin{defi}[Dagger category]
		A dagger category is a category $\mathcal{C}$ such that for any morphism $f:X\to Y$ for $X,Y\in\obj(\mathcal{C})$, there is a specified morphism $f^\dagger:Y\to X$ such that	
			\begin{align*}
				(g\circ f)^\dagger = f^\dagger\circ  g^\dagger
			\end{align*}
		for every pair of composable morphisms $f$ and $g$ and 
			\begin{align*}
				(f^\dagger)^\dagger = f
			\end{align*}
		for every morphism $f$.
	\end{defi}

	\begin{ex}[Hilbert spaces]
		To make \textbf{Hilb} a dagger category, we proceed as follows: For a given linear operator $f:X\to Y$, we can take its Hilbert space adjoint, which yields an operator $f^\dagger:Y \to X$ that is defined as
			\begin{equation*}
				\braket{f^\dagger \psi,\phi}=\braket{\psi,f\phi}
			\end{equation*}
		for all states $\phi\in X$ and $\psi\in Y$.
	\end{ex}

	\begin{ex}[Sets]
		The category \textbf{Set} is an example which we cannot turn into a dagger category. The problem is that there is a function $f:\{\}\to\{\bullet\}$ from the empty set to the set containing only one element, but no function the other way round. 
	\end{ex}

	\begin{ex} However, some of the examples we have already studied can be turned into dagger categories easily:
		\begin{itemize}
			\item In \textbf{Tang}$_k$ the dagger structure $f^\dagger:Y\to X$ can be obtained for any $k$ by reflecting $f:X\to Y$ in the vertical direction and then switching the orientations of arcs and circles. For instance, in \textbf{Tang}$_2$ with orientations:
				
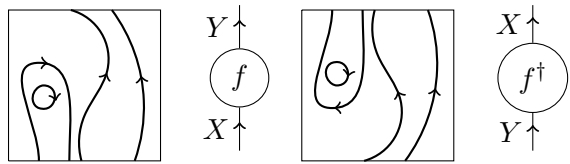
\begin{figure}[H]
\centering
\begin{tikzpicture}[use Hobby shortcut,scale=2,%
x={(1cm,0cm)},
y={(0cm,1cm)},
z={({0.5*cos(45)},{0.5*sin(45)})},decoration={markings,mark=at position 0.5 with {\arrow[scale=1,thick]{<}}}
]
\coordinate (A) at (0,0); 
\coordinate (B) at (1,0) ;
\coordinate (C) at (1,1); 
\coordinate (D) at (0,1); 
\node[left= 1pt of A]{};
\node[right= 1pt of B]{};
\node[right= 1pt of C]{};
\node[left= 1pt of D]{};
\draw[] (A)-- (B) -- (C) -- (D) -- (A);
\draw[thick,postaction={decorate}] (0.4,0)to[out=90,in=0](0.2,0.65)to[out=180,in=90](0.2,0);
\draw[thick,postaction={decorate}] (.68,1)..(.89,.4).. (.83,0);
\draw[thick,postaction={decorate}] (.41,1).. (.65,.6).. (.5,.3)..(.48,0);
\draw[thick,postaction={decorate},closed] (0.16,.4)..(0.2,.35).. (0.23,.5)..(0.2,.49) ;
\end{tikzpicture}
\begin{tikzpicture}[scale=1.1,decoration={markings,mark=at position 0.5 with {\arrow[scale=1,thick]{<}}}] 
\node (start) at (0,0) {};
\node (f) at (0,1) [circle,draw] {$f$};
\node (g) at (0,2) [] {};
\draw[postaction={decorate}] (f) -- (start) node[midway,left] {$X$};
\draw[postaction={decorate}] (g) -- (f) node[midway,left] {$Y$};
\end{tikzpicture}
\begin{tikzpicture}[yscale=-1, use Hobby shortcut, scale=2,%
x={(1cm,0cm)},
y={(0cm,1cm)},
z={({0.5*cos(45)},{0.5*sin(45)})},decoration={markings,mark=at position 0.5 with {\arrow[scale=1,thick]{>}}}
]
\coordinate (A) at (0,0); 
\coordinate (B) at (1,0) ;
\coordinate (C) at (1,1); 
\coordinate (D) at (0,1); 
\node[left= 1pt of A]{};
\node[right= 1pt of B]{};
\node[right= 1pt of C]{};
\node[left= 1pt of D]{};
\draw[] (A)-- (B) -- (C) -- (D) -- (A);
\draw[thick,postaction={decorate}] (0.4,0)to[out=90,in=0](0.2,0.65)to[out=180,in=90](0.2,0);
\draw[thick,postaction={decorate}] (.68,1)..(.89,.4).. (.83,0);
\draw[thick,postaction={decorate}] (.41,1).. (.65,.6).. (.5,.3)..(.48,0);
\draw[thick,postaction={decorate},closed] (0.16,.4)..(0.2,.35).. (0.23,.5)..(0.2,.49) ;
\end{tikzpicture}
\begin{tikzpicture}[scale=1.1,decoration={markings,mark=at position 0.5 with {\arrow[scale=1,thick]{<}}}] 
\node (start) at (0,0) {};
\node (f) at (0,1) [circle,draw] {$f^\dagger$};
\node (g) at (0,2) [] {};
\draw[postaction={decorate}] (f) -- (start) node[midway,left] {$Y$};
\draw[postaction={decorate}] (g) -- (f) node[midway,left] {$X$};
\end{tikzpicture}
\end{figure}
			\item In \textbf{$n$Cob} the strategy is the same: We obtain $f^\dagger: Y\to X$ by switching input and output of $f:X\to Y$ and then switching the orientation of each connected component of $f$:
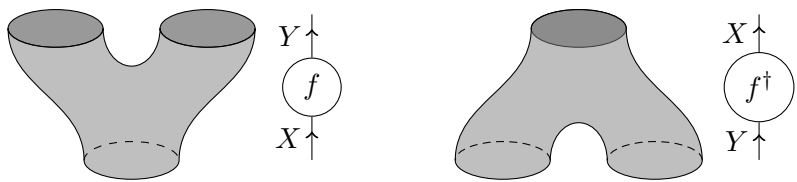
				\begin{figure}[H]
\centering
\begin{tikzpicture}[scale=0.5]
\gcobo
\end{tikzpicture}
\begin{tikzpicture}[scale=1.1,decoration={markings,mark=at position 0.5 with {\arrow[scale=1,thick]{<}}}] 
\node (start) at (0,0) {};
\node (f) at (0,1) [circle,draw] {$f$};
\node (g) at (0,2) [] {};
\draw[postaction={decorate}] (f) -- (start) node[midway,left] {$X$};
\draw[postaction={decorate}] (g) -- (f) node[midway,left] {$Y$};
\end{tikzpicture}
\hspace{35pt}
\begin{tikzpicture}[scale=0.5,yscale=-1]
\filldraw[draw=black,fill=gray,fill opacity=0.8] (-0,-3.5) ellipse (1.25 and 0.5);
\filldraw[draw=black,fill=gray,fill opacity=0.5] (-3.25,0) to[in=90,out=270] (-1.25,-3.5) arc (180:360:1.25 and 0.5) to[in=270,out=90] (3.25,0) arc (0:180:1.25 and 0.5)-- (0.75,0) arc (0:180:0.75 and -1)-- (-0.75,0) arc (0:180:1.25 and 0.5);
\draw[densely dashed](3.25,0) arc (0:180:1.25 and -0.5);
\draw[densely dashed](-0.75,0) arc (0:180:1.25 and -0.5);
\end{tikzpicture}
\begin{tikzpicture}[scale=1.1,decoration={markings,mark=at position 0.5 with {\arrow[scale=1,thick]{<}}}] 
\node (start) at (0,0) {};
\node (f) at (0,1) [circle,draw] {$f^\dagger$};
\node (g) at (0,2) [] {};
\draw[postaction={decorate}] (f) -- (start) node[midway,left] {$Y$};
\draw[postaction={decorate}] (g) -- (f) node[midway,left] {$X$};
\end{tikzpicture}
\end{figure}
		\end{itemize}
	\end{ex}


\section{Modular Tensor Categories}
\label{Ch4}

	The study of modular tensor categories is motivated by their applications in physical theories. For instance, a modular tensor category uniquely determines a $(2+1)d$-topological quantum field theory (TQFT) \cite{Tu1}. Furthermore, modular tensor categories find applications in topological quantum computing (see e.g.~\cite{TopQC,Toric,GolCh,NSSFD08}) and in the development of a physical theory of topological phases of matter (\cite{wilczek1990fractional,MooreRead,LevinWen,Kitaev2005}).

\subsection{Definition}

	So far, we have already encountered different types of categories and seen many physically relevant examples. Since the definition of a modular tensor category is rather long and abstract, we will go through every part of it in detail. The given definition follows the one in \cite{Rowell04} and also uses definitions given in \cite{QGroup}.

		\begin{defi}[Modular tensor category]
			\label{def:mtc}
			A modular tensor category is a category $\mathcal{C}$ that has the following properties:
			\begin{enumerate}
				\item It is monoidal (see \cref{def:monoidal}).
				\item It is braided (see \cref{def:braided}).
				\item It is compact (see \cref{def:compact}).
				\item It is balanced, which means that it is equipped with a twist, i.e.~a family of natural isomorphisms
						\begin{equation*}
							\theta_X\colon X\to X
						\end{equation*}
					that satisfy
						\begin{equation*}
							\theta_{X\otimes Y}=b_{Y,X}\circ b_{X,Y}\circ \left(\theta_X\otimes\theta_Y\right),
						\end{equation*}
		or graphically,
						
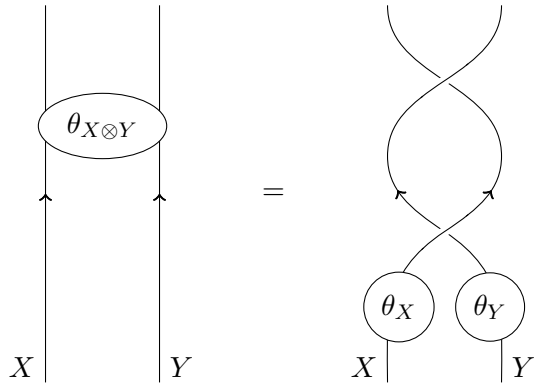
\begin{figure}[H]
\centering
\begin{tikzpicture}[xscale=1.5,decoration={markings,mark=at position .5 with {\arrow[scale=1,thick]{>}}}]
\node (b) at (2,2.5) {$=$};
\node (Y) at (1,.2)[right] {$Y$};
\node (X) at (0,.2)[left] {$X$};
\begin{knot}[flip crossing/.list={}]
\strand[{postaction=decorate}] (0,0)--(0,5);
\strand[{postaction=decorate}] (1,0)--(1,5);
\end{knot}
\node (c) at (0.5,3.4) [ellipse,draw,fill=white]{$\theta_{X\otimes Y}$};
\begin{scope}[shift={(3,0)}]
\node (Y) at (1,.2)[right] {$Y$};
\node (X) at (0,.2)[left] {$X$};
\begin{knot}[flip crossing/.list={2}]
\strand[{postaction=decorate}] (0,0)to[out=90,in=225](.5,2)to[out=45,in=270](1,3)to[out=90,in=315](.5,4)to[out=135,in=270](0,5);
\strand[{postaction=decorate}] (1,0)to[out=90,in=315](.5,2)to[out=135,in=270](0,3)to[out=90,in=225](.5,4)to[out=45,in=270](1,5);
\end{knot}
\node (c) at (0.1,1) [circle,draw,fill=white]{$\theta_X$};
\node (c) at (.9,1) [circle,draw,fill=white]{$\theta_Y$};
\end{scope}
\end{tikzpicture}
\end{figure}
					Another way to present this identity which displays the action of the twist slightly better is the following:
						
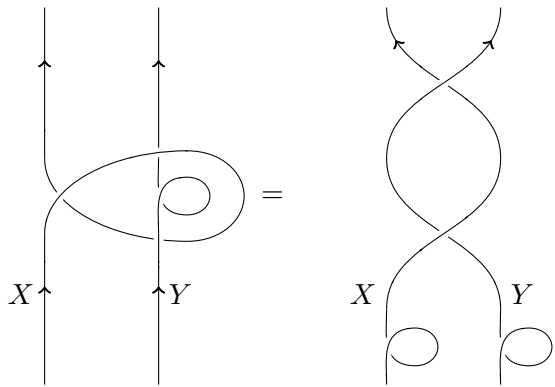
\begin{figure}[H]
\centering
\begin{tikzpicture}[decoration={markings,mark=at position .65 with {\arrow[scale=1,thick]{>}}},xscale=1.5]
\node (b) at (2,1.5) {$=$};
\node (Y) at (1,.2)[right] {$Y$};
\node (X) at (0,.2)[left] {$X$};
\begin{knot}[flip crossing/.list={4,5},consider self intersections=true,ignore endpoint intersections=false]
\strand[{postaction=decorate}] (0,-1)--(0,1);
\strand (0,1)to[out=90,in=180](1.25,2.1)to[out=0,in=90](1.75,1.5)to[out=270,in=0](1.25,.9)to[out=180,in=270](0,2);
\strand[{postaction=decorate}] (0,2)--(0,4);
\strand[{postaction=decorate}] (1,-1)--(1,1);
\strand (1,1)to[out=90,in=180](1.25,1.75)to[out=0,in=90](1.45,1.5)to[out=270,in=0](1.25,1.25)to[out=180,in=270](1,2);
\strand[{postaction=decorate}] (1,2)--(1,4);
\end{knot}
\begin{scope}[shift={(3,0)}]
\node (Y) at (1,.2)[right] {$Y$};
\node (X) at (0,.2)[left] {$X$};
\begin{knot}[flip crossing/.list={8},consider self intersections=true,ignore endpoint intersections=false]
\strand (0,0)to[out=90,in=225](.5,1)to[out=45,in=270](1,2)to[out=90,in=315](.5,3);
\strand[{postaction=decorate}] (.5,3)to[out=135,in=270](0,4);
\strand (1,0)to[out=90,in=315](.5,1)to[out=135,in=270](0,2)to[out=90,in=225](.5,3);
\strand[{postaction=decorate}] (.5,3)to[out=45,in=270](1,4);
\strand (1,-1)to[out=90,in=180](1.25,-.25)to[out=0,in=90](1.45,-.5)to[out=270,in=0](1.25,-.75)to[out=180,in=270](1,0);
\strand (0,-1)to[out=90,in=180](0.25,-.25)to[out=0,in=90](0.45,-.5)to[out=270,in=0](0.25,-.75)to[out=180,in=270](0,0);
\end{knot}
\end{scope}
\end{tikzpicture}
\end{figure}
					Equivalently, we can say that the following diagram commutes:
						\begin{figure}[H]
							\centering
							\begin{tikzpicture}[scale=1.5,decoration={markings,mark=at position 1 with {\arrow[scale=1,thick]{>}}}]
								\node (XoY) at (0,0) {$X\otimes Y$};
								\node (YoX) at (2,0) {$Y\otimes X$};
								\node (XoY2) at (0,-1.5) {$X\otimes Y$};
								\node (YoX2) at (2,-1.5) {$Y\otimes X$};
								\draw[{postaction=decorate}] (XoY)--(YoX) node[midway,above] {$b_{X,Y}$};
								\draw[{postaction=decorate}] (XoY)--(XoY2) node[midway,left] {$\theta_{X\otimes Y}$};
								\draw[{postaction=decorate}] (YoX)--(YoX2) node[midway,right] {$\theta_X\otimes \theta_Y$};
								\draw[{postaction=decorate}] (YoX2)--(XoY2) node[midway,above] {$b_{Y,X}$};
							\end{tikzpicture}
						\end{figure}
					Note that in the case $\theta_X=\mathrm{id}_X$, a balanced category is equivalent to a symmetric monoidal category, i.e.~$X\otimes Y\cong Y\otimes X$.
				\item The properties $1.-4.$ are compatible, i.e.
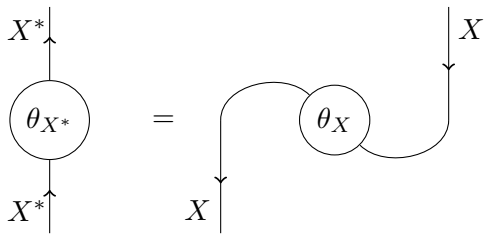
						\begin{equation*}
							\theta_{X^*}=\left(\theta_X\right)^*,
						\end{equation*}
						\begin{figure}[H]
							\centering
							\begin{tikzpicture}[scale=1.5,decoration={markings,mark=at position .5 with {\arrow[scale=1,thick]{<}}}]
							\node (X*) at (0,0) [circle,draw] {$\theta_{X^*}$};
							\draw[{postaction=decorate}] (0,1) to (X*);
							\draw[{postaction=decorate}] (X*) to (0,-1);
							\node (=) at (1,0) {$=$};
							\node (X) at (2.5,0) [circle,draw] {$\theta_X$};
							\draw (X)to[out=315,in=270](3.5,0);
							\draw[{postaction=decorate}] (3.5,0) to (3.5,1);
							\draw (1.5,0)to[out=90,in=135](X);
							\draw[{postaction=decorate}] (1.5,-1) -- (1.5,0);
							\node (X1) at (1.3,-0.8) {$X$};
							\node (X2) at (3.7,0.8) {$X$};
							\node (X*1) at (-0.2,0.8) {$X^*$};
							\node (X*2) at (-0.2,-0.8) {$X^*$};
							\end{tikzpicture}
						\end{figure}
			\end{enumerate}
			\myemph{Remark:} If properties $1.-5.$ are fulfilled, the category is called a \myemph{ribbon category}.
			\begin{enumerate}\setcounter{enumi}{5}
				\item All morphism spaces are $k$-vector spaces, and the composition and tensor product of morphisms are bilinear (this is called an $Ab$-category).
				\item It is \myemph{semisimple}, i.e.~every object $X\in\obj(\mathcal{C})$ is isomorphic to a finite direct sum of simple objects. An object $X$ is a simple object if and only if
					\begin{equation*}
						\mathrm{End}(X)=\myhom(X,X)\cong k\ \mathrm{id}_{X}
					\end{equation*}
				and
					\begin{equation*}
						\myhom(X,Y)=0\hspace{5pt}\mathrm{ for }\hspace{5pt}X\neq Y .
					\end{equation*}
				\item It is equipped with a $k$-linear trace of endomorphisms: Let $f\in\mathrm{End}(X)=\myhom(X,X)$ for any object $X\in\obj(\mathcal{C})$. The trace is defined as
					\begin{align*}
						\tr(f)&=e_X'\circ\left(f\otimes\mathrm{id}_{X^*}\right)\circ i_X\\
							&=e_X\circ b_{X,X^*}\circ \left(\theta_X\circ f\otimes\mathrm{id}_{X^*}\right)\circ i_X,
					\end{align*}
				where the second equality comes from the way we expressed $e_X'$ in \cref{eq:leftrightduality}. The right hand side of the equation is an element of $\mathrm{End}(\mathbb{I})\cong k$. Graphically, this definition of the trace corresponds to
				
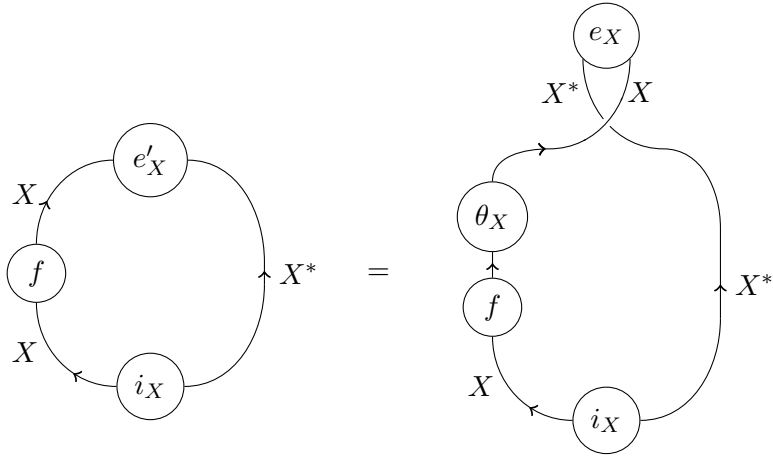
\begin{figure}[H]
				\centering
				\begin{tikzpicture}[scale=1.5,decoration={markings,mark=at position .7 with {\arrow[scale=1,thick]{<}}}]
				\node (f) at (0,0.3) [circle,draw] {$f$};
				\node (i') at (1,1.3) [circle,draw] {$e_X'$};
				\node (e) at (1,-0.7) [circle,draw] {$i_X$};
				\draw[{postaction=decorate}] (i') to [out=180,in=90] (f);
				\draw[{postaction=decorate}] (f) to [out=270,in=180] (e);
				\draw (e) to [out=0,in=270] (2,0.2);
				\draw (2,0.4) to [out=90,in=0] (i');
				\draw[{postaction=decorate}] (2,0.4) to (2,0.2);
				\node (X1) at (-0.1,1) {$X$};
				\node (X2) at (-0.1,-0.4) {$X$};
				\node (X*1) at (2.3,0.3) {$X^*$};
				\node (=) at (3,0.3) {$=$};
				\node (f2) at (4,0) [circle,draw] {$f$};
				\node (th) at (4,0.8) [circle,draw] {$\theta_X$};
				\node (e2) at (5,-1) [circle,draw] {$i_X$};
				\draw[{postaction=decorate}] (f2) to [out=270,in=180] (e2);
				\draw[{postaction=decorate}] (th) to (f2);
				\draw (e2) to [out=0,in=270] (6,-0.1);
				\draw[{postaction=decorate}] (6,0.7) to (6,-0.1);
				\node (i) at (5,2.4) [circle,draw] {$e_X$};
				\node (X3) at (3.9,-0.7) {$X$};
				\node (X5) at (5.3,1.9) {$X$};
				\node (X*2) at (4.6,1.9) {$X^*$};
				\node (X*3) at (6.3,0.2) {$X^*$};
				\begin{knot}[flip crossing/.list={3}]
				\strand[{postaction=decorate}] (i.south east)to[out=270,in=0](4.5,1.4)to[out=180,in=90](th.north);
				\strand (i.south west)to[out=270,in=180](5.5,1.4)to[out=0,in=90](6,.7);
				\end{knot}
				\end{tikzpicture}
				\end{figure}

				The value of $\tr(\mathrm{id}_X)$ is called \myemph{categorical dimension} (or \myemph{quantum dimension}) of $X$, denoted $\dim(X)$ or $d_X$.
				\item As a semisimple ribbon $Ab$-category, it has finitely many isomorphism classes of simple objects enumerated as $\left(X_0=\mathbb{I},X_1,X_2,\dots,X_n\right)$ and the so-called \myemph{S-matrix} is invertible. The S-matrix is defined as follows: Let
					\begin{equation*}
						\tilde{s}_{ij}=\tr\left(b_{X_j^*,X_i}\circ b_{X_i,X_j^*}\right),
					\end{equation*}
				which in string diagram notation is
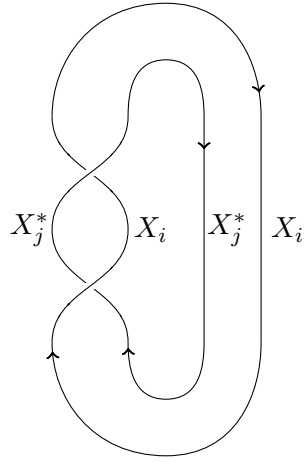
\begin{figure}[H]
\centering
\begin{tikzpicture}[decoration={markings,mark=at position .35 with {\arrow[scale=1,thick]{>}},mark=at position .99 with {\arrow[scale=1,thick]{>}}},yscale=.75]
\begin{knot}[flip crossing/.list={9},consider self intersections=true,ignore endpoint intersections=false]
\strand (0,0)to[out=90,in=225](.5,1)to[out=45,in=270](1,2)to[out=90,in=315](.5,3)to[out=135,in=270](0,4);
\strand (1,0)to[out=90,in=315](.5,1)to[out=135,in=270](0,2)to[out=90,in=225](.5,3)to[out=45,in=270](1,4);
\strand[{postaction=decorate}] (1,4)to[out=90,in=180](1.5,5)to[out=0,in=90](2,4)--(2,0)to[out=270,in=0](1.5,-1)to[out=180,in=270](1,0);
\strand[{postaction=decorate}]  (0,4)to[out=90,in=180](1.5,6)to[out=0,in=90](2.75,4)--(2.75,0)to[out=270,in=0](1.5,-2)to[out=180,in=270](0,0);
\end{knot}
\node (xj1) at (-0.3,2) {$X_j^*$};
\node (xi1) at (1.3,2) {$X_i$};
\node (xj2) at (2.3,2) {$X_j^*$};
\node (xi2) at (3.1,2) {$X_i$};
\end{tikzpicture}
\end{figure}
				Then the entries of the $S$-matrix are
					\begin{equation*}
						s_{ij}=\frac{1}{D}\tilde{s}_{ij}
					\end{equation*}
				with $D^2=\sum_{X_i} d_{X_i}^2$. Note that $S$ is a symmetric matrix.
			\end{enumerate}
		\end{defi}

\subsection{Fusion Rings and MTCs}

	Suppose we have a modular tensor category $\mathcal{C}$ whose (finitely many) isomorphism classes of simple objects are enumerated by $\{X_0=\mathbb{I},X_1,\dots,X_n\}$. Then these classes form a \myemph{fusion ring} under the operation $\otimes$, which is defined as follows \cite{Fuchs94}:

		\begin{defi}[Fusion ring]
			A fusion ring is a ring over the integers $\mathbb{Z}$ such that the following axioms are fulfilled:
				\begin{enumerate}
					\item Commutativity.
					\item Associativity.
					\item Positivity, i.e.~existence of a basis with non-negative structure constants is required.
					\item Conjugation, i.e.~there exists an element of the basis (required by $3.$), such that the evaluation of the product with respect to this basis yields an involutive automorphism.
				\end{enumerate}
		\end{defi}

	Note that a fusion ring $A$ is isomorphic to the lattice $\mathbb{Z}^{\dim A}$: Consider the basis
	\begin{equation*}
		\{\phi_j\,|\,j\in I\}
	\end{equation*}
	with $|I|=\dim A$ and $I$ an index set. In our case, we take $I=\{0,1,...n\}$ and $\phi_j\cong X_j$. We denote the structure constants in this basis by $N_{jk}^l$, i.e.~we write
		\begin{equation}
			\phi_j\otimes\phi_k = \sum_l N_{jk}^l\ \phi_l
			\label{eq:fusionrule}
		\end{equation}
	\noindent
	(these are also sometimes called a \myemph{fusion rule}). We will analyse the properties $1.- 4.$ in this context:
		\begin{enumerate}
			\item Commutativity means that the structure constants fulfil $N_{jk}^l = N_{kj}^l$.
			\item Associativity is given by
					\begin{align*}
						\left(\phi_j\otimes\phi_k\right)\otimes\phi_l&=\phi_j\otimes\left(\phi_k\otimes\phi_l\right)\\
						\left(\sum_m N_{jk}^m\ \phi_m\right)\otimes\phi_k&=\phi_j\otimes\left(\sum_n N_{kl}^n\ \phi_n\right)\\
						\sum_{m,n} N_{jk}^m N_{ml}^n\ \phi_n&=\sum_{m,n} N_{kl}^m N_{jm}^n\ \phi_n,
					\end{align*}
					i.e.~$\sum_{m,n} N_{jk}^m N_{ml}^n=\sum_{m,n} N_{kl}^m N_{jm}^n$.
			\item Positivity simply demands $N_{jk}^l\in\mathbb{Z}^+$.
			\item Conjugation is a bit more complicated. It demands, for the choice of basis we made in $3.$, the existence of an index $i_0\in I$ such that the conjugation map
					\begin{equation*}
						\phi_j\mapsto\sum_{k\in I}c_{jk}\phi_k \equiv \phi_j^*
					\end{equation*}
					with $c_{jk}=N_{jk}^{i_0}$ is an involution, i.e.~the matrix $C$ with entries $c_{jk}$ satisfies $C^2=\mathbbm{1}$. This implies that we can write this map as
					\begin{align*}
						\phi_i\mapsto\phi_i^*\equiv\phi_{i^*}\coloneqq\sum_{j\in I}c_{ij}\phi_j
					\end{align*}
					for some $i^*\in I$ which fulfils $\left(i^*\right)^*=i$. Secondly, $4.$ requires that this map is an automorphism, i.e.~it satisfies
					\begin{align*}
						\phi_{i^*}\otimes \phi_{j^*}=\sum_i N_{ij}^k\phi_{k^*},
					\end{align*}
					which yields $N_{i^* j^*}^{k^*}=N_{ij}^k$. $\phi_{i^*}$ is called the element conjugate of $\phi_i$ and $C$ the conjugation matrix.
		\end{enumerate}

	We now elaborate a bit more on the connection between the structure constants of a fusion ring (given by \cref{eq:fusionrule}) and the $S$-matrix of a modular tensor category (as given in \cref{def:mtc}). First, note that the first column of the $S$-matrix consists of the categorical dimension of the simple objects, i.e.~$S_{i,\mathbb{I}}=\tr(X_i)=\dim(X_i)$. This can easily be seen when we consider the string diagram notation of the matrix entries and the fact that strings with label $\mathbb{I}$ can be neglected:
	
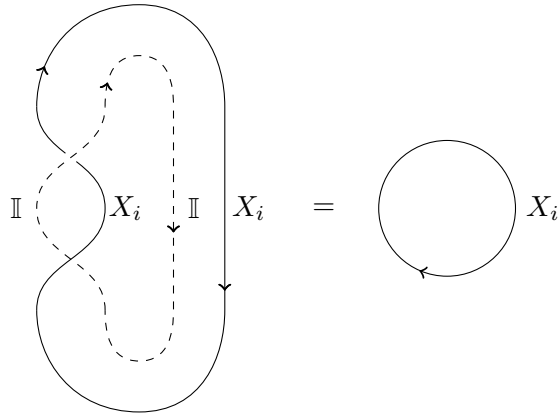
\begin{figure}[H]
	\centering
	\begin{tikzpicture}[scale=0.9,yscale=.75]
		\draw (0,0)to[out=90,in=225](.5,1)to[out=45,in=270](1,2)to[out=90,in=315](.5,3)to[out=135,in=270](0,4);
\node[circle,fill=white,scale=0.5] (w) at (.5,3) {};
\draw[ dashed] (1,0)to[out=90,in=315](.5,1)to[out=135,in=270](0,2)to[out=90,in=225](.5,3)to[out=45,in=270](1,4);
	\begin{knot}[flip crossing/.list={9},consider self intersections=true,ignore endpoint intersections=false]
\strand[{postaction=decorate},decoration={markings,mark=at position .05 with {\arrow[scale=1,thick]{>}},mark=at position .55 with {\arrow[scale=1,thick]{>}}}, dashed] (1,4)to[out=90,in=180](1.5,5)to[out=0,in=90](2,4)--(2,0)to[out=270,in=0](1.5,-1)to[out=180,in=270](1,0);
\strand[{postaction=decorate},decoration={markings,mark=at position .05 with {\arrow[scale=1,thick]{>}},mark=at position .6 with {\arrow[scale=1,thick]{>}}}]  (0,4)to[out=90,in=180](1.5,6)to[out=0,in=90](2.75,4)--(2.75,0)to[out=270,in=0](1.5,-2)to[out=180,in=270](0,0);
\end{knot}
\node (xj1) at (-0.3,2) {$\mathbb{I}$};
\node (xi1) at (1.3,2) {$X_i$};
\node (xj2) at (2.3,2) {$\mathbb{I}$};
\node (xi2) at (3.1,2) {$X_i$};
\node (=) at (4.2,2) {$=$};
\begin{scope}[yscale=1.333333333,shift={(0,-1.5)}]
\draw[{postaction=decorate},decoration={markings,mark=at position .7 with {\arrow[scale=1,thick]{<}}}] (6,3) circle (1cm);
\node (Xi3) at (7.4,3) {$X_i$};
\end{scope}
\end{tikzpicture}
\end{figure}

	\noindent
	Furthermore, since the endomorphism spaces of the simple objects $X_i\in\mathcal{C}$ are one-dimensional, the twists $\theta_{X_i}\in\mathrm{End}(X_i)$ are given by
		\begin{equation*}
			\theta_{X_i}=\theta_i\ \mathrm{id}_{X_i}
		\end{equation*}
	with scalars $\theta_i\in\mathbb{C}$. It can be shown (see~\cite{Bakalov2001}) that the entries of the $S$-matrix are entirely determined by the categorical dimensions (or fusion dimensions), the structure coefficients and the twists on the classes of simple objects:
		\begin{equation}
			\label{eq:Smatform}
			\tilde{s}_{ij}=\frac{1}{\theta_i\theta_j}\sum_k N_{i j^*}^k d_k\theta_k.
		\end{equation}
	The argument behind this formula is simple: it exploits the properties of the twist given in \cref{def:mtc}, i.e.
\begin{figure}[H]
\centering
\begin{tikzpicture}[decoration={markings,mark=at position .1 with {\arrow[scale=1,thick]{>}},mark=at position .65 with {\arrow[scale=1,thick]{>}}},yscale=.75]
\begin{knot}[flip crossing/.list={9},consider self intersections=true,ignore endpoint intersections=false]
\strand (0,0)to[out=90,in=225](.5,1)to[out=45,in=270](1,2)to[out=90,in=315](.5,3)to[out=135,in=270](0,4);
\strand (1,0)to[out=90,in=315](.5,1)to[out=135,in=270](0,2)to[out=90,in=225](.5,3)to[out=45,in=270](1,4);
\strand[{postaction=decorate}] (1,4)to[out=90,in=180](1.5,5)to[out=0,in=90](2,4)--(2,0)to[out=270,in=0](1.5,-1)to[out=180,in=270](1,0);
\strand[{postaction=decorate}]  (0,4)to[out=90,in=180](1.5,6)to[out=0,in=90](2.75,4)--(2.75,0)to[out=270,in=0](1.5,-2)to[out=180,in=270](0,0);
\end{knot}
\node (xj1) at (-0.3,2) {$X_j^*$};
\node (xi1) at (1.3,2) {$X_i$};
\node (xj2) at (2.3,2) {$X_j^*$};
\node (xi2) at (3.1,2) {$X_i$};
\node (=) at (4.2,2) {$=$};
\begin{scope}[shift={(5.5,0)},xscale=1.5]
\begin{knot}[flip crossing/.list={9},consider self intersections=true,ignore endpoint intersections=false]
\strand (0,0)--(0,4);
\strand (1,0)--(1,4);
\strand[{postaction=decorate}] (1,4)to[out=90,in=180](1.5,5)to[out=0,in=90](2,4)--(2,0)to[out=270,in=0](1.5,-1)to[out=180,in=270](1,0);
\strand[{postaction=decorate}]  (0,4)to[out=90,in=180](1.5,6)to[out=0,in=90](2.75,4)--(2.75,0)to[out=270,in=0](1.5,-2)to[out=180,in=270](0,0);
\end{knot}
\node (xj1) at (-0.3,2) {$X_i$};
\node (xi1) at (1.3,2) {$X_j^*$};
\node (xj2) at (2.3,2) {$X_j^*$};
\node (xi2) at (3,2) {$X_i$};
\node (f) at (0.5,.5) [ellipse,draw,fill=white] {$\theta_{X_i\otimes X_j^*}$};
\node (t) at (0,3.5) [circle,draw,fill=white] {$\theta_{X_i}^{-1}$};
\node (s) at (1,3.5) [circle,draw,fill=white] {$\theta_{X_j^*}^{-1}$};
\end{scope}
\end{tikzpicture}
\end{figure}
\noindent
	which yields, along with \cref{eq:fusionrule}:
		\begin{align*}
			\tilde{s}_{ij}&=\frac{1}{\theta_i\theta_j}\tr\left(\theta_{X_i\otimes X_j^*}\right)\\
				  &=\frac{1}{\theta_i\theta_j}\sum_k N_{ij^*}^k\tr\left(\theta_{X_k}\right)\\
				  &=\frac{1}{\theta_i\theta_j}\sum_k N_{ij^*}^k d_k\theta_k.
		\end{align*}

\subsection{Example: Fibonacci Anyons}

	We now study our first example of a modular tensor category. We have already described the nature of indistinguishable point-like particles in two dimensions, called anyons, in \cref{ex:Anyons}, but we did not give the exact category that represents them mathematically. The reason is that we had to give the definition of a modular category first, since these are exactly the categories that describe anyon models. Here we are going to study one simple example of a modular tensor category that describes an anyon model, the so-called \myemph{Fibonacci anyons}:

	\begin{ex}[Fibonacci category]
		\label{ex:FibCat}
		Fibonacci anyons describe the physics of the fractional quantum hall effect at filling factor $\nu=\frac{5}{2}$ (see e.g.~\cite{GolCh,FQHFib}). From the point of view of quantum information, Fibonacci anyons are interesting for topological quantum computation, since they allow for universal quantum computation: any circuit acting on $n$ qubits can be realized by using $4n$ physical Fibonacci anyons (\cite{LectureNotes,FreedmanStuff}). In this category, the simple objects are the \myemph{anyonic charges} or particle types that occur in the model, denoted $\{X_0=\mathbb{I}, X_1=\tau\}$ for the Fibonacci anyon model. We now go through the definition of a modular tensor category and see how the different properties are realized in the Fibonacci category.
			\begin{enumerate}
				\item The category is monoidal: The tensor product is given by the following set of fusion rules:
						\begin{align*}
							\mathbb{I}\otimes\mathbb{I}&= \mathbb{I}\\
							\mathbb{I}\otimes \tau &= \tau\otimes\mathbb{I}=\tau\\
							\tau\otimes\tau &= \mathbb{I}+\tau
						\end{align*}
					Now, the computation (braidings, twists, \dots) will occur within the space that encodes the various ways a set of anyons (i.e.~the tensor product of several objects) can fuse together according to the fusion rules.
					We can represent this graphically by using the string diagram notation: For instance, the tensor product $\tau\otimes\mathbb{I}$ corresponds to
						
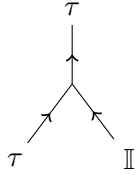
\begin{figure}[H]
							\centering
							\begin{tikzpicture}[scale=1,decoration={markings,mark=at position .5 with {\arrow[scale=1,thick]{>}}}]
								\node (tau1) at (-0.25,0) {$\tau$};
								\node (unit) at (1.25,0) {$\mathbb{I}$};
								\node (tau2) at (0.5,2) {$\tau$};
								\draw[{postaction=decorate}] (tau1) -- (0.5,1);
								\draw[{postaction=decorate}] (unit) -- (0.5,1);
								\draw[{postaction=decorate}] (0.5,1) -- (tau2);
							\end{tikzpicture}
						\end{figure}
					since it has only one outcome. These graphs are called \myemph{fusion trees}. They correspond to states in the \myemph{fusion space}, i.e.~a vector space over $\mathbb{C}$ that we assign to a tensor product with given outcome, which is denoted $V_{\tau\mathbb{I}}^\tau\equiv \hom(\tau\otimes \mathbb{I},\tau)$ in our example. In case a tensor product has several outcomes, e.g.~$\tau\otimes\tau$, we assign several fusion spaces to it, one for each outcome, which are $V_{\tau\tau}^\mathbb{I}\equiv \hom(\tau\otimes \tau,\mathbb{I})$ and $V_{\tau\tau}^\tau\equiv \hom(\tau\otimes \tau,\tau)$. Note that the fusion rules imply that each of these spaces is one-dimensional. There is a dual space to each fusion space, the so-called \myemph{splitting space}. States in the splitting space $V_\tau^{\tau\mathbb{I}}$, for example, are depicted
					as
						
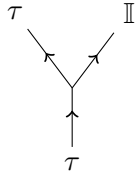
\begin{figure}[H]
							\centering
							\begin{tikzpicture}[scale=1,decoration={markings,mark=at position .5 with {\arrow[scale=1,thick]{<}}}]
								\node (tau1) at (-0.25,0) {$\tau$};
								\node (unit) at (1.25,0) {$\mathbb{I}$};
								\node (tau2) at (0.5,-2) {$\tau$};
								\draw[{postaction=decorate}] (tau1) -- (0.5,-1);
								\draw[{postaction=decorate}] (unit) -- (0.5,-1);
								\draw[{postaction=decorate}] (0.5,-1) -- (tau2);
							\end{tikzpicture}
						\end{figure}
					Splitting and fusion space can be used interchangeably.	Throughout the literature, the splitting space notation is widely used, although the expression \myemph{fusion space} is more popular.

					The vector spaces for tensor products of multiple objects can be built from these fusion and splitting spaces: Let $V_{ABC}^U\equiv \hom(A\otimes B\otimes C,U)$ be the tensor product of three objects $A,B,C$ fusing to the object $U$. This vector space can be constructed from the direct sum
						\begin{equation*}
							V_{ABC}^U=\bigoplus_{E}V_{AB}^E\otimes V_{EC}^U.
						\end{equation*}
					This construction corresponds to the parenthesis structure $\left(A\otimes B\right)\otimes C$, depicted as
						
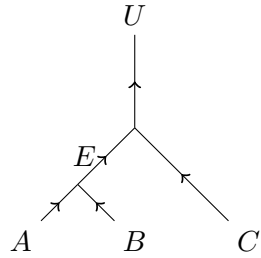
\begin{figure}[H]
							\centering
							\begin{tikzpicture}[scale=1.5,decoration={markings,mark=at position .5 with {\arrow[scale=1,thick]{>}}}]
								\node (A) at (0,0) {$A$};
								\node (B) at (1,0) {$B$};
								\node (C) at (2,0) {$C$};
								\node (D) at (1,2) {$U$};
								\draw[{postaction=decorate}] (A) -- (0.5,0.5);
								\draw[{postaction=decorate}] (B) -- (0.5,0.5);
								\draw[{postaction=decorate}] (C) -- (1,1);
								\draw[{postaction=decorate}] (0.5,0.5) -- (1,1) node[midway,left] {$E$};
								\draw[{postaction=decorate}] (1,1) -- (D);
							\end{tikzpicture}
						\end{figure}
					Equivalently, we can use a different parenthesis structure for this tensor product, namely $A\otimes\left( B\otimes C\right)$, which corresponds to
						\begin{equation*}
							V_{ABC}^U=\bigoplus_{F}V_{AF}^U\otimes V_{BC}^F
						\end{equation*}
					and is depicted as
						
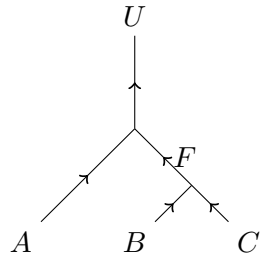
\begin{figure}[H]
							\centering
							\begin{tikzpicture}[scale=1.5,decoration={markings,mark=at position .5 with {\arrow[scale=1,thick]{>}}}]
								\node (A) at (0,0) {$A$};
								\node (B) at (1,0) {$B$};
								\node (C) at (2,0) {$C$};
								\node (D) at (1,2) {$U$};
								\draw[{postaction=decorate}] (A) -- (1,1);
								\draw[{postaction=decorate}] (B) -- (1.5,0.5);
								\draw[{postaction=decorate}] (C) -- (1.5,0.5);
								\draw[{postaction=decorate}] (1.5,0.5) -- (1,1) node[midway,right] {$F$};
								\draw[{postaction=decorate}] (1,1) -- (D);
							\end{tikzpicture}
						\end{figure}
					With this depiction, it becomes clear why the different parenthesis structures of the tensor product are not equal in the first place, so we need the associator that is defined below to switch between them.

					Before we define the associator, we want to show why this category is called \myemph{Fibonacci} category: Consider the tensor product of several objects, for instance, three $\tau$-objects:
						\begin{align*}
							\tau\otimes\tau\otimes\tau&=\tau\otimes\left(\mathbb{I}+\tau\right)\\
							&= \tau+ \left(\tau\otimes\tau\right)\\
							&= \tau+\left(\mathbb{I}+\tau\right)\\
							&= \mathbb{I}+ 2\tau
						\end{align*}
					When working out different tensor powers of $\tau$, one quickly recognizes a pattern:
						\begin{align*}
							\tau&= 0\mathbb{I}+ 1\tau\\
							\tau^{\otimes 2}&= 1\mathbb{I}+ 1\tau\\
							\tau^{\otimes 3}&= 1\mathbb{I}+ 2\tau\\
							\tau^{\otimes 4}&= 2\mathbb{I}+ 3\tau\\
							\tau^{\otimes 5}&= 3\mathbb{I}+ 5\tau
						\end{align*}
					The series of coefficients of $\tau$ is the \myemph{Fibonacci series}, hence the name of the category.

					As mentioned before, the associator is crucial for a monoidal category. This is a natural isomorphism
						\begin{equation}
							\label{eq:nattrans}
							\alpha_{X,Y,Z}:\left(X\otimes Y \right)\otimes Z\to X\otimes\left(Y\otimes Z\right),
						\end{equation}
					
					which is a family of maps, one map for each object $U$ the tensor product $X\otimes Y)\otimes Z$ is mapped to (see \cref{def:nattrans}). In the definitions and theorems we have presented the object the tensor product is mapped to is not specified, for instance in the definition of a monoidal category (\cref{def:monoidalcat}).
					In contrast to this, the concept of fusion and splitting spaces that we have introduced above is widely used in the description of anyon models (see e.g.\ \cite{Kitaev2005,Bonderson,IntFib}). Therefore, we need to translate the operators and equations that appear in the definition of a monoidal category to the language of fusion and splitting spaces. This works as follows: Instead of objects $(X\otimes Y)\otimes Z$ we now consider linear spaces
						\begin{equation*}
							V_U^{(X\otimes Y)\otimes Z}=\hom\left(U,(X\otimes Y)\otimes Z\right).
						\end{equation*}
					Here, we use the notation $V_U^{(X\otimes Y)\otimes Z}$ instead of $V_U^{XYZ}$ to make the parenthesis structure of the tensor product clear. Analogously to \cref{eq:nattrans} we are looking for a map
						\begin{equation*}
							\hom(U,(X\otimes Y)\otimes Z)\to\hom(U,X\otimes (Y\otimes Z)).
						\end{equation*}
					When we apply $\alpha_{X,Y,Z}$ to an element $f\in V_U^{(X\otimes Y)\otimes Z}$, this is exactly the space it is mapped to:
						\begin{equation*}
							\alpha_{X,Y,Z}\circ f\in\hom(U,(X\otimes Y)\otimes Z)=V_U^{X\otimes (Y\otimes Z)}.
						\end{equation*}
					Now choose a basis $\{\ket{j}\}\in V_U^{(X\otimes Y)\otimes Z}$ and a basis $\{\ket{k}\}\in V_U^{X\otimes(Y\otimes Z)}$. Vectors in the first space can be expressed as vectors in the latter one as follows:
						\begin{equation*}
							\ket{j}=\sum_j\left[\alpha_{X,Y,Z}\right]_{kj}\ket{k}\equiv\sum_k\left(F_U^{XYZ}\right)_{kj}\ket{j}.
						\end{equation*}
					Hence, it is now clear that the $F$-symbols, which are widely used in physics, are the matrix representation of the natural transformation $\alpha$, which is usually used in mathematical papers, for a specific basis. Graphically, the pentagon equation in \cref{fig:pentagon} then becomes
						\begin{figure}[H]
							\centering
							\begin{tikzpicture}[scale=.75,baseline=(current bounding box.center),,decoration={markings,mark=at position .5 with {\arrow[scale=1,thick]{>}}}]
								\node (u1) at (0,-0.25) {$U$};
								\node (x1) at (-1.5,2.25) {$X$};
								\node (y1) at (-0.5,2.25) {$Y$};
								\node (z1) at (0.5,2.25) {$Z$};
								\node (w1) at (1.5,2.25) {$W$};
								\draw[{postaction=decorate}] (0,0) -- (0,0.5);
								\draw[{postaction=decorate}] (0,0.5) -- (-0.5,1) node[pos=0.2,left] {$B\ $};
								\draw[{postaction=decorate}] (0,0.5) -- (1.5,2);
								\draw[{postaction=decorate}] (-0.5,1) -- (-1,1.5) node[pos=0.2,left] {$A\ $};
								\draw[{postaction=decorate}] (-0.5,1) -- (0.5,2);
								\draw[{postaction=decorate}] (-1,1.5) -- (-0.5,2);
								\draw[{postaction=decorate}] (-1,1.5) -- (-1.5,2);										
							\end{tikzpicture}
							\begin{tikzpicture}[scale=.9,baseline={([yshift=-6ex]current bounding box.center)}]
								\draw[->] (0,0) -- (2,1) node[pos=0.5,above] {$F_U^{AZW}$};
							\end{tikzpicture}
							\begin{tikzpicture}[scale=.75,baseline={([yshift=-12ex]current bounding box.center)},,decoration={markings,mark=at position .5 with {\arrow[scale=1,thick]{>}}}]
								\node (u1) at (0,-0.25) {$U$};
								\node (x1) at (-1.5,2.25) {$X$};
								\node (y1) at (-0.5,2.25) {$Y$};
								\node (z1) at (0.5,2.25) {$Z$};
								\node (w1) at (1.5,2.25) {$W$};
								\draw[{postaction=decorate}] (0,0) -- (0,0.5);
								\draw (0,0.5) -- (-0.5,1);
								\draw[{postaction=decorate}] (0,0.5) -- (1.5,2) node[pos=0.4,right] {$\ C$};
								\draw[{postaction=decorate}] (-0.5,1) -- (-1,1.5) node[pos=0.2,left] {$A\ $};
								\draw[{postaction=decorate}] (-1,1.5) -- (-0.5,2);
								\draw[{postaction=decorate}] (-1,1.5) -- (-1.5,2);
								\draw[{postaction=decorate}] (1,1.5) -- (0.5,2);
							\end{tikzpicture}			
							\begin{tikzpicture}[scale=.9,baseline={([yshift=-6ex]current bounding box.center)}]
								\draw[->] (0,0) -- (2,-1) node[pos=0.5,above] {$\ F_U^{XYC}$};
							\end{tikzpicture}
							\begin{tikzpicture}[scale=.75,baseline=(current bounding box.center),,decoration={markings,mark=at position .5 with {\arrow[scale=1,thick]{>}}}]
								\node (u1) at (0,-0.25) {$U$};
								\node (x1) at (-1.5,2.25) {$X$};
								\node (y1) at (-0.5,2.25) {$Y$};
								\node (z1) at (0.5,2.25) {$Z$};
								\node (w1) at (1.5,2.25) {$W$};
								\draw[{postaction=decorate}] (0,0) -- (0,0.5);
								\draw (0,0.5) -- (-0.5,1);
								\draw[{postaction=decorate}] (0,0.5) -- (1.5,2) node[pos=0.4,right] {$\ C$} node[pos=0.1,right] {$\ D$};
								\draw[{postaction=decorate}] (-0.5,1) -- (-1,1.5);
								\draw (-1,1.5) -- (-1.5,2);
								\draw[{postaction=decorate}] (1,1.5) -- (0.5,2);
								\draw[{postaction=decorate}] (0.5,1) -- (-0.5,2);
							\end{tikzpicture}\\
							\begin{tikzpicture}[scale=.9,baseline={([yshift=4ex]current bounding box.center)}]
								\draw[->] (0,0) -- (1,-1) node[pos=0.6,left] {$F_B^{XYZ}\ $};
							\end{tikzpicture}
							\begin{tikzpicture}[scale=.75,baseline={([yshift=15ex]current bounding box.center)},,decoration={markings,mark=at position .5 with {\arrow[scale=1,thick]{>}}}]
								\node (u1) at (0,-0.25) {$U$};
								\node (x1) at (-1.5,2.25) {$X$};
								\node (y1) at (-0.5,2.25) {$Y$};
								\node (z1) at (0.5,2.25) {$Z$};
								\node (w1) at (1.5,2.25) {$W$};
								\draw[{postaction=decorate}] (0,0) -- (0,0.5);
								\draw[{postaction=decorate}] (0,0.5) -- (-0.5,1) node[pos=0.2,left] {$B\ $};
								\draw[{postaction=decorate}] (0,0.5) -- (1.5,2);
								\draw (-0.5,1) -- (-1,1.5);
								\draw[{postaction=decorate}] (-0.5,1) -- (0.5,2) node[pos=0.2,right] {$\ E$};
								\draw[{postaction=decorate}] (-1,1.5) -- (-1.5,2);
								\draw[{postaction=decorate}] (0,1.5) -- (-0.5,2);
							\end{tikzpicture}		
							\begin{tikzpicture}[scale=.9,baseline={([yshift=15ex]current bounding box.center)}]
								\draw[->] (0,0) -- (2,0) node[pos=0.5,above] {$F_U^{XEW}$};
							\end{tikzpicture}
							\begin{tikzpicture}[scale=.75,baseline={([yshift=15ex]current bounding box.center)},,decoration={markings,mark=at position .5 with {\arrow[scale=1,thick]{>}}}]
								\node (u1) at (0,-0.25) {$U$};
								\node (x1) at (-1.5,2.25) {$X$};
								\node (y1) at (-0.5,2.25) {$Y$};
								\node (z1) at (0.5,2.25) {$Z$};
								\node (w1) at (1.5,2.25) {$W$};
								\draw[{postaction=decorate}] (0,0) -- (0,0.5);
								\draw (0,0.5) -- (-0.5,1);
								\draw[{postaction=decorate}] (0,0.5) -- (1.5,2) node[pos=0.1,right] {$\ D$};
								\draw[{postaction=decorate}] (-0.5,1) -- (-1,1.5);
								\draw (-1,1.5) -- (-1.5,2);
								\draw[{postaction=decorate}] (0.5,1) -- (-0.5,2)  node[pos=0.2,left] {$E\ $};
								\draw[{postaction=decorate}] (0,1.5) -- (0.5,2);
							\end{tikzpicture}
							\begin{tikzpicture}[scale=.9,baseline={([yshift=4ex]current bounding box.center)}]
								\draw[->] (0,-1) -- (1,0) node[pos=0.4,right] {$\ F_D^{YZW}$};
							\end{tikzpicture}		
							\label{fig:pentagonstrdiag}	
							\caption{Pentagon equation in terms of string diagrams.}
						\end{figure}
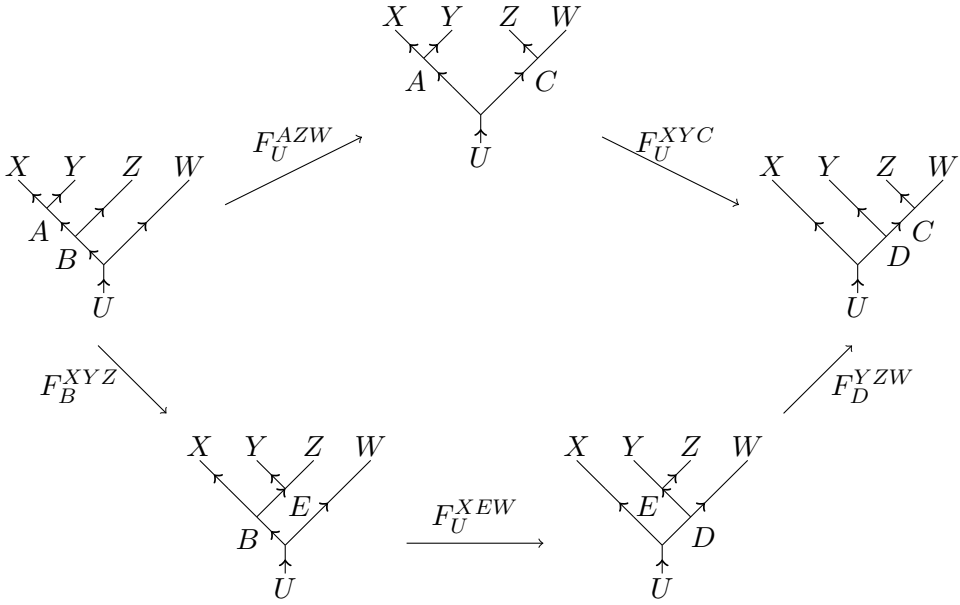
					Therefore, we will give the associator in terms of $F$-matrices $F_U^{XYZ}$ that act on splitting spaces $V_U^{XYZ}$ (this works analogously in the fusion space $V_{XYZ}^U$ and $F$-matrices $F_{XYZ}^U$). In the Fibonacci category, whenever one of the objects is the unit object $\mathbb{I}$, the $F$-matrix is given by the identity. The only non-trivial case is the one when $X=Y=Z=U=\tau$. This means that $\tau$ splits to $\tau\otimes \tau\otimes \tau$ (or, equivalently,  $\tau\otimes\tau\otimes\tau$ fuse to $\tau$) and it is given by a $2\times 2$ matrix:
						\begin{equation*}
							F_\tau^{\tau\tau\tau}=\begin{pmatrix}
								\frac{1}{\phi} & \frac{1}{\sqrt{\phi}}\\
								\frac{1}{\sqrt{\phi}} & -\frac{1}{\phi}
							\end{pmatrix}
						\end{equation*}
					where $\phi=\frac{1+\sqrt{5}}{2}$ is the golden ratio.
				\item The category is braided, i.e.~it has a family of natural isomorphisms
						\begin{equation*}
							b_{X,Y}:X\otimes Y\to Y\otimes X.
						\end{equation*}
					As explained above, we need to translate this operator to one that acts on a splitting or fusion space, which means it is a map
						\begin{align*}
							R_U^{XY}:V_U^{XY}&\to V_U^{YX}\\
							f&\mapsto b_{X,Y}\circ f,
						\end{align*}
					where $V_U^{XY}=\hom(U,X\otimes Y)$ and $V_U^{YX}=\hom(U,Y\otimes X)$. In terms of string diagrams the braiding operator is depicted as
						\begin{figure}[H]
						\centering
							\begin{tikzpicture}[scale=1,baseline={([yshift=15ex]current bounding box.center)},decoration={markings,mark=at position .8 with {\arrow[scale=1,thick]{>}}}]
							\node (x1) at (-.5,1.25) {$X$};
							\node (u1) at (0,-0.25) {$U$};
							\node (w1) at (.5,1.25) {$Y$};
							\draw[{postaction=decorate}] (0,0) -- (0,0.5);
							\draw[{postaction=decorate}] (0,0.5) -- (-0.5,1);
							\draw[{postaction=decorate}] (0,0.5) -- (.5,1);
							\end{tikzpicture}		
							\begin{tikzpicture}[scale=1,baseline={([yshift=15ex]current bounding box.center)}]
							\draw[->] (0,0) -- (2,0) node[pos=0.5,above] {$R_U^{XY}$};
							\end{tikzpicture}
							\begin{tikzpicture}[scale=1,baseline={([yshift=15ex]current bounding box.center)},decoration={markings,mark=at position .8 with {\arrow[scale=1,thick]{>}}}]
							\node (x1) at (-.5,3.25) {$Y$};
							\node (u1) at (0,-0.25) {$U$};
							\node (w1) at (.5,3.25) {$X$};
							\draw[{postaction=decorate}] (0,0) -- (0,0.5);
							\draw[{postaction=decorate}] (0,0.5) -- (-0.5,1);
							\draw[{postaction=decorate}] (0,0.5) -- (.5,1);
							\begin{scope}[shift={(-.5,1)}]
							\begin{knot}[
							flip crossing/.list={}  ]
							\strand[{postaction=decorate}] (0,0) to[out=90,in=225](.5,1)to[out=45,in=270](1,2);
							\strand[{postaction=decorate}] (1,0)to[out=90,in=315](.5,1)to[out=135,in=270](0,2);
							\end{knot}
							\end{scope}
							\node (x1) at (-.5,.6) {$X$};
							\node (w1) at (.5,.6) {$Y$};
							\end{tikzpicture}
						\caption{Graphical description of the braiding operator.}
						\end{figure}
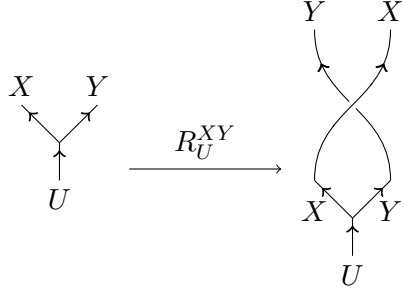
					In the Fibonacci category, braiding always yields a complex phase since all non-trivial vector spaces $V_U^{XY}$ are one-dimensional. The braiding operator is trivial (i.e.\ the identity) if one of the objects $X,Y$ is the unit object $\mathbb{I}$. If $X=Y=\tau$, braiding is given by \cite{Wang10}
						\begin{align*}
							R^{\tau\tau}_\mathbb{I}=e^{-\frac{4\pi i}{5}},\\
							R^{\tau\tau}_\tau=e^{\frac{3\pi i}{5}}.
						\end{align*}
				\item It is compact: Both objects are their own left and right dual, as can be seen from the fusion rules:
						\begin{align*}
							\tau\otimes\tau&=\mathbb{I}+\tau,\\
							\mathbb{I}\otimes\mathbb{I}&=\mathbb{I}.
						\end{align*}
					This point is important for expressing the notion of conjugate anyonic charges (i.e.~conjugate particles).
				\item It is balanced: As stated in part $7$ of the definition, the endomorphism spaces of the simple objects are one-dimensional, hence the twist operators are~\cite{Wang10}
						\begin{align*}
							\theta_\tau&=\vartheta_\tau\ \mathrm{id}_\tau=e^{\frac{4\pi i}{5}}\mathrm{id}_\tau,\\
							\theta_\mathbb{I}&=\vartheta_\mathbb{I}\ \mathrm{id}_\mathbb{I}=\mathrm{id}_\mathbb{I}.
						\end{align*}
					We use $\vartheta_X$ to denote the scalar corresponding to the twist operator $\theta_X$ to avoid confusion.
				\item The properties $1.-4.$ are compatible:
							\begin{figure}[H]
								\centering
								\begin{tikzpicture}[scale=1.5,decoration={markings,mark=at position .5 with {\arrow[scale=1,thick]{<}}}]
								\node (X*) at (0,0) [circle,draw] {$\theta_{\tau^*}$};
								\draw[{postaction=decorate}] (0,1) to (X*);
								\draw[{postaction=decorate}] (X*) to (0,-1);
								\node (X*1) at (-0.2,0.8) {$\tau^*$};
								\node (X*2) at (-0.2,-0.8) {$\tau^*$};
								\begin{scope}[shift={(1,0)},xscale=.5]
								\node (=) at (-.5,0) {$=$};
								\draw[{postaction=decorate}] (1,1) to (1,-1);
								\node (X*1) at (.5,0) {$\vartheta_\tau^*$};
								\node (X*2) at (1.5,0) {$\tau^*$};
								\end{scope}
								\begin{scope}[shift={(2.5,0)},xscale=.5]
								\node (=) at (-.5,0) {$=$};
								\draw[{postaction=decorate}] (1,-1) to (1,1);
								\node (X*1) at (.5,0) {$\vartheta_\tau$};
								\node (X*2) at (1.5,0) {$\tau$};
								\end{scope}
								\begin{scope}[shift={(4.5,0)},xscale=.5]
								\node (=) at (-1,0) {$=$};
								\draw (1,-.3)to[out=270,in=270](2,-.3);
								\draw[{postaction=decorate}] (2,-.3) to (2,1);
								\draw (0,.3)to[out=90,in=90](1,.3);
								\draw[{postaction=decorate}] (0,-1) -- (0,.3);
								\draw (1,.3)--(1,-.3);
								\node (X1) at (-.2,-0.8) {$\tau$};
								\node (X2) at (2.2,0.8) {$\tau$};
								\node (X) at (1,0) [circle,draw,fill=white] {$\theta_\tau$};
								\end{scope}
								\end{tikzpicture}
							\end{figure}
					For the unit object $\mathbb{I}$, this is trivial.
				\item From the construction in $1.$, it is clear that the morphism spaces are $\mathbb{C}$-linear vector spaces. We will now take a closer look at how the composition and the tensor product of morphisms work: Diagrammatically, taking the tensor product of two morphisms is the horizontal
				disjoint union of the diagrams. For instance, take the tensor product of the morphisms $f:\tau\otimes\tau\to\mathbb{I}$ and $g:\tau\to\tau$, the fusion of two $\tau$-type anyons to the unit object and the identity on a $\tau$-object:
					
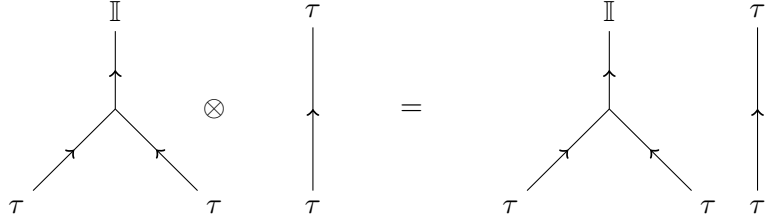
\begin{figure}[H]
						\centering
						\begin{tikzpicture}[scale=1.3,decoration={markings,mark=at position .5 with {\arrow[scale=1,thick]{>}}}]
							\node (A) at (0,0) {$\tau$};
							\node (C) at (2,0) {$\tau$};
							\node (D) at (1,2) {$\mathbb{I}$};
							\draw[{postaction=decorate}] (A) -- (1,1);
							\draw[{postaction=decorate}] (C) -- (1,1);
							\draw[{postaction=decorate}] (1,1) -- (D);
							\node (D) at (2,1) {$\otimes$};
							\begin{scope}[shift={(2,0)}]
							\node (C) at (1,0) {$\tau$};
							\node (D) at (1,2) {$\tau$};
							\draw[{postaction=decorate}] (C) -- (D);
							\end{scope}
							\begin{scope}[shift={(5,0)}]
							\node (D) at (-1,1) {$=$};
							\node (A) at (0,0) {$\tau$};
							\node (C) at (2,0) {$\tau$};
							\node (D) at (1,2) {$\mathbb{I}$};
							\draw[{postaction=decorate}] (A) -- (1,1);
							\draw[{postaction=decorate}] (C) -- (1,1);
							\draw[{postaction=decorate}] (1,1) -- (D);
							\begin{scope}[shift={(1.5,0)}]
							\node (C) at (1,0) {$\tau$};
							\node (D) at (1,2) {$\tau$};
							\draw[{postaction=decorate}] (C) -- (D);
							\end{scope}
							\end{scope}
						\end{tikzpicture}
					\end{figure}
			The composition of two morphisms corresponds to vertical stacking of the diagrams, e.g.\ 
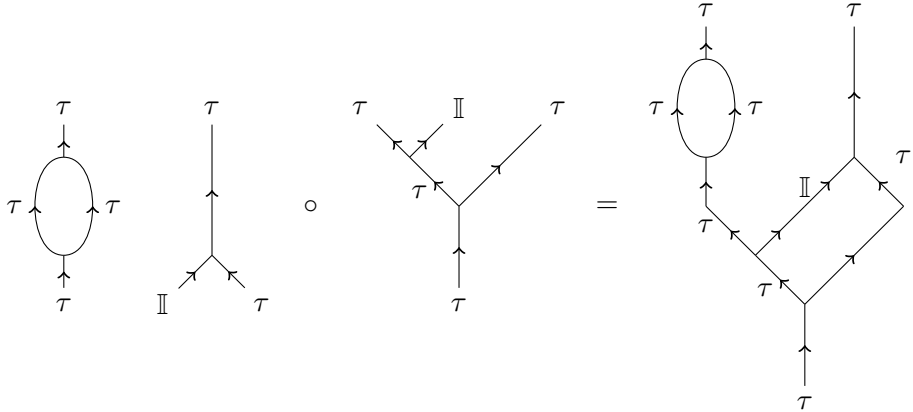
				\begin{figure}[H]
					\centering
					\begin{tikzpicture}[scale=1.3,decoration={markings,mark=at position .5 with {\arrow[scale=1,thick]{>}}}]
					\node (C) at (0.5,1) {$\tau$};
					\node (D) at (1.5,1) {$\tau$};
					\node (C) at (1,0) {$\tau$};
					\node (D) at (1,2) {$\tau$};
					\draw[{postaction=decorate}] (1,.5) to[out=180,in=180] (1,1.5);
					\draw[{postaction=decorate}] (1,.5) to[out=0,in=0] (1,1.5);
					\draw[{postaction=decorate}] (C) -- (1,.5);\draw[{postaction=decorate}] (1,1.5) -- (D);\begin{scope}[shift={(2,0)}]
					\node (A) at (0,0) {$\mathbb{I}$};
					\node (C) at (1,0) {$\tau$};
					\node (D) at (.5,2) {$\tau$};
					\draw[{postaction=decorate}] (A) -- (.5,.5);
					\draw[{postaction=decorate}] (C) -- (.5,.5);
					\draw[{postaction=decorate}] (.5,.5) -- (D);
					\node (D) at (1.5,1) {$\circ$};
					\end{scope}
					\begin{scope}[shift={(4,0)}]
					\node (A) at (0,2) {$\tau$};
					\node (C) at (2,2) {$\tau$};
					\node (D) at (1,0) {$\tau$};
					\node (B) at (1,2) {$\mathbb{I}$};
					\node (F) at (.6,1.15) {$\tau$};
					\draw[{postaction=decorate}] (.5,1.5)--(A);
					\draw[{postaction=decorate}] (1,1)--(C);
					\draw[{postaction=decorate}] (D)--(1,1);
					\draw[{postaction=decorate}] (.5,1.5)--(B);
					\draw[{postaction=decorate}] (1,1)--(.5,1.5);
					\end{scope}
					\node (D) at (6.5,1) {$=$};
					\begin{scope}[shift={(7.5,-1)}]
					\node (A) at (1,0) {$\tau$};
					\node (F) at (.6,1.15) {$\tau$};
					\draw[{postaction=decorate}] (.5,1.5)--(0,2);
					\draw[{postaction=decorate}] (1,1)--(2,2);
					\draw[{postaction=decorate}] (A)--(1,1);
					\draw[{postaction=decorate}] (.5,1.5)--(1,2);
					\draw[{postaction=decorate}] (1,1)--(.5,1.5);
					\end{scope}
					\begin{scope}[shift={(6.5,1)}]
					\node (C) at (0.5,1) {$\tau$};
					\node (D) at (1.5,1) {$\tau$};
					\node (C) at (1,-.2) {$\tau$};
					\node (D) at (1,2) {$\tau$};
					\draw[{postaction=decorate}] (1,.5) to[out=180,in=180] (1,1.5);
					\draw[{postaction=decorate}] (1,.5) to[out=0,in=0] (1,1.5);
					\draw[{postaction=decorate}] (1,0) -- (1,.5);\draw[{postaction=decorate}] (1,1.5) -- (D);\begin{scope}[shift={(2,0)}]
					\node (A) at (0,.2) {$\mathbb{I}$};
					\node (C) at (1,.5) {$\tau$};
					\node (D) at (.5,2) {$\tau$};
					\draw[{postaction=decorate}] (0,0) -- (.5,.5);
					\draw[{postaction=decorate}] (1,0) -- (.5,.5);
					\draw[{postaction=decorate}] (.5,.5) -- (D);
					\end{scope}\end{scope}
					\end{tikzpicture}
				\end{figure}
			These operations are bilinear by construction.

				\item The category is semisimple: All objects in this category can be expressed as a sum of simple objects. We have seen examples of this above, e.g.
						\begin{align*}
							\tau\otimes\tau\otimes\tau= \mathbb{I}+ 2\tau.
						\end{align*}
				\item It is equipped with a $\mathbb{C}$-linear trace as defined in \cref{def:mtc}. Furthermore, the quantum dimensions are given by
						\begin{align*}
							d_\tau&=\phi,\\
							d_\mathbb{I}&=1.
						\end{align*}
				\item The isomorphism classes are labelled by $\mathbb{I}$ and $\tau$, and the $S$-matrix can be calculated using \cref{eq:Smatform}:
						\begin{equation*}
							S=\frac{1}{\sqrt{2+\phi}}\begin{pmatrix}
								1 & \phi\\
								\phi & -1
							\end{pmatrix}
						\end{equation*}
						which is invertible with the inverse
						\begin{equation*}
							S^{-1}=\sqrt{2+\phi}\begin{pmatrix}
								\frac{1}{\phi ^2+1} & \frac{\phi }{\phi ^2+1}\\
								\frac{\phi}{\phi ^2+1} & -\frac{1}{\phi ^2+1}
							\end{pmatrix}.
						\end{equation*}

						Why is it important that this matrix is invertible? We can find the answer in its definition: The entries of the S-matrix are defined via the braiding operator, i.e.
							\begin{equation*}
								\tilde{s}_{ij}=\tr\left(b_{X_j^*,X_i}\circ b_{X_i,X_j^*}\right).
							\end{equation*}
						The invertibility of this matrix ensures that it has full rank. What does this imply for the braiding operation? It ensures that for each label $X\neq\mathbb{I}$, there is some label $A$ such that $b_{AX}\circ b_{XA}$ is not the identity. If this is fulfilled, braiding is said to be \myemph{nondegenerate} and it is exactly what distinguishes anyonic systems from fermions and bosons.
			\end{enumerate}
		Next, we will study the fusion ring corresponding to the Fibonacci modular tensor category in more detail. We proceed in the same way as we did when studying the lattice corresponding to a fusion ring: Since our simple objects are $\{X_0=\mathbb{I}, X_1=\tau\}$, we have two basis elements, $\phi_0=\mathbb{I}$ and $\phi_1=\tau$. The fusion rules are similar to the way the tensor product was given above:
			\begin{align*}
				\phi_0\times\phi_0&=\phi_0\\
				\phi_1\times\phi_0&=\phi_1=\phi_0\times\phi_1\\
				\phi_1\times\phi_1&=\phi_0+\phi_1.
			\end{align*}
		Hence, the structure constants are given by
			\begin{align*}
				&N_{00}^0=1, N_{00}^1=0\\
				&N_{11}^0=1, N_{11}^1=1\\
				&N_{10}^0=0, N_{10}^1=1\\
				&N_{01}^0=0, N_{01}^1=1.
			\end{align*}
		We now check whether the axioms of a fusion ring are fulfilled:
			\begin{enumerate}
				\item Commutativity: Given the structure constants, it is obvious that $N_{jk}^l=N_{kj}^l$.
				\item Associativity: We check one example:
						\begin{align*}
							\sum_l N_{01}^l N_{l1}^1&=\sum_l N_{11}^l N_{0l}^1\\
							N_{01}^0 N_{01}^1+N_{01}^1 N_{11}^1&=N_{11}^0 N_{00}^1+N_{11}^1 N_{01}^1\\
							1&=1
						\end{align*}
					The remaining ones work analogously.
				\item Positivity: This axiom is fulfilled by construction, since we have only used positive structure constants.
				\item Conjugation: To prove that this axiom is fulfilled, we need to find a label $i_0$ such that $c_{jk}\equiv N_{jk}^{i_0}$ satisfies $C^2=\mathbbm{1}$. Choose $i_0=0$:
						\begin{align*}
							C=\begin{pmatrix}
								1 & 0\\
								0 & 1
							\end{pmatrix}
							\Rightarrow C^2=\mathbbm{1}.
						\end{align*}
			\end{enumerate}
	\end{ex}

	We have now seen an example of a specific modular tensor category and have investigated its connection to fusion rings. More information about Fibonacci anyons can be found in \cite{IntFib} and in the references we have already mentioned. There are also many more interesting examples to study, for instance other anyon models like Ising anyons.

\section{Outlook and Further Reading}
\label{Ch5}

There have been many attempts to investigate the connection between categories and physics. We have already mentioned \cite{Baez2011}, where the authors make an effort to create a network of analogies between physics, topology, logic and computation by using the language of category theory. 

There is also a mature body of work that aims at reconstructing finite-dimensional quantum theory purely from postulates in category-theoretic terms with an emphasis on the diagrammatic description, see for example \cite{CK17} and \cite{SSC18}. Different aspects of the usage of categories in physics are also described in \cite{Coecke2011}, together with an introductory chapter on categories in general for readers from different backgrounds.

Introductions to category theory that focus purely on mathematical aspects can be found in \cite{Awodey,Leinster} and, of course, in the classical book on category theory \cite{MacLane} from Saunders Mac~Lane, who was one of the inventors of this field.

More information about modular tensor categories can be found in \cite{RSW09} and \cite{Mue03}.

\section*{Acknowledgements}
This work was supported by the DFG through SFB 1227 (DQ-mat) and the RTG 1991.

\bibliographystyle{alpha}
\bibliography{Literature}

\end{document}